\documentstyle [peer,12pt] {article}
\newcommand{\bye}{\end{document}}

\newcommand{\baufg}{\begin{description}}
\newcommand{\eaufg}{\end{description}}
\newcommand{\bteilaufg}{\begin{description}}
\newcommand{\eteilaufg}{\end{description}}

\newcommand{\be}{\begin{equation}}
\newcommand{\ee}{\end{equation}}
\newcommand{\bes}{\begin{eqnarray}}
\newcommand{\ees}{\end{eqnarray}}
\newcommand{\bma}{\left( \begin {array}}
\newcommand{\ema}{\end {array} \right)}
\newcommand{\pslash}{\kern 0.2 em p\kern -0.45em /}
\newcommand{\dslash}{\kern 0.2 em \delta\kern -0.45em /}
\newcommand{\sla}[1]{\kern 0.2 em #1\kern -0.45em /}

\newcommand{\bt}{\begin{tabbing}
            \hskip 7.1 true cm \=\hskip 7.1 true cm \kill}
\newcommand{\et}{\end{tabbing}}
\newcommand{\bfig}{\begin{figure}}
\newcommand{\efig}{\end{figure}}

\begin{document}

\newread\fpsfilein    % file to \read
\newif\iffpsfileok    % continue looking for the bounding box?
\newif\iffpsbbfound   % success?
\newif\iffpsverbose   % report what you're making?
\newdimen\fpsxsize    % horizontal size after scaling before rotation
\newdimen\fpsysize    % vertical size after scaling before rotation
\newdimen\fpstsize    % horizontal size before and after scaling/rotation
\newdimen\fpsrsize    % vertical size before and after scaling/rotation
\newdimen\fpshosize   % horizontal offset
\newdimen\fpsvosize   % vertical offset
\newdimen\fpshskip    % additional horizontal offset
\newdimen\fpsvskip    % additional vertical offset
\newdimen\fpstmp      % register for arithmetic manipulation
\newdimen\pspoints    % conversion factor
%%%%%%%%%%%%%%%%%%%%%%%%%%%%%%%%%%%%%%%%%%%%%%%%%%%%%%%%%%%%%%%%%%%%%%%%%
%  DEFAULT sizes
%%%%%%%%%%%%%%%%%%%%%%%%%%%%%%%%%%%%%%%%%%%%%%%%%%%%%%%%%%%%%%%%%%%%%%%%%
\pspoints=1bp          % Adobe points are `big'
\fpsxsize=0pt          % Default value, means `use natural size'
\fpsysize=0pt          % ditto
\fpshskip=0pt          % Default value, means `no additional hoffset'
\fpsvskip=0pt          % Default value, means `no additional voffset'
%%%%%%%%%%%%%%%%%%%%%%%%%%%%%%%%%%%%%%%%%%%%%%%%%%%%%%%%%%%%%%%%%%%%%%%%%
%  The \fpsbox command
%%%%%%%%%%%%%%%%%%%%%%%%%%%%%%%%%%%%%%%%%%%%%%%%%%%%%%%%%%%%%%%%%%%%%%%%%
\def\fpsbox#1{\global\def\fpsllx{36}\global\def\fpslly{36}%
   \global\def\fpsurx{756}\global\def\fpsury{576}%
   \def\lbracket{[}\def\testit{#1}\ifx\testit\lbracket%
   \let\next=\fpsgetlitbb\else\let\next=\fpsnormal\fi\next{#1}}%
%  \ifx#1[\let\next=\fpsgetlitbb\else\let\next=\fpsnormal\fi\next{#1}}%
%
\def\fpsgetlitbb#1#2 #3 #4 #5]#6{\fpsgrab #2 #3 #4 #5 .\\%
   \fpssetgraph{#6}}%
\def\fpsnormal#1{\fpsgetbb{#1}\fpssetgraph{#1}}%
\def\fpsgetbb#1{%
%%%%%%%%%%%%%%%%%%%%%%%%%%%%%%%%%%%%%%%%%%%%%%%%%%%%%%%%%%%%%%%%%%%%%%%%%
%
%   The first thing we need to do is to open the
%   PostScript file, if possible.
%
%%%%%%%%%%%%%%%%%%%%%%%%%%%%%%%%%%%%%%%%%%%%%%%%%%%%%%%%%%%%%%%%%%%%%%%%%
\openin\fpsfilein=#1
\ifeof\fpsfilein\message{I couldn't open #1, will ignore it}\else
%%%%%%%%%%%%%%%%%%%%%%%%%%%%%%%%%%%%%%%%%%%%%%%%%%%%%%%%%%%%%%%%%%%%%%%%%
%
%   Okay, we got it. Now we'll scan lines until we find one that doesn't
%   start with %. We're looking for the bounding box comment.
%
%%%%%%%%%%%%%%%%%%%%%%%%%%%%%%%%%%%%%%%%%%%%%%%%%%%%%%%%%%%%%%%%%%%%%%%%%
{\fpsfileoktrue \chardef\other=12
    \def\do##1{\catcode`##1=\other}\dospecials \catcode`\ =10
    \loop
       \read\fpsfilein to \fpsfileline
       \ifeof\fpsfilein\fpsfileokfalse\else
%%%%%%%%%%%%%%%%%%%%%%%%%%%%%%%%%%%%%%%%%%%%%%%%%%%%%%%%%%%%%%%%%%%%%%%%%
%
%   We check to see if the first character is a % sign;
%   if not, we stop reading (unless the line was entirely blank);
%   if so, we look further and stop only if the line begins with
%   `%%BoundingBox:'.
%
%%%%%%%%%%%%%%%%%%%%%%%%%%%%%%%%%%%%%%%%%%%%%%%%%%%%%%%%%%%%%%%%%%%%%%%%%
          \expandafter\fpsaux\fpsfileline:. \\%
       \fi
   \iffpsfileok\repeat
   \iffpsbbfound\else
    \iffpsverbose\message{No bounding box comment in #1;
                           using defaults}\fi\fi
}\closein\fpsfilein\fi}%
%%%%%%%%%%%%%%%%%%%%%%%%%%%%%%%%%%%%%%%%%%%%%%%%%%%%%%%%%%%%%%%%%%%%%%%%%
%
%   Now we have to calculate the scale and offset values to use.
%   First we compute the natural sizes.
%
%%%%%%%%%%%%%%%%%%%%%%%%%%%%%%%%%%%%%%%%%%%%%%%%%%%%%%%%%%%%%%%%%%%%%%%%%
\def\fpssetgraph#1{%
   \fpsrsize=\fpsury\pspoints
   \advance\fpsrsize by-\fpslly\pspoints
   \fpstsize=\fpsurx\pspoints
   \advance\fpstsize by-\fpsllx\pspoints
%%%%%%%%%%%%%%%%%%%%%%%%%%%%%%%%%%%%%%%%%%%%%%%%%%%%%%%%%%%%%%%%%%%%%%%%%
%
%   If `fpsxsize' is 0, we default to the natural size of the picture.
%   Otherwise we scale the graph to be \fpsxsize wide.
%
%%%%%%%%%%%%%%%%%%%%%%%%%%%%%%%%%%%%%%%%%%%%%%%%%%%%%%%%%%%%%%%%%%%%%%%%%
   \fpsxsize\fpssize\fpsrsize\fpstsize
   \ifnum\fpsxsize=0 \ifnum\fpsysize=0
      \fpsxsize=\fpstsize \fpsysize=\fpsrsize
%%%%%%%%%%%%%%%%%%%%%%%%%%%%%%%%%%%%%%%%%%%%%%%%%%%%%%%%%%%%%%%%%%%%%%%%%
%
%   We have a sticky problem here:  TeX doesn't do floating point arithmetic!
%   Our goal is to compute y = rx/t. The following loop does this reasonably
%   fast, with an error of at most about 16 sp (about 1/4000 pt).
%
%%%%%%%%%%%%%%%%%%%%%%%%%%%%%%%%%%%%%%%%%%%%%%%%%%%%%%%%%%%%%%%%%%%%%%%%%
     \else\fpstmp=\fpstsize \divide\fpstmp\fpsrsize
       \fpsxsize=\fpsysize \multiply\fpsxsize\fpstmp
       \multiply\fpstmp\fpsrsize \advance\fpstsize-\fpstmp
       \fpstmp=\fpsysize
       \loop \advance\fpstsize\fpstsize \divide\fpstmp 2
       \ifnum\fpstmp>0
          \ifnum\fpstsize<\fpsrsize\else
             \advance\fpstsize-\fpsrsize \advance\fpsxsize\fpstmp \fi
       \repeat
     \fi
   \else\fpstmp=\fpsrsize \divide\fpstmp\fpstsize
     \fpsysize=\fpsxsize \multiply\fpsysize\fpstmp
     \multiply\fpstmp\fpstsize \advance\fpsrsize-\fpstmp
     \fpstmp=\fpsxsize
     \loop \advance\fpsrsize\fpsrsize \divide\fpstmp 2
     \ifnum\fpstmp>0
        \ifnum\fpsrsize<\fpstsize\else
           \advance\fpsrsize-\fpstsize \advance\fpsysize\fpstmp \fi
     \repeat
   \fi
%%%%%%%%%%%%%%%%%%%%%%%%%%%%%%%%%%%%%%%%%%%%%%%%%%%%%%%%%%%%%%%%%%%%%%%%%
%
%  We rotate as described above
%
%%%%%%%%%%%%%%%%%%%%%%%%%%%%%%%%%%%%%%%%%%%%%%%%%%%%%%%%%%%%%%%%%%%%%%%%%
   \fpshosize=0.0pt \fpsvosize=0.0pt
   \def\testangled{0}\def\testangler{90}%
   \def\testanglel{270}\def\testangleu{180}%
   \ifx\fpsangle\testangled
      \fpsrsize=\fpsysize \fpstsize=\fpsxsize\else
   \ifx\fpsangle\testangler
      \fpshosize=\fpsysize \fpsvosize=0.0pt
      \fpsrsize=\fpsxsize \fpstsize=\fpsysize\else
   \ifx\fpsangle\testanglel
      \fpshosize=0.0pt \fpsvosize=\fpsxsize
      \fpsrsize=\fpsxsize \fpstsize=\fpsysize\else
   \ifx\fpsangle\testangleu
      \fpshosize=\fpsxsize \fpsvosize=\fpsysize
      \fpsrsize=\fpsysize \fpstsize=\fpsxsize\else
      \fpsrsize=0.0pt \fpstsize=0.0pt
   \fi\fi\fi\fi
% \iffpsverbose\message{#1: width=\the\fpsxsize, height=\the\fpsysize}\fi
 \iffpsverbose\message{#1: width=\the\fpstsize, height=\the\fpsrsize}\fi
 \iffpsverbose\message{hoffset=\the\fpshosize, voffset=\the\fpsvosize}\fi
 \iffpsverbose\message{angle=\fpsangle, pspoints=\the\pspoints}\fi
 \iffpsverbose\message{llx=\fpsllx, lly=\fpslly}\fi
 \iffpsverbose\message{urx=\fpsurx, ury=\fpsury}\fi
%%%%%%%%%%%%%%%%%%%%%%%%%%%%%%%%%%%%%%%%%%%%%%%%%%%%%%%%%%%%%%%%%%%%%%%%%
%
%  Finally, we make the vbox and stick in a \special that dvips can parse.
%
%%%%%%%%%%%%%%%%%%%%%%%%%%%%%%%%%%%%%%%%%%%%%%%%%%%%%%%%%%%%%%%%%%%%%%%%%
   \advance\fpshosize by\fpshskip \advance\fpsvosize by\fpsvskip
   \divide\fpshosize\pspoints     \divide\fpsvosize\pspoints
%   \fpshosize=0.000015253\fpshosize
   \fpstmp=10\fpsxsize \divide\fpstmp\pspoints
   \vbox to\fpsrsize{\vfil\hbox to\fpstsize{%
   \includegraphics{#1}%
\hfil}}%
\fpsxsize=0pt\fpsysize=0pt}%
%%%%%%%%%%%%%%%%%%%%%%%%%%%%%%%%%%%%%%%%%%%%%%%%%%%%%%%%%%%%%%%%%%%%%%%%%
%
%   We still need to define the tricky \fpsaux macro. This requires
%   a couple of magic constants for comparison purposes.
%
%%%%%%%%%%%%%%%%%%%%%%%%%%%%%%%%%%%%%%%%%%%%%%%%%%%%%%%%%%%%%%%%%%%%%%%%%
{\catcode`\%=12 \global\let\fpspercent=%\global\def\fpsbblit{%BoundingBox}}%
%%%%%%%%%%%%%%%%%%%%%%%%%%%%%%%%%%%%%%%%%%%%%%%%%%%%%%%%%%%%%%%%%%%%%%%%%
%
%   So we're ready to check for `%BoundingBox:' and to grab the
%   values if they are found.
%
%%%%%%%%%%%%%%%%%%%%%%%%%%%%%%%%%%%%%%%%%%%%%%%%%%%%%%%%%%%%%%%%%%%%%%%%%
\long\def\fpsaux#1#2:#3\\{\ifx#1\fpspercent
   \def\testit{#2}\ifx\testit\fpsbblit
      \fpsgrab #3 . . . \\%
      \fpsfileokfalse
      \global\fpsbbfoundtrue
   \fi\else\ifx#1\par\else\fpsfileokfalse\fi\fi}%
%%%%%%%%%%%%%%%%%%%%%%%%%%%%%%%%%%%%%%%%%%%%%%%%%%%%%%%%%%%%%%%%%%%%%%%%%
%
%   Here we grab the values and stuff them in the appropriate definitions.
%
%%%%%%%%%%%%%%%%%%%%%%%%%%%%%%%%%%%%%%%%%%%%%%%%%%%%%%%%%%%%%%%%%%%%%%%%%
\def\fpsgrab #1 #2 #3 #4 #5\\{%
   \global\def\fpsllx{#1}\ifx\fpsllx\empty
      \fpsgrab #2 #3 #4 #5 .\\\else
   \global\def\fpslly{#2}%
   \global\def\fpsurx{#3}\global\def\fpsury{#4}\fi}%
%%%%%%%%%%%%%%%%%%%%%%%%%%%%%%%%%%%%%%%%%%%%%%%%%%%%%%%%%%%%%%%%%%%%%%%%%
%
%   We default the fpssize macro:
%
%      \fpsxsize   % just leave the old value alone
%      0pt         % use the natural sizes
%      #1          % use the natural sizes
%      \hsize      % scale to full width
%      0.5#1       % scale to 50% of natural size
%      \ifnum#1>\hsize\hsize\else#1\fi  % smaller of natural, hsize
%%%%%%%%%%%%%%%%%%%%%%%%%%%%%%%%%%%%%%%%%%%%%%%%%%%%%%%%%%%%%%%%%%%%%%%%%
\def\fpssize#1#2{\fpsxsize}%
%\def\fpssize#1#2{0.5#2}
%%%%%%%%%%%%%%%%%%%%%%%%%%%%%%%%%%%%%%%%%%%%%%%%%%%%%%%%%%%%%%%%%%%%%%%%%
%
%   DEFAULT rotation and message mode
%
%%%%%%%%%%%%%%%%%%%%%%%%%%%%%%%%%%%%%%%%%%%%%%%%%%%%%%%%%%%%%%%%%%%%%%%%%
\def\fpsangle{90}%
\fpsverbosetrue
%%%%%%%%%%%%%%%%%%%%%%%%%%%%%%%%%%%%%%%%%%%%%%%%%%%%%%%%%%%%%%%%%%%%%%%%%
%
%   Finally, other definitions for compatibility with older macros.
%
%%%%%%%%%%%%%%%%%%%%%%%%%%%%%%%%%%%%%%%%%%%%%%%%%%%%%%%%%%%%%%%%%%%%%%%%%
\let\fpsfile=\fpsbox
\let\epsfxsize=\fpsxsize
\let\epsfysize=\fpsysize
\let\epsfverbosetrue=\fpsverbosetrue
\let\epsfverbosefalse=\fpsverbosefalse
\def\epsfbox#1{\fpsangle{0}\fpsbox{#1}}
   % ps-figures
 
%%Ich kann mich dunkel erinnern, das ich mit latex und floats
%%auchmal so Probleme hatte bei meinen Schladming lecturenotes.
%%Da hatte ich nur Leerplaetze fuer figs.
%%Ich habe mit folgenden Aenderungen der defaults rumgespielt:
%\renewcommand{\topfraction}{0.8}
%\renewcommand{\bottomfraction}{0.8}
%\renewcommand{\floatpagefraction}{0.8}
 
\def \fhat{\hat{F}}
\def \element{\in}
\def \eps{\epsilon}
\def \fm{\rm fm}
\def \Re{\rm Re}
\def \MeV{\rm MeV}
\def \twiggle{\tilde}
\def \kc{K_{crit}}
\newcommand{\vx}{\mbox{$\vec{x}$}}
\newcommand{\vy}{\mbox{$\vec{y}$}}
\def \v0{\vec{0}}
\newcommand{\lae}{\raisebox{-0.3ex}
{$\renewcommand{\arraystretch}{0.4}\begin{array}{c} < \\ \sim \end{array}$}}
\newcommand{\gae}{\raisebox{-0.3ex}
{$\renewcommand{\arraystretch}{0.4}\begin{array}{c} > \\ \sim \end{array}$}}

\begin{flushleft}
DESY 94--011 \\
WUB 94--04  \\
January 1994 \\
\end{flushleft}
 
\begin{center}
{\bf
\large Beauty Physics in Lattice Gauge Theory\\
}
\end{center}
\normalsize
\medskip
 
\begin{center}
R. Sommer \\
Deutsches Elektronen-Synchrotron DESY, Hamburg
     \\
\end{center}

\centerline{{\bf Abstract}}

\noindent
We summarize the present status of lattice gauge theory computations
of the leptonic decay constants $f_D$ and $f_B$.  The various  
sources of systematic errors are explained in a manner
easily understood by the  non--expert. The results obtained by the 
different groups are then systematically compared. 
As a result, we derive estimates for $f_D$ and $f_B$
in the quenched approximation through  an appropriate
combination of the data available from the different groups. 
Since we account for a possible
lattice spacing dependence, the final errors are quite large. 
However, it is now well known
how these uncertainties can be reduced.\\ 
For the decay constant of heavy--light pseudoscalar mesons 
with  masses of 1-2~GeV, an
interesting comparison of  a full QCD result with the corresponding
simulation in the quenched approximation can be done. Effects of sea quarks
of mass $m_s$ are below the statistical accuracy of these simulations. \\
Related quantities, like $B$--parameters, the spectrum of beauty--hadrons
and
the breaking of the QCD string  are discussed briefly. 
 
\newpage 
 
\tableofcontents
 
\newpage

\section{Introduction}
 
During the past decades, experiments in high energy physics  have resulted
in  quite an indepth understanding of the interactions between
quarks and leptons. It is mathematically formulated as the Standard Model
of particle physics.
The fundamental parameters
that characterize the overall strengths of  the electromagnetic, weak
and strong interactions in the Standard Model
have been determined precisely   in
experiments. Also, the masses of the matter fields -- apart
from the postulated Higgs-boson and top-quark -- are known.
However, the detailed structure of the weak interactions of quarks
is not determined well. Let us discuss this in more detail.
 
In the Standard Model, quark fields $q$ couple
to the charged weak interaction vector bosons $W_+^\mu$
through a term
$$
\sum_{U=u,c,t} \sum_{D=d,s,b} \bar{q}_U ~\gamma_{\mu} \frac{1-\gamma_5}{2}~
V_{UD} ~q_D~~ W_+^{\mu}~.
$$
The Cabibbo--Kobayashi--Maskawa (CKM) matrix\cite{CKM} $V_{UD}$ originates from
the transformation from weak interaction eigenstates to mass eigenstates.
Assuming that there are only three generations,
it is unitary and can be written in terms of four observable parameters.
 
Given the strength of the weak interactions through
$\mu$ -- decays, in principle one needs to determine
the rate of decays of quarks such as $ d \rightarrow u $ to
obtain the matrix elements of $V$. However, quarks do not
exist as free particles and we can only observe
the decay of hadrons. In  general,
the full knowledge of the hadron wave function is  needed
in order to relate a measured
decay rate of a hadron to the parameter $V_{UD}$ in the Lagrangian.
 
Fortunately, the neutron and the
proton are related by
 isospin symmetry which is well tested experimentally.
Hence, the vector part of the decay amplitude
of the ordinary $\beta$ decay ($n \rightarrow p$)
is given
by a  Clebsch--Gordan coefficient times the desired
parameter.
So $|V_{ud}|$ can be determined with almost no theoretical input.
 
Analogously, $|V_{us}|$ can be extracted from
$K \rightarrow \pi$ decays starting
from an approximate SU(3)$_{flavor}$ symmetry. Here, it is already
important to include symmetry breaking corrections, which can be
done reliably using chiral perturbation theory\cite{LeRo}.
 
With these two matrix elements determined from experiment\cite{PDG}
( $|V_{ud}|=0.9744(10)$, $|V_{us}|=0.2205(18)$ )
and imposing the constraints of unitarity, Wolfenstein observed that
mixing appears hierarchically\cite{Wolf}. The hierarchy is parametrized
by a small parameter $\lambda=|V_{us}|$:
$$
V =  \left( \begin{array}{ccc}
   V_{ud} &V_{us} &V_{ub} \\
   V_{cd} &V_{cs} &V_{cb} \\
   V_{td} &V_{ts} &V_{tb} \\  \end{array} \right)
 = \left( \begin{array}{ccc}
  1-\lambda^2/2 & \lambda                        & A\lambda^3(\rho-i\eta) \\
  -\lambda     & 1-\lambda^2/2-iA^2\lambda^4 \eta & A\lambda^2 \\
A\lambda^3(1-\rho-i\eta) & -A\lambda^2           & 1\\  \end{array} \right)  .
$$
The other parameters $A, \rho$ and $\eta$ are of order $O(\lambda^0)$.
 
The parameter $A$ has been determined from semileptonic
decays $B\rightarrow D^* l \nu$.
As theoretical input,
one needs some information on the formfactors of these transitions.
This has previously been extracted using model wave functions for the
$B$-- and the $D$--mesons \cite{formf}. It has been discovered, however,
that the theoretical
description of these formfactors is  considerably simplified in the limit
when the masses of both the $b$ and the $c$ quark are large compared
to typical hadronic scales. In this limit, there appears an
approximate
SU$(2)_{flavor}
\times $SU$(2)_{spin}$ symmetry relating the $b$--quark and the $c$--quark.
Consequently, the number of formfactors
describing these transitions is reduced to one, the Isgur--Wise
function\cite{IsWi}.
The effective theory that starts from this symmetry and tries to
include the corrections of order $O(m_N/m_h)$, with $m_h$ the
heavy quark mass and $m_N$ the nucleon mass representing a typical
QCD scale,
as well as the QCD radiative corrections, is called Heavy Quark
Effective Theory (HQET).
Apart from radiative corrections\cite{Mann},
the normalization of the Isgur--Wise function at the
point $v . v'=1$ ($v$ and $v'$ are the 4--velocities of the initial
and final state meson respectively) is given by the symmetry up to
corrections of order $O((m_N/m_c)^2)$ \cite{Luke}.
Assuming that the latter terms are small,
one extrapolates
the experimental data to that kinematical point
and one obtains\cite{Argus_vcb,AlLo} $A=0.90(12)$.
 
Again, a symmetry has helped to circumvent the full solution of
the bound state problem of QCD. It is important to note, however,
that in the case of the HQET, we have to date a far more limited
understanding of the size of symmetry breaking terms
than in the case of SU(3)$_{flavor}$. In the latter case, there is
a large amount of experimental information on symmetry breaking,
which helped to develop a comprehensive theoretical treatment of
these effects.
Therefore,
it is  of interest to quantify or bound the $O((m_p/m_c)^2)$
corrections in the above analysis. A promising approach to this problem
are lattice gauge theory calculations of the semileptonic
form factors\cite{SLD,euroSL}.
In addition, QCD--sumrules allow to estimate the size of the $O((m_p/m_c)^2)$
corrections\cite{FaNe}.
 
There are two additional unknowns in the CKM-matrix: $\rho$ and $\eta$.
They are of particular interest, since they are a (and probably {\it
the}) source of
CP--violation in the Standard Model\cite{CP}.
Their values are mainly constrained by three experimental observations.
 
$B$--meson
decays together with model calculations determine \cite{AlLo}
$\sqrt{\rho^2 + \eta^2} = 0.36(9)$.
Moreover, the CP--violation parameter $|\epsilon|$ in the $K$--system\cite{CP}
and the $B_0$--$\bar{B_0}$ mixing parameter $x_d$ restrict $\rho$ and
$\eta$. However, in order to extract $\rho$ and
$\eta$ from these measurements, one needs to know
the mixing matrix elements
$$
\frac{8}{3} B_B f_B^2 M_B^2=
< B_0| (\bar{d}\gamma_{\mu}(1-\gamma_5) b)
       (\bar{d}\gamma_{\mu}(1-\gamma_5) b) | \bar B_0>
$$
and
$$
\frac{8}{3} B_K f_K^2 M_K^2=
< K_0| (\bar{d}\gamma_{\mu}(1-\gamma_5) s)
       (\bar{d}\gamma_{\mu}(1-\gamma_5) s) | \bar K_0>~~,
$$
where $f_B$ and $f_K$ are the leptonic decay constants of
the $B$-- and the $K$--mesons and $B_B$, $B_K$ parametrize the
matrix elements relative to the vacuum insertion
``approximation''\cite{CP}.
 
The dependence of the allowed domain in the $(\rho,\eta)$ plane
on the theoretical uncertainty of $B_K$ ($f_K$ is known from experiment)
is not very large. The allowed region depends sensitively on the
value of the product $B_B f_B^2$, however\cite{Roos,Lusi,AlLo}.
 
Until fairly recently, phenomenological analysis of the allowed
region in the $(\rho,\eta)$ plane used
the ``old prejudice''
$f_B\sqrt{B_B}=(110 - 160)$~MeV. Such values favor  a negative $\rho$
and lead to small predictions of CP-asymmetries in $B$ decays, which
one would like to measure in future experiments at $B$--factories\cite{bfac}.
First lattice estimates\cite{Allt1,fb1} of these quantities that were
done within the
HQET approximation, indicated much larger values of $f_B$. Also,
QCD sum rule calculations performed in the  HQET limit have
subsequently yielded such large values\cite{Ne92,BBBD}.
Together with $B_B\simeq 1$ which is supported by a number of theoretical
investigations, this indicates that positive values of $\rho$ are
favored by the experimental constraints.
If this is the case, much larger signals for CP--violation in $B$ decays are
expected -- an interesting perspective for $B$--factories.

We emphasize, that the value of  $B_B f_B^2$ is the central
theoretical uncertainty on our way from the experimental data
to the determination of the CP-violating part of the CKM-matrix.
It is therefore important to determine  $B_B f_B^2$ with a good precision
in order
to narrow down the allowed region of $(\rho,\eta)$. To this end, precise
lattice calculations of these quantities are necessary and one has to take
account of the full spectrum of the statistical {\it and}
systematic uncertainties.  Besides this number (and similar
other ones that are of interest), lattice gauge theory calculations
also offer  the possibility to quantify the corrections to the
HQET limit. Starting from the simpler case of the $D$--meson,
we review the present status of these
efforts and their future perspective in sections \ref{s_fb}, \ref{s_Bp}
and \ref{s_corr}.
A discussion of some general features of lattice gauge theory calculations
-- as given in section \ref{s_LQCD} -- is needed to understand the
different sources of systematic errors in the computations.
Other sections cover related topics.
 
\newpage

\section{~Lattice QCD \label{s_LQCD}}
Lattice QCD\cite{Wils,LQCD} is a
 regularization of the
fundamental theory of strong interactions that allows nonperturbative
calculations.
 In principle, it involves no
approximations. The latter arise because our numerical possibilities
of evaluating the path integral by a Monte Carlo process are limited.
This means that computations have to be performed at parameters that are
different from their values in nature. Therefore, extrapolations have to be
performed e.g. in the lattice spacing, quark masses and the space-time -
volume $V$. Furthermore, we  are
limited by the statistical errors of the Monte Carlo calculation.
 
In order to be able to discuss the various systematic errors in
current calculations, we need to introduce some definitions
and summarize  the crucial points in a practical lattice QCD
computation. This section primarily addresses  readers that are not
very familiar with lattice QCD. The specialist may, however, be interested in
 the
discussion about universality in the quenched approximation, which is given in
section~\ref{s_QA}.

\subsection{Path Integral Representation of Greens Functions}
 
The starting point of lattice QCD  is
to define the field theory on a space-time lattice  with  spacing $a$.
This is one way of regularizing the ultraviolet divergences that are
intrinsic to a quantum field theory: the  lattice spacing serves as
a cutoff for the high frequency modes of the fields. As in the case of any
cutoff,
the field theory is finally defined in the limit $a^{-1} \rightarrow \infty$.
 
The basic variables of lattice QCD are the dimensionless quark fields
$
q_f (x)
$
($f$ labels the different flavors of quarks and we omit the color--
and Dirac--indices) and the gluon field variables
$
U_{\mu}(x)
$
which are SU(3) matrices.
The coordinate $x$ resides on the
4-dimensional space--time lattice: $x=(x_0,x_1,x_2,x_3),~~ x_{\mu} / a ~
\epsilon~ N$.
The gluon variables $U_{\mu}(x)$ are the gauge connection
 of nearest neighbor points $x$ and $x+a\hat{\mu}$.
 
The dynamics of the theory (i.e. bound state masses and properties, transition
probabilities) is contained in the euclidean Greens functions
or correlation functions. They are defined through the Feynman path integral
\bes
<O[q,\bar{q},U]> &=& Z^{-1} \int {\rm D}[q]  ~ \int {\rm D} [\bar{q}]
 ~ \int {\rm D} [U] \exp(-S[q,\bar{q},U] ) ~~ O [q,\bar{q},U]  \nonumber \\
Z &=& \int {\rm D} [q]  ~ \int {\rm D} [\bar{q}] ~ \int {\rm D} [U]
\exp(-S[q,\bar{q},U] ) . \label{green}
\ees
Here, $\int {\rm D} [q] ~ \int {\rm D} [\bar{q}]$ stand for  multiple
Grassmann integrals over each component of
the quark variables and  $\int {\rm D} [U]$ represents an integral
over the gluon variables
on each link with the SU(3) Haar measure.
$O [q,\bar{q},U]$ is a functional of
any number of variables of the theory and $S[q,\bar q,U]$
is the euclidean action
\be
S[q,\bar q,U] = S_G[U] + S_F[q,\bar q,U]
\ee
with a gluonic part
\be
S_{G}[U] = \beta \sum_x \sum_{\mu > \nu = 0}^3 \{
           1 - \frac{1}{6}{\rm Re}\; {\rm tr}\; P_{\mu,\nu}(x) \}~;
~~ \beta=\frac{6}{g_0^2}
\ee
and a fermionic part
\bes
S_{F}[\bar{q},q,U] &=& \sum_x \bar{q}(x) {\cal D} q(x)~;  \label{fermact} \\
    {\cal D} &=& M_0 + \frac{1}{2} [~\sum_{\mu = 0}^3 \gamma_{\mu}
                       ( \nabla_{\mu} + \nabla_{\mu}^*)
            - \sum_{\mu = 0}^3 \nabla_{\mu}\nabla_{\mu}^*
     + i \frac{c_{SW}(g_0^2)}{2} \sum_{\mu , \nu = 0}^3
                    F_{\mu  \nu}\sigma_{\mu  \nu}~] ~. \nonumber
\ees
Here,
$P_{\mu, \nu}(x)$ is the product of gauge parallel transporters $U_\mu(x)$
 around an elementary plaquette
\be
P_{\mu,\nu}(x) = U_{\mu}(x) U_{\nu}(x+a\hat{\mu})
                U_{-\mu}(x+a\hat{\mu}+a\hat{\nu})
                U_{-\nu}(x+a\hat{\nu})~
;~~~U_{-\mu}(x) \equiv U_{\mu}^{\dagger}(x-a\hat{\mu})~,
\ee
the covariant finite differences are
\be
 \nabla_{\mu}q(x) = U_\mu(x)q(x+a\hat{\mu}) - q(x)~,~~
 \nabla_{\mu}^* =  - \nabla_{-\mu}
\ee
and the euclidean gamma matrices fulfill
\be
\{\gamma_{\mu},\gamma_{\nu}\} = 2 \delta_{\mu,\nu}~,~~~
\gamma_{\mu}^{\dagger} = \gamma_{\mu}~,~~~
\sigma_{\mu  \nu} = \frac{i}{2}[\gamma_{\mu},\gamma_{\nu}] .
\ee
A lattice approximation of the field strength  is given by
\bes
F_{\mu \nu}(x) = \frac{1}{8}
              [ & P_{\mu,\nu}(x)  + P_{-\mu,-\nu}(x)
              + P_{\nu,-\mu}(x) + P_{-\nu,\mu}(x) & \nonumber  \\
              - & P_{\nu, \mu}(x) - P_{-\nu,-\mu}(x)
              - P_{-\mu,\nu}(x)  - P_{\mu,-\nu}(x) & ]~.
\ees
The lattice action $S$ depends on the bare parameters of the theory,
namely $g^2_0$, the gauge coupling, and the hopping parameters $K_f$ that are
collected in the bare quark mass matrix
$M_0=\hbox{diag}(1/(2K_1),~...~,1/(2K_{n_f})) -4$.
The last term in the fermion matrix\cite{SW} $\cal D$,
 which is accompanied by the
coefficient function $c_{SW}(g^2_0)$ will be discussed at the end of
section~\ref{s_CL}.
 
Eq. (\ref{green}) looks identical to a statistical mechanics thermal average.
It can be evaluated by Monte Carlo importance sampling,
once the space time grid is restricted to a finite number of points\cite{LQCD}.
Such a Monte Carlo evaluation is the main nonperturbative method
employed in lattice gauge theory calculations.
There is an essential difference to a simulation of a spin model,
however, namely the fermionic nature of the quark fields.
Since the fermionic action is a quadratic form,
the Grassmann integrals
can be performed analytically, e.g.
\be
\int {\rm D} [q] ~ \int {\rm D} [\bar{q}]   ~~
q^{c'}_{\alpha'}(x')
\bar q^c_{\alpha}(x)
= {\cal D}^{-1}_{x,f,\alpha,c;x',f',\alpha',c'}[U]~ \det {\cal D}[U]~
         \exp(- S_G[U] ) ~,
       \label{fermint}
\ee
with $\alpha,\alpha'$ Dirac indices and $c,c'$ color indices.
In general, one obtains
\be
<O[q,\bar{q},U]> =  Z^{-1} ~ \int {\rm D} [U] ~
 \exp(-S_G[U] )~ \det {\cal D}[U] ~ F_O [U] \label{partf}
\ee
with a functional $F_O[U]$ that is easily derived using
eq.~(\ref{fermint}) and
the property that the Grassmann fields are anticommuting.
It is eq.(\ref{partf}) which is evaluated by
Monte Carlo important sampling, generating ``configurations'' $U$
with the probability  $P[U] \propto \exp(-S_G[U] ) ~ \det {\cal D}[U]$
and then averaging
$F_O[U]$ over these configurations.
 
\subsection{A Pseudoscalar Correlation Function\label{s_PCF}}
We discuss here the most simple (and for the purpose of this review almost
sufficient) observable that is well calculable in lattice QCD.
We choose a lattice with a finite number of points $L/a$ in the three
space directions and $L_t/a$ points in euclidean time.
The gauge fields are periodic functions: $U_{\mu}(x)=U_{\mu}(x+\hat{k}L),$
$k=1,2,3$; $U_{\mu}(x)=U_{\mu}(x+\hat{0}L_t)$. The quark fields are
taken antiperiodic in time and periodic (or antiperiodic) in space.
We may take the axial vector current
\be
A_\mu^{ff'}(x)=Z_A ~\bar{q}_f(x) \gamma_{\mu} \gamma_5 q_{f'}(x) \label{amu}
\ee
as a convenient interpolating field
for pseudoscalar mesons\footnote{
$|P>$ denotes the pseudoscalar
state with momentum zero, the appropriate flavor quantum numbers and
with a normalization $<P|P>=1$.
}:
\be
<0|A_{0}(0)|P>=F_P \sqrt{M_P/2} ~.\label{fp}
\ee
The finite renormalization $Z_A$ of the axial vector current will
be discussed in section~\ref{s_R}.
The (space--) momentum zero correlation function
of $A_{0}$ has a spectral decomposition
\be
\sum_{\vx}<A_{0}(x) A_{0}^{\dagger}(0)> = \sum_{n \geq 1} |<0|A_{0}|P,n>|^2
              \{ \exp [-E_n x_0]+\exp [E_n (x_0-L_t)]\}~, \label{spec}
\ee
where $|P,n>$ denote the excited states in this channel and
$|P,1> \equiv |P>$. 

With the Wilson action (i.e. $c_{SW}(g_0^2)\equiv0$), the 
transfer matrix, which describes the propagation in the euclidean time,
is positive and hermitian\cite{LuTM}. In this case,  
eq.~(\ref{spec}) is therefore exact up to
irrelevant terms of order $O(\exp[-E_{gap}L_t])$, where
$E_{gap}$ is the lowest state with the quantum numbers of the vacuum.
When $c_{SW}(g_0^2)\neq 0$, one expects eq.~(\ref{spec}) to hold to
a good approximation at moderately large $x_0$.

At sufficiently large values of the euclidean time $x_0$, the
spectral representation shows that the
correlation function eq.~(\ref{spec}) is dominated by the lowest state
and one may extract both the mass $M_P\equiv E_1$ and the leptonic
decay constant
$F_P$ from a Monte Carlo estimate of $<A_{0}(x) A_{0}^{\dagger}(0)>$.
 
\subsection{The Continuum Limit\label{s_CL}}
In this section, we discuss how the
continuum limit $a \rightarrow 0$ is taken in QCD.
 
Lattice gauge theories contain only dimensionless variables and
fields. All quantities are calculated in units of the lattice
spacing $a$. The latter is not a parameter of the calculation
but is determined aposteriori once a dimensionful observable is fixed
to its experimental value.
 
In order to shorten the discussion, consider QCD with only
mass--degenerate quarks,
i.e. $K_f = K$ for all $f$.
Eq.~(\ref{spec}) then determines the dimensionless functions\footnote{
We assume that at each point  $(g_0^2,K)$, the space extent $L$ of the
lattice has been chosen large enough such that finite size effects
are negligible compared to the statistical accuracy of the calculation.
Therefore, the dependence on the third dimensionless parameter $(F_P~L/a)$ is
not written.
}
\be
  F_{P}(g_0^2,K)
\ee
and
\be
  M_{P}(g_0^2,K)~.
\ee
In addition, for example,
the mass of the nucleon in lattice units $M_{N}(g_0^2,K)$ may
be calculated from a
nucleon correlation function.
 
Since the quark masses are free parameters in the Standard Model,
we have to set a nonperturbative renormalization condition to fix
the hopping
parameter $K$ in terms of a renormalized, experimentally observable
quantity. A convenient choice is
\be
M_{P}(g_0^2,K)/F_{P}(g_0^2,K) =m_{\pi}/f_{\pi}~, \label{RGT}
\ee
where $f_{\pi}$ and $m_{\pi}$ refer to the experimentally measured
leptonic decay constant and mass of the pion.
Eq.~(\ref{RGT}) defines a renormalization group trajectory
in the $(g_0^2,K)$ plane.
Along this trajectory, the ratio
\be
R_{M_N,F_P}(F_P) \equiv \frac{M_{N}(g_0^2,K)}{F_{P}(g_0^2,K)}|_{
M_{P}(g_0^2,K)/F_{P}(g_0^2,K) =m_{\pi}/f_{\pi} }
\ee
is a function of $F_P$ only and reaches its
continuum limit at $F_P=0$ since $F_P$ is the physical decay constant
times the lattice spacing.
From asymptotic freedom one infers that the continuum limit is reached
 at the point
$g^2_0 = 0,~K=1/8$.
 
 Equivalently, on the trajectory eq.~(\ref{RGT}), we may
consider the lattice spacing to be determined by
\be
a_{f_{\pi}}(g^2_0)  = F_{P}(g_0^2,K) / f_{\pi}~ \label{afpi}
\ee
and  the nucleon mass in
physical units is given by $m_N=M_N(g_0^2,K)/a_{f_{\pi}}(g^2_0)$.
 
Corrections to the continuum limit constitute a major source of uncertainty
for practical lattice gauge theory determinations of quantities like $f_B$.
These corrections have to be addressed by Monte Carlo simulations
done at different points on the renormalization group trajectory.
Nevertheless, it is instructive to consider perturbation theory and
learn something about the {\it structure}
of the finite lattice spacing corrections.
Symanzik has discussed the cutoff dependence of Feynman diagrams\cite{Symm}.
For the action eq.~(\ref{fermact}), such a discussion\cite{SW}
suggests that the leading  dependence of physical quantities
such as $R_{M_N,F_P}(F_P)$ on the lattice spacing is linear (up to
numerically unimportant logarithmic modifications):
\be
R_{M_N,F_P}(F_P) = \frac{m_N}{f_{\pi}} + \rho_1 F_P+O(F_P^2)~; ~~F_P\equiv f_{\pi}
  a_{f_{\pi}}~.
 \label{CL}
\ee
How small does the lattice spacing have to be such that the finite
$a$ corrections
are small? Again this can only be answered by a nonperturbative calculation
of the function $R_{M_N,F_P}(F_P)$ and for each such a function
separately. It is however worth noting, that
one may expect that the relevant scales are the sizes of the wave functions of
the hadrons considered (for S-wave states),
not their Compton wave length, since we mainly need
a good approximation of the continuum wave function by the lattice one.
We may therefore hope that eq.~(\ref{CL}) can be applied for $a << 0.5 fm$.

{\bf Improving the Approach to the Continuum Limit\\}
In a formal expansion in powers of the lattice spacing,
the fermion action eq.~(\ref{fermact}) breaks up into three pieces.
The first two,
the naively discretized Dirac operator and  the Wilson term,
 are necessary to
obtain the correct continuum limit\cite{Reis}. The Wilson term,
$\bar q \sum_{\mu} \nabla_{\mu}\nabla_{\mu}^* q$
has a classical continuum limit $-a\bar{q} \triangle q+O(a^2)$,
where $\triangle$ is the covariant Laplace operator. This dimension ~5,
irrelevant, operator
 is responsible
for the lattice spacing corrections of order $O(a)$ as e.g. in eq.~(\ref{CL}).
It is expected that for {\it on--shell} matrix elements
these terms can be cancelled by adding {\it one}  dimension~5 operator to the
action\cite{LWimpr}. Most conveniently, one chooses the third term in the action
with coefficient function
$c_{SW}(g_0^2)$, since it does not
involve further derivative terms in the quark fields\cite{SW}.

Symanzik's idea was to calculate coefficients like $c_{SW}(g_0^2)$
in perturbation theory leading to a perturbatively improved action.
Already the tree level expression $c_{SW}(g_0^2)=1$ leads to
$\rho_1=O(g_0^2)$.
Since i) the arguments that lead us to expect that the $O(a)$ lattice
artifacts  can be cancelled in this way are due to perturbation theory
and ii)  the coefficient function $c_{SW}(g_0^2)$ is calculated
perturbatively, improvement needs to be demonstrated through
Monte Carlo results.
At present,  the evidence for improvement with the
action of Sheikholeslami and Wohlert is somewhat indirect but
nevertheless suggests  that
$O(a)$ lattice
artifacts are  significantly reduced compared to the Wilson
action\cite{imprres}.
In this context, it is furthermore of interest  that in
 the case of the pure gauge theory with Dirichlet boundary conditions,
an improvement has been established unambiguously for a one loop
Symmanzik improved action\cite{su2}.
Thus, there is an example which shows that the perturbative improvement
program of Symmanzik works at those values of the gauge coupling that are
needed in realistic simulations.

Below, we will also discuss recent results for heavy-light (HL) mesons
that were obtained with the
perturbatively improved action. We will use the synonyms ``improved action''
and ``SW--action'' for the action eq.~(\ref{fermact}) with $c_{SW}(g_0^2)=1$,
while we use the name ``Wilson action'' when $c_{SW}(g_0^2)=0$.
 
\subsection{Renormalization\label{s_R}}
In the previous section, we have ignored the problem of the determination
of the strength of the axial vector current $Z_A$.
At first sight, it requires a nontrivial renormalization since
it is a composite operator. However, at the formal level, the flavor
offdiagonal axial vector
current $A_{\mu}^{f,f'}(x)$ is the current of chiral symmetry, which is
broken only softly:
\be
\partial_{\mu}A_{\mu}^{f,f'}(x)=(m_f+m_{f'})~P^{f,f'}(x);~~~
 P^{f,f'}(x) = \bar{q}_f(x) \gamma_5 q_{f'}(x)~. \label{pcac}
\ee
This relation (more precisely the chiral ward identities) would normally
insure that $A_{\mu}$ does not get renormalized.
The quantum field theory needs to be defined with a regulator, however.
As the lattice  regularization (eq.~(\ref{fermact})) breaks chiral symmetry
by terms of order $O(a)$, the formal argument does not apply and
$A_{\mu}$ acquires a nontrivial
renormalization\cite{KaSm,Boch}.\footnote{
Only in the case of the vector current, a conserved current can be defined
for most actions.}
 
In the following, we mainly
discuss the case of the Wilson action, which is still
the action for which most results exist.
At the
  very end of this section,
we mention the case  $c_{SW}=1$.
 
We write the relation between the bare lattice current and the
renormalized current $A_\mu^{f,f'}(x)$ in the most
general way:
\be
A_\mu^{f,f'}(x) = Z_A(g^2_0,K_f,K_{f'}) ~~ \bar{q}_f(x) \gamma_{\mu}
\gamma_5 q_{f'}(x)~. \label{za}
\ee
Here, $Z_A$ needs to be defined through an appropriate normalization condition,
such as eq.(\ref{curralg}) which will be discussed below.
We may rewrite eq.~(\ref{za}) in a form that was suggested by
Lepage, Kronfeld and Mackenzie\cite{Lepa,Kron,LM}\footnote{
The argument that is given in ref.~\cite{Lepa} for this normalization
is based on the limit $am_f >> 1$ and is therefore only indicative
in the present
context.  On the other hand, Kronfeld\cite{Kron} starts from
the canonical
normalization of the quark fields in the transfer matrix
representation\cite{LuTM}
of Green functions like eq.~(\ref{spec}).
Subsequently, he uses a mean field approximation {\it and implicitly
assumes} that the quark fields enter with zero space--momentum. One is
therefore well advised to consider this normalization as an alternative
and not as the solution to the problem of handling heavy quarks with masses
of the order of the inverse lattice spacing.
}
\be
A_\mu^{f,f'}(x) = \tilde{Z}_A(g^2_0,K_f,K_{f'})~ \sqrt{{1\over 2K_f}-3\bar{u}}~
            \sqrt{{1\over 2K_{f'}}-3\bar{u}}~~\bar{q}_f(x) \gamma_{\mu}
\gamma_5 q_{f'}(x)~, \label{zaKron}
\ee
where the mean field value $\bar{u}$ of the gauge field is defined through a
gauge invariant quantity, e.g.\cite{Kron}
\be
\bar{u} = 1/(8\kc)~.
\ee
The critical value of the hopping parameter $\kc$ is to be determined
nonperturbatively from the Monte Carlo calculations through $M_P(g_0^2,\kc)=0$.
 
If the renormalization constants $Z_A$ and $\tilde{Z}_A$ are
defined through a
nonperturbative normalization condition at a finite value of the lattice
spacing, eq's.
(\ref{za},\ref{zaKron}) are completely equivalent.
In most applications, the normalization of the currents is taken from
perturbation theory and after dropping the $O(a)$ terms. In this approximation,
there
is a significant difference between the currents
(\ref{za},\ref{zaKron}). Numerically,  the
difference is dominated by a factor $\exp[(m_{f} + m_{{f'}}) a /2]$;
$a m_f\equiv \log[1+(2K_f)^{-1}-(2\kc)^{-1}]$.
Although this factor becomes irrelevant
in the continuum limit, it is sizeable in present HL calculations.

In 1--loop order perturbation theory one obtains\cite{KaSm,Za,LM}
\bes
Z_A(g^2_0,K_f,K_{f'}) &=& 1 - 0.133373 \tilde{g}^2 + O(\tilde{g}^4) +
O(am_f) + O(am_{f'}) \label{ZApert} \\
\tilde{Z}_A(g^2_0,K_f,K_{f'}) &=& 1 - 0.02480~ \tilde{g}^2 + O(\tilde{g}^4) +
O(am_f) + O(am_{f'}) \label{ZApertKron}\\
\tilde{g}^2 &=& g_0^2 / P~; ~~P={1\over 3}<{\rm tr} P_{\mu, \nu}(x)>~.
\ees
In the above equations, we did not need to specify the normalization condition
for the current,
because the $Z$--factors depend on these conditions only
when  $O(a m_f)$--terms are included. Instead of using the bare coupling
as expansion parameter, we have written the perturbative
expansion of $Z_A, \tilde{Z}_A$ in terms of the effective
coupling $\tilde{g}^2$, introduced by
G. Parisi\cite{Paris}. Through
several examples,
Lepage and Mackenzie\cite{LMold,LM} demonstrated  that the series in $\tilde{g}^2$
(at momentum scales of the lattice
cutoff $\pi / a$) converges more rapidly than the one in the bare
coupling\footnote{
Lepage and Mackenzie actually go further\cite{LM} and change from $\tilde{g}^2$
to a physical coupling defined from the interquark potential. In addition, they
give a prescription how to optimize the scale at which this coupling should be
evaluated when one inserts it into a 1--loop perturbative result.
Although this does indeed improve the agreement of 1--loop perturbation theory
with nonperturbative Monte Carlo results for the examples that are considered
in  \cite{LM}, it is important to be aware that the errors remain of order
$O(\alpha^2)$. When perturbation theory is used in this review, we use the
expansion in terms of $\tilde{\alpha}=\tilde{g}^2/(4\pi)$ and try to
indicate the expected size of
the $O(\tilde{\alpha}^2)$ terms.
}
.
In the known examples, the correction terms to
the 1--loop order are compatible with having coefficients
of the order one or less, when the series is written in terms of
$\tilde{\alpha}=\tilde{g}^2/(4\pi)$. In addition, Lepage and Mackenzie
present examples where 1--loop perturbation theory gives quite accurate
results when compared to full nonperturbative computations of
the quantity considered; see also ref.~\cite{grsu3}.
 
It is argued in ref. \cite{Kron} that eq.~(\ref{zaKron}) is
the definition  with the smaller lattice artifacts when it is used together
with the perturbative value of the renormalization constant (as given above).
In subsequent works, this normalization is quoted as the ``correct'' one
and eq.~(\ref{zaKron}) as ``false''. We stress that this is  not the
case as a general statement and give a counter example in the appendix.
The example, as well as the lattice spacing dependence of $f_D$, which we
investigate later,  mainly teaches
us that the size of the lattice artifacts has to be determined by
precise simulations with varying lattice spacings for each case
separately
(A one--loop calculation including the $O(m_fa)$--terms  could shed
some light onto this question as well).
 
The $O(\tilde{g}^4)$ corrections to the one loop estimates of the
renormalization constants are a systematic error.
In principle, the axial current may be renormalized nonperturbatively.
Here we only outline the idea
and refer the reader to ref.~\cite{Boch} for a more detailed discussion.
The idea is to require current algebra relations to be valid for
the renormalized currents. 
For instance, one may require\cite{Boch}
\bes
\frac{1}{2}~\sum_{x}~\langle~\bigl[ ( m_f P^{f,f}(x)  - 
              \partial_\mu A_\mu^{f,f}(x) ) - (f \rightarrow f')
  \bigr]
A_\nu^{f,f'}(y) V^{f',f}_\nu(z) ~\rangle = \nonumber \\
= - \langle V_\nu^{f,f'}(y) V^{f',f}_\nu(z) \rangle
+ \langle A_\nu^{f,f'}(y) A^{f',f}_\nu(z) \rangle ~~~ \label{curralg}
\ees
for the renormalized currents. Taking $\nu$ to be
i) spacelike and ii) timelike, Eq.(\ref{curralg}) constitutes two 
rather independent linear equations for the current renormalization constants. 
As these equations are nonlinear,
the renormalization constants can be computed -- without the
use of perturbation theory.

It is important to note the following points about this equation\footnote{
We would like to thank M. L\"uscher and S. Sint for pointing out this problem.
}.
The axial vector current at point $y$ on the left hand side is changed
into the vector current on the right hand side by the contact with the
pseudoscalar density and analogously for $V(z)$. 
This means that the relevant contributions in the
integrated three--point function on the left hand side originate from
short distances $|x-y|$ and $|x-z|$. At distances of a few lattice spacings,
lattice correlation functions do exhibit a strong dependence on the lattice
spacing. This will translate into potentially large lattice artifacts
of the renormalization constants. A sufficiently small lattice spacing
is therefore required to be sure that the $O(a)$--terms injected through
eq.~(\ref{curralg}) are numerically small. At current values of the
lattice spacings, it is not evident that this requirement is fulfilled.

One can avoid this problem at the price of giving up one prediction:
In ratios of decay constants, the renormalization cancels and only
unavoidable $O(a)$ terms remain. Unfortunately, this option increases
the statistical errors in the decay constants $f_D$ and $f_B$
and consequently the remaining $O(a)$ terms are more difficult to control.
 
We point out that eq.~(\ref{zaKron}) is only an ansatz.
In order to systematically reduce the $O(am_f)$ corrections one has to both
improve the action  through $c_{SW} \neq 0$ and the operators. With
$c_{SW} = 1$,
one can achieve
$O(am_f) \rightarrow O(\alpha am_f)$\cite{Heat,imprres}.
 
In summary, the renormalization of the axial vector current is a
nontrivial problem in practice and introduces an uncertainty in
the determination of pseudo scalar decay constants.
The same holds true for the normalization of four fermion operators
such as the one responsible for $B \bar{B}$ mixing. Furthermore,
when working in the HQET in order to evaluate the limiting behavior
of the decay constant as one of the quark masses becomes very large,
the current develops an anomalous dimension and -- at present --
no practicable nonperturbative method to renormalize the current is
known.
 
\subsection{Reaching Asymptotics\label{s_RA}}
 
In a Monte Carlo simulation one evaluates correlation functions such
as
eq.~(\ref{spec}) at finite values of the euclidean time $x_0$.
In order to be able to extract hadron masses and decay constants,
one needs to insure that the terms with $n>1$ are
negligible in the spectral representation eq.~(\ref{spec}).
For a quantitative investigation of this problem,
we consider an example that is relevant for the physics discussed
in this review, namely the correlation function of a heavy--light
(HL) axial vector
current, composed of one heavy quark (with a mass of the order of
 the charm quark mass)
and one light quark.

\begin{figure}%%[htb]
%%\topinsert
\vbox{
\vskip 0 true cm

\def\fpsangle{0}
\fpshskip=0.8 true cm

\centerline{
\fpsxsize=12.5 true cm
\fpsbox[30 30 275 285]{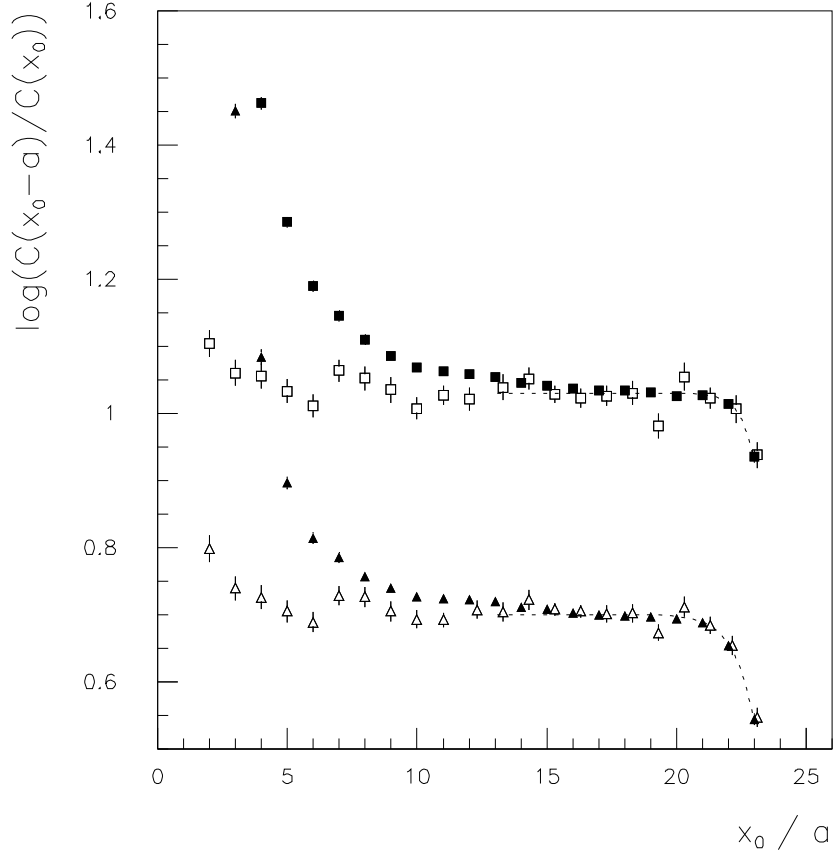}}

\vskip -0.5 true cm

\caption[1]{\footnotesize
 The logarithmic slope of the pseudoscalar correlation
 functions $C^{loc,loc}(x_0)$ (filled symbols) and
 $C^{G,G}(x_0)$ (open symbols)\cite{fb4}. 
 $G$ denotes a gaussian wave function with r.m.s. radius of about
0.3~fm.  
The cutoff is
 $a^{-1}=3.17$GeV ($\beta=6.26$) and the simulation was done in the 
 quenched approximation. 
 The light quark mass was chosen 
 about twice the strange quark mass (K=0.1492), while the heavy quark mass is
 roughly the charm mass (K=0.1350) in the lower part and
 somewhat higher in the upper part (K=0.12).
 The dashed curves illustrate the effect of the finite time extent of the
 lattice according to eq.~(\ref{spectmes}), assuming that only one state
 contributes in this range of $x_0$.
                                                       }\label{f_locmas}
}
%%\endinsert
\end{figure}

It is useful to look at the logarithmic slope
of the correlation
function. This slope, called ``effective mass'', approaches a constant
when the above mentioned
corrections are negligible. In fig.~\ref{f_locmas}~(filled symbols),
one observes that
the  effective mass shows a slow variation with the euclidean time
still at a distance of $x_0=15a \sim 5$GeV$^{-1}$. At larger $x_0$,
a plateau
seems to develop in the effective mass. The region of the
plateau in fig.~\ref{f_locmas} is rather short, however.
 
One may improve the situation by choosing better interpolating fields
for the hadrons.
We briefly describe the approach that has been taken so far for the case
of meson fields.
The general idea is to ``smear'' the local quark fields with
trial wave functions $\Phi ^J$:
\be
q^J(x) = \sum_{y} \Phi ^J(\vx,\vy;{ U}(y_0))~\delta_{x_0 y_0}
   ~q(y) \quad .\label{smear}
\ee
An interpolating field for a meson is then given by
\be
{\cal M}^{J}(x) = \bar{q} (x)
~\gamma_0 \gamma_5 ~q^J(x)\quad ,      \label{meson}
\ee
where the index $J$ denotes the
trial wave function and we have omitted
the flavor indices\footnote{$\Phi$ is the wave function in terms of the
relative coordinate. In addition to the form eq.~(\ref{meson}),
the case where both quark fields are smeared has been considered.}.
$\Phi ^J(\vx,\vy;{ U}(y_0))$ depends either in a gauge covariant
way\cite{Step,ga,fb3,UKQCDfb}
on the
link variables  denoted by ${ U}(y_0)$, or the
construction is done in Coulomb gauge\cite{ape1}.
A meson correlation function is constructed by
\be
C^{I,J}(x_0) = \sum_{\vx} <  {\cal M}^{I}(x)~ [ {\cal M}
^{J}(0)]^\dagger >  \quad .    \label{correla}
\ee
It has the spectral representation
\be
C^{I,J}(x_0) = \sum_{n \geq 1}
 <0|  {\cal M}^{I}(0) |P, n >< P,n| [ {\cal M}^{J}(0)
]^\dagger |0>
           \{ \exp [-E_n x_0]+\exp [E_n (x_0-L_t)]\}
 \label{spectmes} \quad .
\ee
In fig.~\ref{f_locmas} we have included the logarithmic slope
for such a smeared correlation function\cite{fb4}.
It reaches a plateau at values of $x_0 \sim 5a$ -- much earlier than
the correlation function of the local axial vector current.

{\bf A clean test for ground state dominance} is provided by
the ratio
\be
R^{I,J}(x_0)=\frac{C^{I,I}(x_0) C^{J,J}(x_0)}
                  {[C^{I,J}(x_0)]^2}~, \label{ratio}
\ee
which is constructed such that it becomes one
when all correlation functions
are dominated by the ground state\cite{fb4,Eich90}.
This is a very stringent test since
contrary to the search for a plateau in the local mass,
the value of the plateau is known here too.

\begin{figure}[htb]
%%\topinsert
\vbox{
\vskip 0 true cm

\def\fpsangle{0}
\fpshskip=0.8 true cm

\centerline{
\fpsxsize=12.5 true cm
\fpsbox[30 30 275 285]{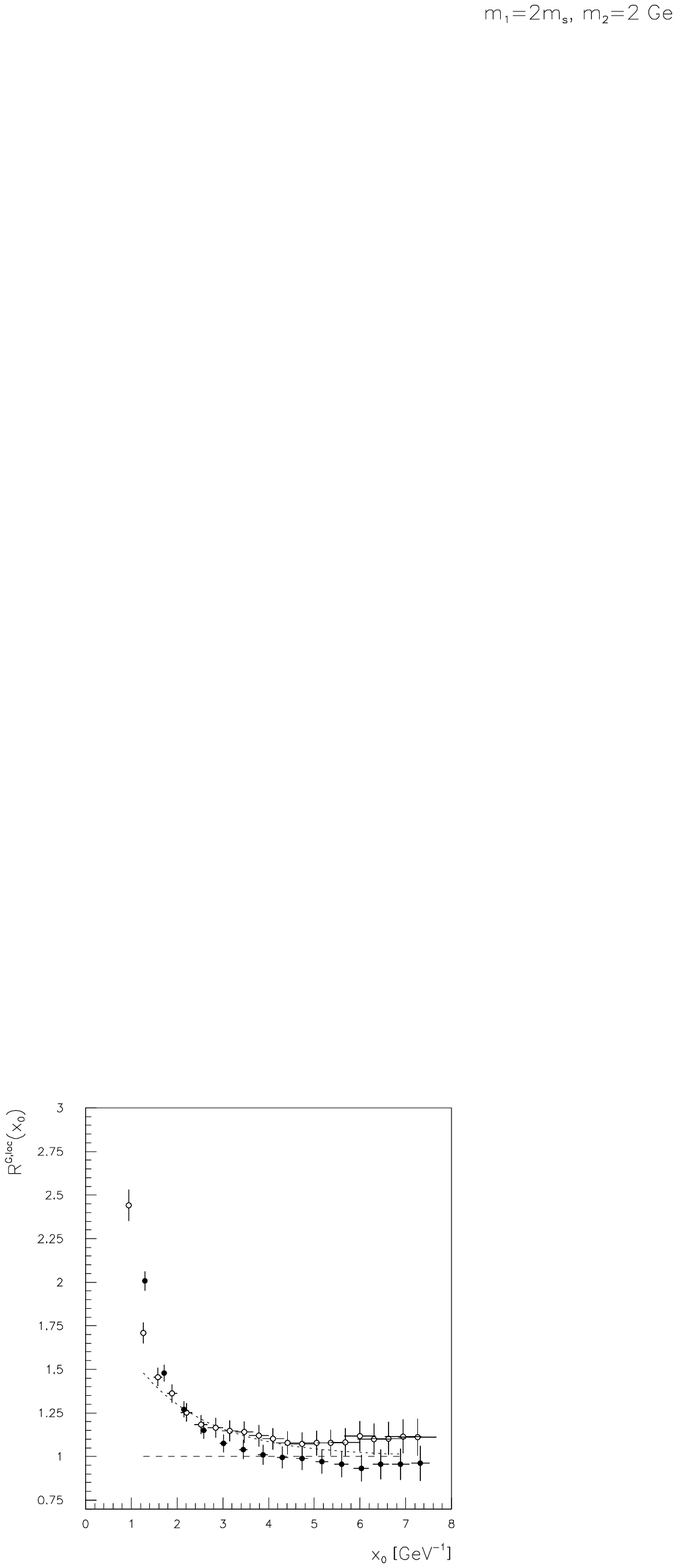}}

\vskip -0.5 true cm

\caption[1]{\footnotesize
 The ratio $R^{G,loc}(x_0)$ is shown for a light quark mass of
 about twice the strange quark mass and a heavy quark mass of
 roughly the charm mass. Open circles are for a cutoff of
 $a^{-1}=3.17$GeV ($\beta=6.26$)
 and full points are for $a^{-1}=2.32$GeV ($\beta=6.00$).
 The lattice spacing and thus the scale on $x_0$ are defined
 through eq.~(\ref{afpi}).
                                                       }\label{f_ratio}
}
%%\endinsert
\end{figure}

Since most computations of $f_D$ have been carried out implementing
 the local
axial vector current, it is of interest to use this ratio
 to further investigate in how far the  correlation function
is dominated by the ground state at accessible values
of $x_0$. So we display
$R^{G,loc}(x_0)$ in fig.~\ref{f_ratio} where $G$ refers to the
gaussian smearing
already used in fig.~\ref{f_locmas} and $loc$ denotes the local
axial vector current.
$R^{G,loc}(x_0)$ is only consistent with one for the largest
values of $x_0$.
Therefore, we consider
the region of $x_0$, where all correlation functions
are dominated by the lowest {\it two} states. There, $R^{I,J}(x_0)$ is
described by
\bes
R^{I,J}(x_0)&=&\frac{[1 + \eps_I^2 \exp(-x_0 \Delta )]
                   [1 + \eps_J^2 \exp(-x_0 \Delta )]}
                  {1 + \eps_I \eps_J \exp(-x_0 \Delta )}~,
                \\ \nonumber
       \eps_I &=& <0|  {\cal M}^{I}(0) |P, 2 >
                 /<0|  {\cal M}^{I}(0) |P, 1 > ~,~~ \Delta=E_2-E_1
~. \label{ratioexp}
\ees
Fig.~\ref{f_locmas} suggests that the smearing wave function was chosen
well resulting in $\eps_G << \eps_{loc}$. We may
therefore simplify further:
$R^{G,loc}(x_0) \sim [1 + \eps_{loc}^2 \exp(-x_0 \Delta)]$.
A fit to this form is shown as dashed curve in  fig.~\ref{f_ratio}.
It describes
the ratio well with a gap $\Delta \sim 600 \MeV$.
We have thus roughly quantified the dominant correction to the 
asymptotic behavior of the local axial vector correlation function.
The size of this correction term  shows us that some of the
earlier
lattice computations that investigated the dependence of
the decay constant of HL mesons on the mass of the meson are
subject to a significant systematic error since they used only
local meson fields.
One should
not regard this investigation  as a quantitative determination of
the gap in this channel.

If one uses $C^{loc,loc}$ and does not
isolate the ground state properly, one obtains an overestimate
of the decay constant. Repeating the above analysis at different
values of the  heavy quark mass shows that
$\eps_{loc}^2$ increases when the
heavy quark mass increases, while $\Delta$ stays roughly
constant\cite{fb4}.
Therefore, the effect becomes stronger with increasing quark mass,
resulting in a misleading quark--mass dependence.
Physically, this effect means that the ratio of
the leptonic decay constant
of the excited state  to the ground state decay constant increases
with the mass of the heavy quark. Unfortunately, at present,  the
precision is not high enough to determine $\eps_{loc}^2$ and its
mass dependence well.
 
A further comment is in order concerning fig.~\ref{f_ratio}:
the figure contains results for two values of the cutoff. The ratio
has a continuum limit only in the region of $x_0$ where it is approximated
well by the truncated form
$R^{G,loc}(x_0) \sim [1 + \eps_{loc}^2 \exp(-x_0 \Delta)]$ that was used in
the fit. At smaller $x_0$ where this is not true, nonuniversal
matrix elements of the trial wave functions appear and there is no reason
to expect an (approximate) independence of the lattice spacing.
Therefore, the a--independence of the ratio that is observed
 for $x_0>1.5$GeV$^{-1}$
is consistent with $\eps_G << \eps_{loc}$, that means a
well chosen trial wave function in refs.~\cite{fb3,fb4}.
 
As a consequence of this investigation,
it appears mandatory to use smeared meson fields in order
to reach asymptotics and to
perform a reliable computation of the decay constant\footnote{
Of course, the problem can also be solved by going to sufficiently
large values of $x_0$ on a large lattice with large statistics
in the Monte Carlo\cite{guido}.}.
In order to extract $f_P$, the matrix element
$<0|  {\cal M}^{I}(0) |P, 1 > $ which is
introduced through the trial wave function, must be cancelled.
As seen from eq.~(\ref{spectmes}) this can
be achieved by combining the results of an analysis
of both $C^{I,I}(x_0)$ and $C^{I,loc}(x_0)$.

\subsection{The Quenched Approximation \label{s_QA}}
As we mentioned in section~(2.1), the first step in a
Monte Carlo evaluation of the
expectation values eq.~(\ref{green}) consists of generating
lattice gauge fields $U$ with probability
$\exp(-S_G[U] ) ~ \det {\cal D}[U]$.
The only  algorithm which  generates this distribution
exactly\footnote{Actually, the HMC algorithm\cite{HMC}
is valid for pairwise mass degenerate quarks only.}
is the Hybrid Monte Carlo algorithm (HMC).
It is numerically quite slow. In particular, the required CPU
time for a given simulation scales\footnote{
This scaling law includes the scaling of the box length $L$ that
is necessary to avoid finite size effects when going to small
quark masses.
}
 approximately like $(m_q a)^{-\eta}$
with $\eta\sim 5$ \cite{GIKP}.
At present, this allows for
exploratory simulations at unphysically heavy masses $m_q$ of
the quarks.
 
In perturbation theory, $\det {\cal D}[U]$ generates quark loops.
One may assume as a starting point, that for the full
path integral at a given value of the cutoff, the effect of
the quark loops is mainly to modify  the renormalization of
the coupling. Therefore, low energy observables may be
approximately represented by a model with
$\det {\cal D}[U] \rightarrow$ const. Also the success of constituent
quark models suggests that neglecting quark loops in this way
may be a sensible approximation to QCD. Clearly, we do
expect significant differences between QCD and its quenched
approximation, in quantities like flavor singlet densities in
hadrons or the connection between the high energy running coupling
and hadron masses. In this sense, it is better to think of
the ``quenched approximation'' as a model. It does not
constitute a systematic approximation to QCD and its accuracy depends
strongly on the quantity that one considers.

\begin{figure}[htb]
%%\topinsert
\vbox{
\vskip 0 true cm

\def\fpsangle{0}
\fpshskip=0. true cm

\centerline{
\fpsxsize=12.5 true cm
%%\fpsbox[80 80 400 400]{GF11_1.ps}}
\fpsbox{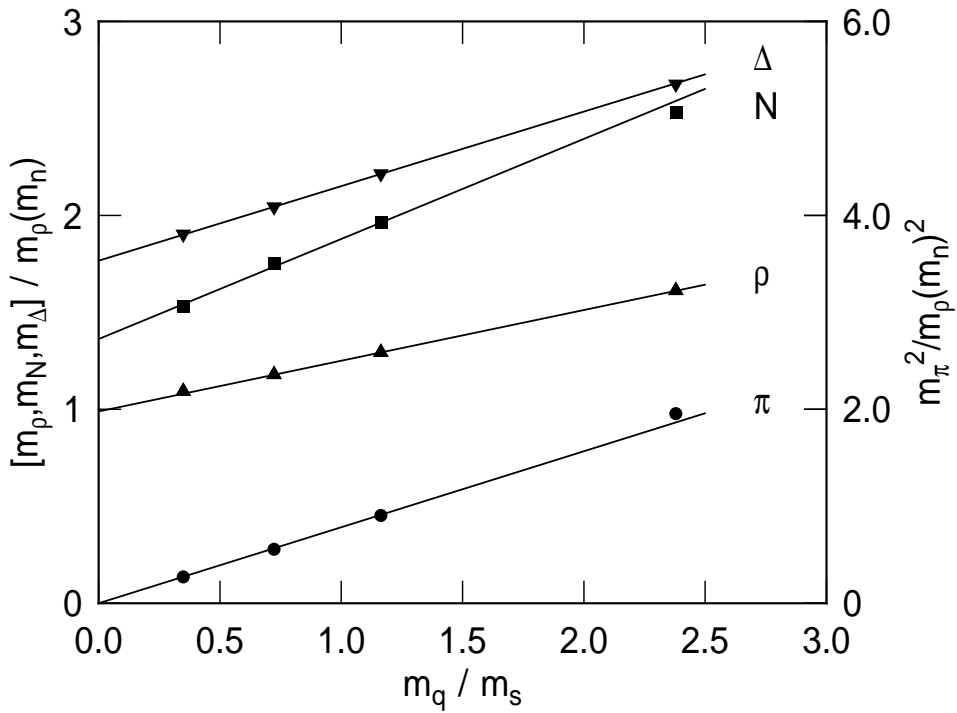}}
%%\vskip -0.5 true cm

\caption[1]{\footnotesize
Extrapolation of hadron masses to the chiral limit.
The extrapolation curves are used to locate the physical point
in the quark mass, that means one point on the renormalization group
trajectory.
The ordinate $m_q/m_s$ is the bare 
quark mass in units of the strange quark mass. $m_{\rho}(m_n)$
denotes the $\rho$--meson mass at the physical point.
The figure is taken from ref.~\cite{GF11_mass};  error bars are smaller
than the size of the symbols.  
                                                       }\label{f_GF11_1}
}
%%\endinsert
\end{figure}

Having emphasized this, we proceed nevertheless
to summarize numerical evidence
that the ``quenched approximation'' is surprisingly close to
the real world. We consider the low lying hadron
mass spectrum.
The large effort that went into computing the hadron
mass spectrum has been reviewed in references \cite{To91,Latt_mass}.
A significant step on the way towards a reliable computation
of the hadron mass spectrum has been taken by the APE--group,
using nonlocal trial wave functions (cf. section~\ref{s_RA})
and quark masses around the strange quark mass\cite{APE}.
 
This has recently been
improved further in the GF11-project\cite{GF11_mass}
reaching a rather high statistical precision.
The latter allowed for a more systematic study of the lattice
spacing dependence and the finite volume effects.
Using the assumption that up and down quark masses are degenerate,
and the quenched approximation, the problem has two parameters
only: $K=K_u=K_d$ and $g_0^2$. To get to the
physical masses in the continuum limit, one proceeds as discussed
in section~\ref{s_CL}.
An important difference to the idealization in section~\ref{s_CL}
is that one cannot compute the quark propagators
at the physical values of $K$ since this corresponds to the
solution of an almost  singular system of linear equations.
Instead, all hadron masses
are calculated at various values of the hopping parameter
$K$ and then extrapolated using forms suggested by
chiral symmetry (for the pion mass) and simple mass perturbation
theory. As an illustration, we show
the extrapolations that were
done in ref.~\cite{GF11_mass} in fig.~\ref{f_GF11_1}\footnote{
There are strong arguments\cite{qu_chir} that the quark--mass dependence
of hadron masses is more complicated in the quenched approximation.
It may be non--analytic at zero quark mass.
It is, however, not obvious at which values of the quark mass such
behavior sets in. Furthermore, such non--analyticities would mean that
observables that are calculated in the quenched approximation
deviate very significantly from the true result when the quark masses
approach zero. Therefore,  one should stay away sufficiently
far from the chiral point and extrapolate using the mass dependencies
expected for full QCD.
}.
 
In ref.~\cite{GF11_mass} the value of $K$ at each $g_0^2$
was fixed requiring $M_{V}(K,g_0^2)/M_P(K,g_0^2)=m_{\rho}/m_{\pi}$,
where $M_V$ is the vector meson mass in lattice units and
$m_{\rho}$ the experimental $\rho$--meson mass (the $\rho$--meson
is a stable particle in the quenched approximation).

\begin{figure}[htb]
%%\topinsert
\vbox{
\vskip 0 true cm

\def\fpsangle{0}
\fpshskip=0. true cm

\centerline{
\fpsxsize=11.5 true cm
%%\fpsbox[80 80 400 400]{GF11_2.ps}}
\fpsbox{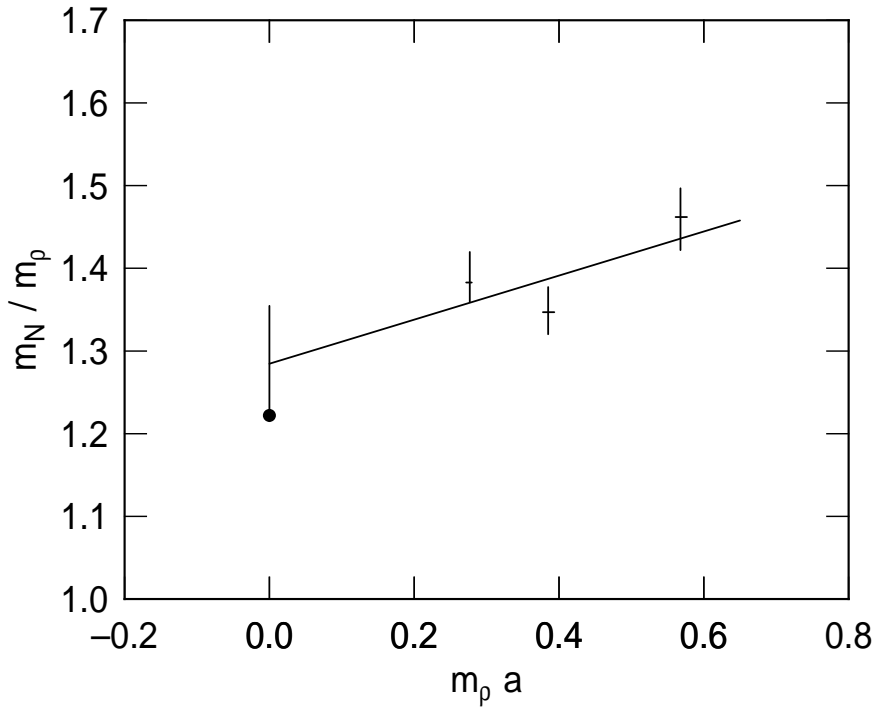}}
%%\vskip -0.5 true cm

\caption[1]{\footnotesize
Extrapolation of the nucleon to $\rho$ mass ratio to the continuum limit.
The figure is taken from ref.~\cite{GF11_mass}.
                                                       }\label{f_GF11_2}
}
%%\endinsert
\end{figure}

The crucial point is now to calculate hadron mass ratios on
the renormalization group trajectory and investigate their
dependence on the lattice spacing measured as
$M_{\rho} \equiv m_{\rho} a$.
Fig.~\ref{f_GF11_2} shows the example of
$\frac{M_N}{M_{\rho}} (M_{\rho})$. A linear dependence on
$M_{\rho}$ (see section~\ref{s_CL}) is in agreement with the data
and gives a value in the continuum limit that is consistent with the
experimental value of the mass ratio.
The same holds true for $m_\Delta/m_{\rho}$.
 
In order to compute
the masses of strange hadrons as well, Butler et al. assumed
first order $SU(3)_{\rm flavor}$ symmetry breaking.
These results,  once they are extrapolated to
the continuum,
are in agreement with the experiments as well.
 
In all these cases, the error bars are somewhat large and the
number of points is too small to check whether
the mass ratios really
depend linearly on the lattice spacing.  Nonlinear terms
could be important in this range of lattice spacings.
To some extent, we can investigate this question by using another
precision--quantity:
the potential $V(r)$ between static quarks
can be calculated very precisely in lattice QCD.
A well calculable length scale $r_0$ may be obtained from the force
$F(r)=\frac{d}{dr} V(r)$  through the implicit definition
$F(r_0)r_0^2 = 1.65$\cite{su2}.
The constant 1.65 is chosen such that one obtains $r_0\sim 0.5$~fm when one
uses
the force derived from potential models that successfully
describe the $b\bar{b}$ spectra (0.5~fm is about the distance, where
the phenomenological potential models are most severely
restricted by the spectrum).
$r_0$ has not been determined at all values of the bare
coupling which we need here. We therefore take the following approach:
at distances
$r\sqrt{\sigma}>0.3$ the potential
is well described by
\be
V(r)=\pi/(12r) + \sigma r ~~,\label{pot}
\ee
with $\sigma$ representing the string tension\footnote{
The subleading term $\pi/(12r)$ is the universal
first correction of an effective bosonic string\cite{Lues_string}. It is
both in agreement with SU(3) gauge theory simulations and with
the force in the continuum limit of the SU(2)-theory\cite{su2}.
}.
We use $\sigma$ as defined in eq.~(\ref{pot}), noting that through the
very parametrization of the potential
one has $\sqrt{\sigma}r_0 \equiv 1.18$. For the precision that is required
in the following, the uncertainty due to the parametrization eq.~(\ref{pot})
is irrelevant. In the future, when hadronic masses with higher precision
are available, it will be important  to use the direct
definition of $r_0$.
 
We have combined the data obtained from the
simulations of ref.~\cite{MTC,BS} as listed in
\cite{fb3} with masses of the $\rho$--meson from ref.~\cite{GF11_mass,
APE,fb4,UKQCD}\footnote{
In three cases (for $L/a M_{\rho}<6$) it was necessary to correct the
mass of the $\rho$--meson
for a slight finite volume effect\cite{UkFS}. The correction was taken
from \cite{UkFS} and its uncertainty was taken into account.
}
 to form the ratio $M_{\rho}/\sqrt{\Sigma}$.
(By $\Sigma$ we denote  the estimate of $\sigma$ at a finite
value of $a$ in lattice units: $\Sigma = a_{\sigma}^2 \sigma$.)

\begin{figure}[htb]
%%\topinsert
\vbox{
\vskip 0 true cm

\def\fpsangle{0}
\fpshskip=0.8 true cm

\centerline{
\fpsxsize=12.5 true cm
\fpsbox[30 30 275 285]{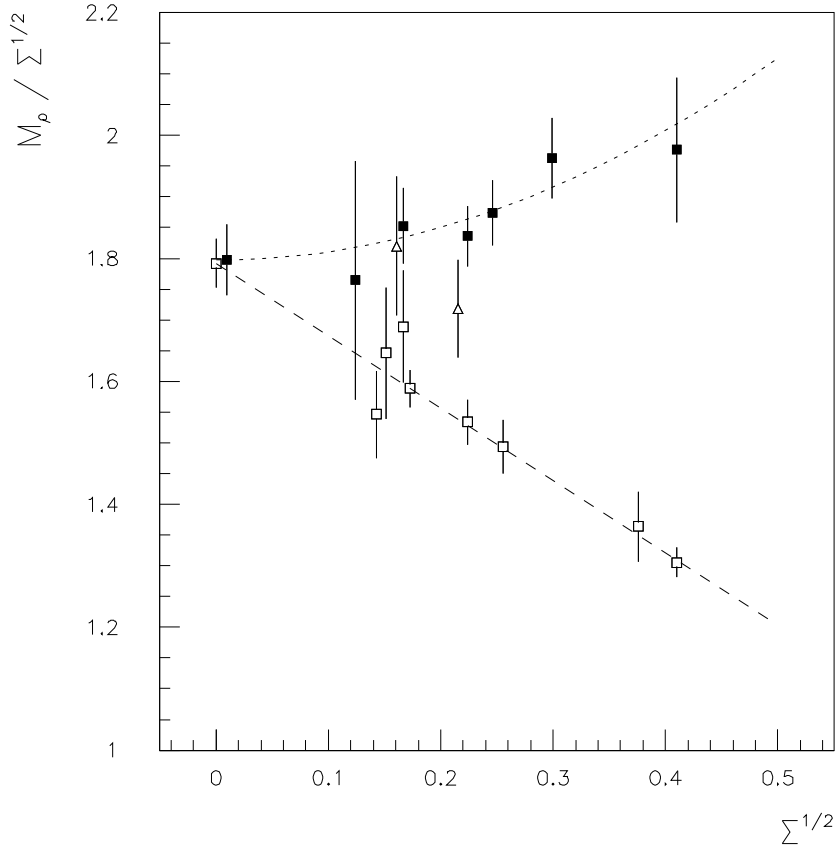}}

\vskip -0.5 true cm

\caption[1]{\footnotesize
 The lattice spacing dependence of $M_{\rho}/\sqrt{\Sigma}$
 for three different fermionic actions: Wilson action (full squares
 \cite{GF11_mass,APE,fb4,UKQCD}), 
 Sheikholeslami Wohlert action 
  (triangles \cite{UKQCD,APEfstat,APEdyn}) and staggered fermion
 action\cite{KS} (filled squares \cite{ape_ks,stag_ks,hemcgc_ks,fmiou_ks}). 
                                                       }\label{f_mrho_sig}
}
%%\endinsert
\end{figure}

The lattice spacing dependence of $M_{\rho}/\sqrt{\Sigma}$
is shown in fig.~\ref{f_mrho_sig} (open squares).
A linear relation describes the data well in the
whole range.
The range of lattice spacings that is covered here
includes the one in the extrapolations
of \cite{GF11_mass} but it extends
further towards $a=0$. The linearity observed in fig.~\ref{f_mrho_sig}
lends support to the extrapolations\cite{GF11_mass} that we discussed above.
 
The extrapolation of $M_{\rho}/\sqrt{\Sigma}$
is further checked by computations that were done
using different forms of the fermion action. 1) The $\rho$--mass
data of various groups\cite{ape_ks,stag_ks,hemcgc_ks,fmiou_ks}
using the staggered fermion action \cite{KS} (filled squares) have
corrections to the continuum limit of
the opposite sign. For this fermion action, mass ratios should approach
their continuum limit with corrections proportional to $a^2$\cite{Shar}.
This form is fitted to the data. As a result we observe
{\bf universality} in the continuum limit despite the quite significant
differences at finite values of the lattice spacing!
The figure underscores the importance of performing such extrapolations
in order to obtain the physical continuum limit numbers.
2) The recent computations by
the UKQCD--group\cite{UKQCD} and the APE--group\cite{APEfstat,APEdyn},
using the  improved action  are represented by the triangles.
 Although the error bars are somewhat large to
 draw definite conclusions, it appears to be close to the above
continuum limit extrapolations indicating small discretization errors.
Further points for larger values of the
lattice spacing in this plot could clearly settle the question to which
extent the improved action of
Sheikholeslami and Wohlert reduces the $O(a)$ scaling
violations in nonperturbative quantities.
 
Altogether the investigation of the lattice spacing dependence and
universality of  $M_{\rho}/\sqrt{\Sigma}$ provides further support
for the continuum limit extrapolations of mass ratios of Butler et al.
As the extrapolated numbers agree with the experimental observations
to within $\sim$10\%, one may conclude
that the quenched approximation is at least a good {\it model}
for QCD. In fact, it appears to be the best model of QCD that we have,
considering that it contains only the fundamental parameters of
QCD as unknowns.
However, we would like to emphasize again that there is an unknown
systematic error involved when we replace QCD by this model for
calculating a new quantity.
 
We do not want to review the numerous efforts that have been made
to directly compare
QCD and its quenched approximation (at a given value of the cutoff).
Instead, we refer the interested reader to review articles on this
subject\cite{To91,Latt_mass}.
Sea quark masses of around the strange
quark mass  and higher have been used so far.
For these quark masses, no significant changes in the physical observables
have been found apart from the aforementioned change in the
coupling constant at the cutoff scale. This is in agreement with
the conclusion drawn in the previous paragraph.

\subsection{The b--Quark on the Lattice \label{s_bL}}
In the previous sections, we have discussed the general problems in practical
lattice QCD calculations. Apart from renormalization of operators, the
dependence of lattice results on the spacing $a$ is most important.
As long as we are considering hadrons consisting of light quarks such as
$u,d$ and $s$, it is sufficient to have a good resolution of the wave--function
in order for $a$--effects to be moderate and to be able to extrapolate to
the continuum limit.
 
When one considers bound states of a heavy quark like the $b$--quark
with other constituents,
it is  necessary to have
\be
  m_h~a << 1~ \label{mhcond}
\ee
in addition. (We denote a generic heavy quark by $h$.)
In a given background gauge field,
the heavy quark only propagates correctly  under this condition.
 
This condition is hard to fulfill for the following reason.
In order to control finite size effects in hadron masses and matrix
elements,
we need to compute with a box size of the order of more than one fermi
(see section \ref{s_FS}).
Even a lattice with $L/a=32$ points has $m_b~a > 1$ for these values
of $L$. Consequently,
a direct simulation of the $b$--quark requires finer lattices than
the ones that have been technologically accessible so far. One had to
search for alternatives to a simulation with fully propagating
$b$--quarks.
 
Intuitively, one expects that (for low energy matrix elements)
the high frequency parts of the
propagation of a heavy quark can be absorbed into local terms
of an effective action\cite{NRQCD}.
It should therefore be possible to work with a modified
action and a cutoff $a^{-1} << m_b$ as long as one is only
interested in long distance observables.
 
Through a formal expansion in $1/m_h$
one obtains the non--relativistic QCD action\cite{NRQCD}
\be
 S_{NRQCD}=\phi^\dagger [D_0 - {{\vec{D}^2}\over{2m_h}}
                             - {{\vec{\sigma} \vec{B} }\over{2m_h}} ]\phi
                             + O(1/m_h^2) , \label{NRQCD}
\ee
written here in terms of the covariant derivative $D_{\mu}=(D_0,\vec{D})$ 
\cite{LQCD} and
the chromo magnetic field strength, $\vec{B}$. In this formulation,
the heavy quark field $\phi$
is  a 2--component spinor 
field and $\vec{\sigma}$ is a vector composed of the three
Pauli matrices.
 
Beyond the formal level, the use of the effective action eq.~(\ref{NRQCD})
is nontrivial.
The theory defined through $S_{NRQCD}$ and its appropriate regularization
is nonrenormalizable. A finite cutoff has to be kept. The
condition $a^{-1} << m_b$ requires that computations have to be done
just in the opposite range from eq.~(\ref{mhcond}).
This means the lattice spacing is in the middle of the range
in fig.~\ref{f_mrho_sig}. For such resolutions, one has to worry about
the effects of
order $O(a)$ as it is clearly demonstrated in that figure.
These cannot be removed by extrapolation $a \rightarrow 0$
since that limit does not exist.
Instead, $O(a)$--effects have to be reduced by adding new (higher dimensional)
operators to the action. Their coefficients need to be determined by
matching a number of observables in the effective theory
to the observables in the full theory.
One possibility is to match directly to experimental observations.
This procedure both reduces the predictability of the theory and introduces
statistical errors into the coefficients.
The second possibility is to estimate the coefficients through mean field
theory or by matching perturbatively\cite{PT_NRQCD}.
 
Concerning the second choice,
it is likely that incalculable nonperturbative contributions
in the coefficients induce finite terms into the final physical matrix
elements  that are 
of the same order as the physical result that one wants
to calculate\cite{MaMaSa}. This possibility originates from the
powerlaw divergencies of the effective theory\cite{MaMaSa}.
It is argued, however, that such terms should be small
numerically\cite{Mack,Lepa}.

According to our judgment, it will be
difficult to quantify the
uncertainty due to either missing higher order terms in the
action or the incalculable nonperturbative contributions to the
coefficients of the effective action.
Currently, non--relativistic QCD on the lattice
is under intense
investigation\cite{Latt93}.
 
It is important to notice, that these problems of non--relativistic QCD
do not exist when one truncates the action eq.~(\ref{NRQCD}) after the
first term\cite{Zstat1,Zstat2}. It has been suggested by Eichten\cite{Eich} to
apply this $O((1/m_h)^0)$ approximation to the b--quark on the lattice.
This treatment of a heavy quark  is called the {\bf static approximation},
since the heavy quark is represented
by a static color source \cite{Eich,EF}. Its propagator $S_h(x,y)$ in a
given background field is just a Wilson line:
\bes
S_h(x,y) &=&
 \delta_{\vx,\vy} ~~(1 +\gamma_0 )/2~~W^{\dagger}(x_0,y_0;\vec x)
~~~{\mbox{for}} ~~ x_0>y_0~~; \label{statprop} \\
 W(x_0,y_0;\vec x) &=& \prod_{z_0=x_0}^{y_0-a}~U_0(z=(z_0,x_1,x_2,x_3))~.
\nonumber
\ees
In the static approximation, the only powerlaw divergent -- and
therefore incalculable in perturbation theory -- renormalization
is the renormalization of the quark mass. The latter  is of limited
physical interest. Most importantly, the renormalization of the
axial vector current can be calculated in perturbation
theory\cite{Zstat1,Zstat2} allowing for a computation of the decay constant
in this approximation.
 
One can therefore compute properties of  B--mesons
 through the following strategy. As a first step, one determines their
limiting behavior for $m_h\rightarrow \infty$ using the
static approximation.
Then one investigates their dependence  on the mass of the heavy quark
for  masses as large as possible and
finally one interpolates between the results at finite values
of the meson mass and in the static approximation.
The interpolation is done in the form suggested by the $1/m_h$--expansion,
but the coefficients of the different powers of $1/m_h$ are not
computed explicitly since they suffer from
powerlaw divergencies\cite{MaMaSa}. Rather they are taken from the
phenomenological
matching between finite mass and infinite mass results.
\newpage

\section{~The Leptonic Decay Constants $f_{D}$, $f_{D_s}$ \label{s_fd}}
 
This section is a review of
 the computations of heavy light decay constants for
heavy quark masses around the mass of the c--quark.
As a general orientation, we list in table~\ref{t_fd} the
results that have been quoted in the literature. Note that the different
computations have used different renormalization constants
and different ways to determine the lattice spacing. At finite values
of the lattice spacing and in the quenched approximation
this can introduce quite significant variations.
This explains the spread of the values given in
table~\ref{t_fd}.
 
% ----------------------------------------------------------------------
\begin{table}[htb]
 
\centering
\begin{tabular} { c c c c c }
Ref.   & $f_D$[MeV]       & $f_{D_s}/f_D$  & $a^{-1}_{\sigma}$[GeV]  & $n_f$\\
\hline
\cite{BDHS} & $174\pm 26\pm 46$    & ($f_{D_s} = 234\pm 46\pm 55$ MeV) & 2.2 & 0 \\
\cite{GMPMP}& $197\pm 14$          &                             & 1.9 & 0 \\
\cite{GMPMP}& $181\pm 27$          &                             & 2.6 & 0 \\
\cite{GL}   & $190\pm 45$          & $1.17 \pm 0.22$               & 1.9 & 0 \\
\cite{fb1}  & $198\pm 17$          & ($f_{D_s} = 209\pm 18  $ MeV) & 1.9 & 0 \\
\cite{euro} & $210\pm 15$          &   $1.08 \pm 0.02$                          & 1.9-3.4 & 0 \\
\cite{BLS}  & $208\pm 9\pm 35\pm 12$ & $1.11\pm 0.05$              & 2.9 & 0 \\
\cite{UKQCDfb} & $185^{+4~+42}_{-3~~-7}$ & $1.18\pm 0.02$             & 1.9-2.6 & 0 \\
\cite{fb4} & $170\pm 30$ & $1.09\pm 0.02\pm 0.05$ & 1.1 - 2.8& \\
 & & & & \\
\cite{HH}   & $283\pm 28$          &                             & $\sim 1.5$ & 2 \\
\cite{GC}   & 130 - 300          & ($f_{D_s} = 170 - 395 $ MeV)  & $\sim 1.5$ & 2 \\
\hline
 
\end{tabular}
\caption[t_fd]{\footnotesize
A compilation of published values of the decay constants in the D--system.
Note that the treatment of systematic errors is very different in the
different works( In particular, the large range quoted in \cite{GC} is
due to a conservative estimate of the systematic errors,
while \cite{HH} did not estimate systematic errors.)
All computations use the Wilson action ($c_{SW}=0$) except for
ref.~\cite{UKQCDfb}, which uses the treelevel $O(a)$ improved
action($c_{SW}=1$)\cite{SW}.
The forth column gives the value of the cutoff
estimated from the string tension as
$a^{-1}_{\sigma}$[GeV] = 0.420~/$\sqrt{\Sigma}$ .
         }\label{t_fd}
\end{table}
% ----------------------------------------------------------------------
 
In order to combine the information from the different publications,
it is necessary to separately investigate  each of the possible sources
of systematic errors. A good statistical accuracy is needed before
systematic errors can be investigated in a meaningful way.
The pioneering computations\cite{BDHS,GMPMP,GL,fb1} did not have a
sufficient accuracy for that purpose.
Therefore, we will  only use the results of 
refs.~\cite{BLS,UKQCDfb,fb4,GC,APEdyn}
in the following. It also would have been of great  use to include
the computation that was performed at the smallest value of the lattice
spacing\cite{euro}. Unfortunately, applying what we learned in
section~\ref{s_RA} about the effects of excited
states to that computation shows
that systematic errors due to excited states are of the order of the
statistical errors. Consequently it is  not included in the following.
The same holds true for part of the results of ref.~\cite{fb2};
the conclusions of that paper have to be revised.
The data that we do
include in the following discussion\cite{BLS,UKQCDfb,fb4,GC}
 either was
obtained by appropriate smearing techniques or at sufficiently
large separations $x_0$. Uncertainties due to  excited state
contributions should be well below the statistical errors.

\subsection{Extrapolation to Light Quark Masses  \label{s_ELQM} }
In the work mentioned above,
all extrapolations are performed as follows. Motivated by
chiral perturbation theory\cite{GaLe} one fits
\be
M^2_P(g_0^2,K_f,K_{f'}) = A(g_0^2)
  ({1 \over {2K_f}} +  {1 \over {2K_{f'}}} - {1 \over {\kc}})
\ee
to determine $A(g_0^2)$ and $\kc$. Then the points on the renormalization
group trajectory corresponding to the s-- quark and the u/d--quark
may be fixed by requiring the pion and the Kaon to acquire their
physical mass in units of the string tension or another dimensionful
observable such as $f_{\pi}$
\bes
A(g_0^2)
  ({1 \over {K_u}}  - {1 \over {\kc}})/\Sigma =
m^2_{\pi} / \sigma \\
A(g_0^2)
  ({1 \over {2K_u}} +  {1 \over {2K_s}} - {1 \over {\kc}})/\Sigma =
m^2_{K} / \sigma
\ees
with $m_{\pi} = 137$~MeV, $m_{K} = 490$~MeV and $\sigma \sim 420$~MeV.
 
Other quantities, such as masses and decay constants
of HL--mesons, have a much  weaker dependence on the bare quark mass.
They are extrapolated linearly in $1/K_f$ to the corresponding points
$K_f=K_u$ or $K_f=K_s$. The linear behavior is confirmed
in all simulations (within the errors). So, for the quantities discussed
here, the
 extrapolations appear to be quite
reliable and are not discussed any further\footnote{
We note that the choice of scale (namely $\sigma$ in the above
equations) influences the precise values of $K_u$ and $K_s$.
Numerically, it is not important in the following, however.
}.
 
We proceed now to investigate finite size effects, the
lattice spacing dependence and renormalization in the quenched
approximation to arrive at an estimate of
the decay constant in that model. In a final subsection, we will try to
see to what extent light sea quarks alter that result.

\subsection{Finite Size Effects \label{s_FS}}

At asymptotically large lengths $L$ of the box, the dominant finite size
effects originate from a self interaction of the meson through an exchange of
a particle with the quantum numbers of the vacuum. The particle can
propagate once around the world, introducing an $L$--dependence.
 In the quenched approximation the relevant lightest particle
is the $0^{++}$ glueball. The effect is suppressed exponentially in $L$ with a
decay rate of the inverse glueball mass. For lattice sizes that are being used
in the simulations of HL--mesons the product $L m_{0^{++}}$ is larger than
ten\cite{MiTe,UKQCDmg}.
This mechanism for finite size effects is therefore irrelevant in the present
context.
 
The dominant finite size effects
within an intermediate
regime should originate from a distortion of the
wave function in the finite volume.
In a nonrelativistic potential model, the rms--radius
of an  HL--meson is significantly smaller than the rms--radius of a
LL--meson since the reduced mass is smaller by almost a factor of two.
Taking a look at LL--meson decay constants that are published in
the literature, a size--dependence is not visible for
$L \sqrt{\Sigma} \sim 3$ and larger. As a result, we
clearly do not expect relevant finite size effects for  HL--mesons
once $L$ is in that range.

\begin{figure}[htb]
%%\topinsert
\vbox{
\vskip 0 true cm

\def\fpsangle{0}
\fpshskip=0.8 true cm

\centerline{
\fpsxsize=12.5 true cm
\fpsbox[30 30 275 285]{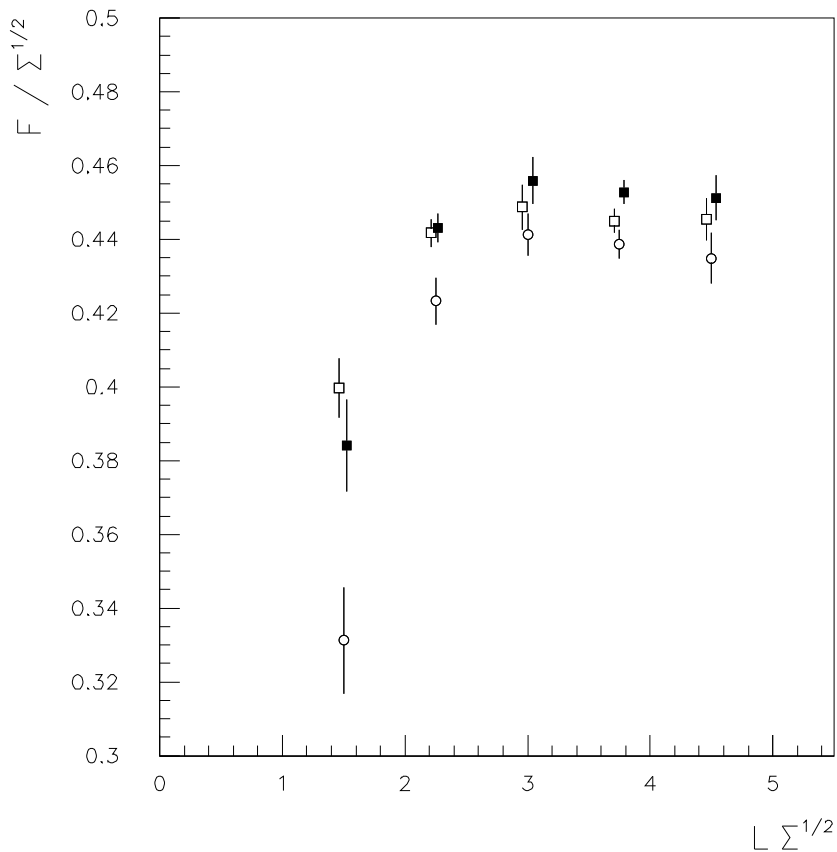}}

\vskip -0.5 true cm

\caption[1]{\footnotesize
 The dependence of the HL pseudo scalar decay constant $F$ on the
box size $L$\cite{fb4}. The light--quark mass is fixed at about twice the strange quark
mass and the meson mass varies from about 0.8 GeV (circles) to 1.5 GeV (filled
boxes).
              }\label{f_f_fs}
}
%%\endinsert
\end{figure}

Indeed, as shown in fig.~\ref{f_f_fs},
numerical investigations with moderate light--quark masses
and varying mass of the heavy quark, confirm the expectation\cite{fb4}.

One may speculate further that  the finite size effects
seen within the
regime of $L \sqrt{\Sigma} \sim 1.5~-~2$ originate from a distortion of the
light-quark wave function in the finite volume.
For an exponentially decaying wave function,
the finite size effects on the wave function at the
origin are again exponential in $L$.
The characteristic scale is the coefficient in the
falloff of the wave function. A rough estimate of 1.5 in units of the
string tension
is derived from the Bethe--Salpeter
wave function of HL and static-light mesons
in Landau gauge\cite{BLS_wf}.
 The data in fig.~\ref{f_f_fs} suggests an even
larger coefficient of the falloff if one fits to an exponential law in the
whole range. Consequently, finite size effects are expected to be smaller
than the statistical errors when $L \sqrt{\Sigma} \sim 3$.
They are neglected in the following. This statement is checked by the data
on the level of $\sim5$\%\cite{fb4}. Note that
finite size effects must be reevaluated
once  future investigations reach higher precisions.

\subsection{Lattice Spacing Dependence and Renormalization \label{s_LSDR}}
The most important systematic uncertainty of the decay constant
originates from the dependence
on the lattice spacing. In principle, the cleanest way of investigating
this dependence and altogether removing the uncertainty through an
extrapolation to the continuum is the following.
 
One fixes the hopping parameter of the light quark $K_u$ as described above
and  the hopping parameter of the charm quark $K_C$ is determined through
\be
M_P(K_C,K_u,g^2_0) /  F_P(K_u,K_u,g^2_0) = m_D / f_{\pi}~.
\ee
At these values of the hopping parameters
one considers the ratio
\be
Q(g^2_0) =  F_P(K_C,K_u,g^2_0) / F_P(K_u,K_u,g^2_0)~.
\ee
Up to lattice artifacts, this is the ratio of the D--meson decay constant to the
pion decay constant $f_{\pi}$:
\be
Q(g^2_0) * f_{\pi} = f_D + O(a)~. \label{Qcont}
\ee
In the literature this is called ``setting the scale through $f_{\pi}$''.
The advantage over every other method is that the renormalization of the
axial vector current (see section \ref{s_R}) cancels, apart from its
dependence on the mass in lattice units, which represents a
lattice artifact. This means that the axial vector
current is renormalized  nonperturbatively.
 
With increased statistical precision,
this method should give the most reliable determination
of $f_D$. At present, the error bars of  $F_P(K_u,K_u,g^2_0)$ are
relatively large and dominate the errors of the ratio $Q$.
Therefore,
the $O(a)$ terms cannot be detected well enough to allow
for a continuum extrapolation using eq.~(\ref{Qcont}).
 
Consequently, it is advantageous to follow a detour when one wants
to exploit the  results presently available.
We split $Q$ into
\bes
Q(g^2_0) &=&  q_1(g^2_0) / q_2(g^2_0), \\
q_1(g^2_0) &=& F_P(K_C,K_u,g^2_0) / \sqrt{\Sigma}~,~~
q_2(g^2_0) = F_P(K_u,K_u,g^2_0) / \sqrt{\Sigma}~~.
\ees
Then $q_1(g^2_0)$ and $q_2(g^2_0)$ can be extrapolated separately  to the
continuum limit and $f_D$ is calculated by their ratio in the continuum limit.
 
Since the renormalization of the axial vector current is known to one--loop
order only, there are in addition to the terms of order $O(a)$,
errors of
order $O(\twiggle{\alpha}^2)$ in $q_1$ and $q_2$.
From the work of Lepage and
Mackenzie\cite{LM}, it appears justified to assume that these missing
terms have coefficients of order one. A coefficient of four would mean
$4~ \twiggle{\alpha}^2  \sim$ 2\% to 8\% in the range of cutoffs
that are  considered below.  As it will become evident below,
an uncertainty of that order does not need to be too disquieting.
In addition, the
same systematic uncertainty is present in $q_1$ and $q_2$ and the effect
of the uncertainty on the extrapolation will cancel partly when one
takes the ratio.

\begin{figure}[htb]
%%\topinsert
\vbox{
\vskip 0 true cm

\def\fpsangle{0}
\fpshskip=0.8 true cm

\centerline{
\fpsxsize=12.5 true cm
\fpsbox[30 30 275 285]{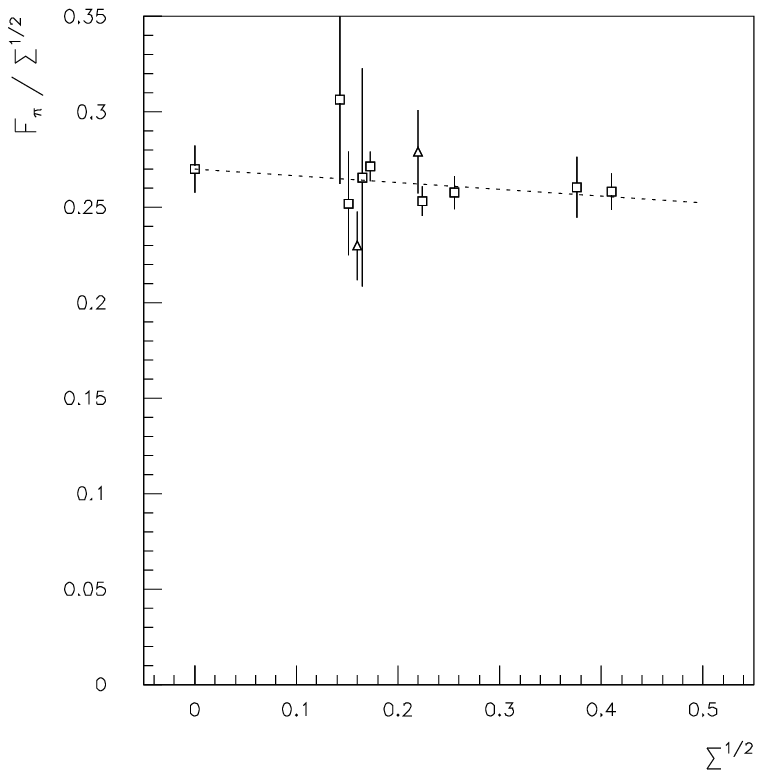}}

\vskip -0.5 true cm

\caption[1]{\footnotesize
The lattice spacing dependence of the pion decay constant.
The data for the pion decay constant was obtained from several simulations.
For the Wilson fermion action (squares): enumerating the points from left 
to right,
ref.~\cite{BLS} contributes to points 1,5,8; ref.~\cite{GF11_f} to 4,6,8;
ref.~\cite{fb4} to 2,5,7  and ref.~\cite{APE} to point 5. Point number 3 as 
well as
the points with the SW action (triangles) are taken from 
ref.~\cite{UKQCD}.
              }\label{f_fpi_sig}
}
%%\endinsert
\end{figure}

The two advantages of this procedure are that i) the lattice spacing
dependence of $q_1$ can be studied with small statistical errors and
ii) additional data at other values of $g^2_0$ are available
for the pion decay constant. Due to this additional data,
most notably the one of ref.~\cite{GF11_f},
$q_2$ can be extrapolated quite well.
To do this, we have taken the data for $f_\pi$ from the
literature\cite{BLS,GF11_f,fb4,APE,UKQCD} and applied the perturbative
renormalization of the axial vector current eq.~(\ref{ZApert}).
 
We show the extrapolation of $q_2$
to the continuum in fig.~\ref{f_fpi_sig}. The lattice spacing dependence
is very weak and we obtain in the continuum limit of the quenched
approximation: $f_{\pi} / \sqrt{\sigma} =
0.270(12)$. The error does not contain the (maybe around 5\%) uncertainty
due to the missing terms of order $O(\twiggle{\alpha}^2)$ in the
perturbative renormalization.
 
Fig.~\ref{f_fpi_sig} also contains results obtained with the improved action.
Within their uncertainty they coincide with the Wilson--data. This is in
agreement with the assumption that both lattice spacing effects and
higher order perturbative terms in the
 renormalization are small.
 
The extrapolation of $q_1$ is a
 more difficult task.  Fig.~\ref{f_fd_sig}
summarizes the available data of sufficient precision.
The filled squares are results with the Wilson action, the normalization
eq.~(\ref{za}) and 1--loop perturbative $Z_A$. The open squares are the same
data with normalization eq.~(\ref{zaKron}). It is more than
apparent
that the lattice spacing dependence is stronger for this normalization.
This matches completely with the case of the local vector current
renormalization discussed in the appendix.
From a theoretical point of view, one can only speculate about this surprising
result. It is helpful to recall that lattice artifacts originate
both from the operators in a correlation function and
from the correction terms in
the action. While eq.~(\ref{zaKron}) is designed to improve the operator at
treelevel, the action remains unchanged. A possible interpretation of
fig.~\ref{f_fd_sig} is that without this ``improvement'' the lattice artifacts
originating from the operator and the one from the  action partly compensate
each other, such that the   full squares show little
a--dependence. On the other hand, the open squares show the lattice
artifacts originating from the action. In this interpretation the latter
 show the
``natural'' size of lattice artifacts which  appears to be somewhat reduced
when one systematically improves both the action and the operators
(triangles).

\begin{figure}[htb]
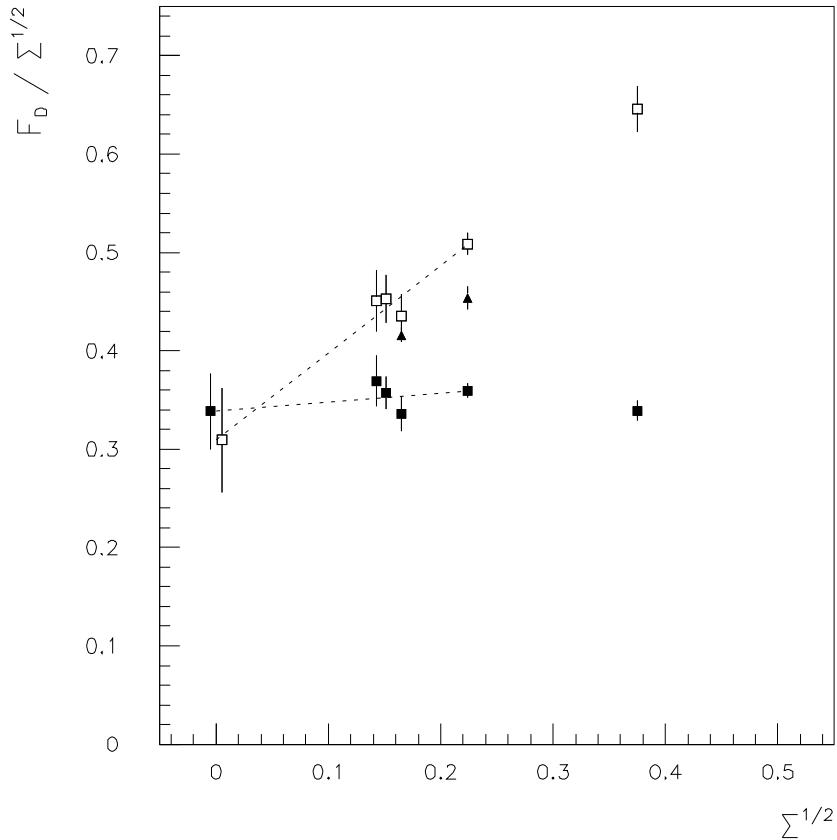

%%\topinsert
\vbox{
\vskip 0 true cm

\def\fpsangle{0}
\fpshskip=0.8 true cm

\centerline{
\fpsxsize=12.5 true cm
\fpsbox[30 30 275 285]{fd_sig.psn}}

\vskip -0.5 true cm

\caption[1]{\footnotesize
The lattice spacing dependence of the D-meson decay
constant. The data with the Wilson action are from ref.~\cite{BLS} (first point)
ref.~\cite{APEdyn} (third point) and ref.~\cite{fb4}. 
The triangles represent the results with the improved action. They
are from ref.~\cite{APEdyn} (first point) and
 \cite{UKQCDfb} (second point).\label{f_fd_sig}
}
}
%%\endinsert
\end{figure}

Without performing any fit, by mere inspection of the data points, all three
data sets are consistent with extrapolating to the same value in the
continuum limit $\Sigma=0$.
The question is, whether any one of the data sets can be extrapolated to
$\Sigma=0$ reliably. The large difference of the different data sets
simply reflects that the mass of the charm quark in lattice units
is quite large at least for  $\sqrt{\Sigma}>0.3$.
Therefore, we omit the  point with the largest value of the lattice spacing
and extrapolate the others linearly
in $\sqrt{\Sigma}$. The points at $\Sigma=0$ in fig.~\ref{f_fd_sig}
show the results of the
extrapolations. They are consistent and furthermore, results within those
error bars would be obtained if one included the points at the largest
value of the lattice spacing.
We take this as evidence  that the error bars include the systematic error
from the extrapolation. In particular, the full squares extrapolate to
$f_D/\sqrt{\sigma} = 0.338(39)$. The uncertainties due to the perturbative
renormalization are as above.
 
\subsection{Estimate in the Quenched Approximation \label{s_EQA}}
 
We combine $f_D/\sqrt{\sigma} = 0.338(39)$ and
$f_{\pi}/\sqrt{\sigma} = 0.270(12)$ to attain
\be
f_D/f_{\pi} = 1.25(15)~\label{fD}.
\ee
Here, we have neglected the $O(\tilde{\alpha}^2)$ renormalization error.
This uncertainty, which may be sizeable in the individual ratios $q_i$,
should cancel out to a large degree when one forms $f_D/f_{\pi}$.
In particular, we may also regard the fitted function in fig.~\ref{f_fpi_sig}
as a legitimate interpolation of the data. Since it does not have any
significant $a$--dependence, the ratio $F_D/\sqrt{\Sigma}$ contains to a
reasonable
approximation the full $a$--dependence of the ratio $Q$.
As explained above, $Q$ has only $O(a)$ effects and no $O(\tilde{g}^4)$
corrections. Consequently, the linear extrapolation of $q_1$ and $q_2$
separately seems justified at the present level of precision.
 
The above value corresponds to $f_D=164(20)$~MeV.
 
Regrettably, the computations with the improved action do not yet cover a
sufficient range of lattice spacings to perform such an analysis.
As seen in figs.~\ref{f_fpi_sig},~\ref{f_fd_sig}, the data are in agreement
with the ones obtained with the Wilson action so far.
 
\subsection{Full QCD \label{s_FQCD}}
 
Simulations of full QCD have so far mainly been carried out using the
staggered fermion action\cite{KS}.
In this case, one has fewer degrees of freedom,
which renders the simulations more tractable. Unfortunately, no results for
HL--mesons exist with that discretization of QCD. It appears to be particularly
difficult to find operators that have sufficient projection onto the ground
state in the HL channel when using this action\cite{Greg}.
 
In two publications, listed as the last two lines in table~\ref{t_fd},
decay constants of HL--mesons are computed with the Wilson action.
Ref.~\cite{HH} presents a very
exploratory simulation with an approximate algorithm to account for the fermion
determinant.  On the other hand, ref.~\cite{GC} simulates with the
Hybrid Monte Carlo algorithm with two different sea quark masses,
one of which is some 25\% below the strange quark mass and the second one
is $\sim$15\% above the strange quark mass. The lattice spacing determined
from the mass of the $\rho$--meson is around
$a = 1/(1.5~$GeV). This corresponds to $\sqrt{\Sigma} \sim 0.38$, the
 point with the largest lattice spacing in fig.~\ref{f_fd_sig}.
Consequently, the systematic errors
due to a finite lattice spacing are  estimated to be rather large\cite{GC},
which results in the wide range for $f_D,~f_{D_s}$ that is
quoted in table~\ref{t_fd}.

\begin{figure}[htb]
%%\topinsert
\vbox{
\vskip 0 true cm

\def\fpsangle{0}
\fpshskip=0.8 true cm

\centerline{
\fpsxsize=12.5 true cm
\fpsbox[30 30 275 285]{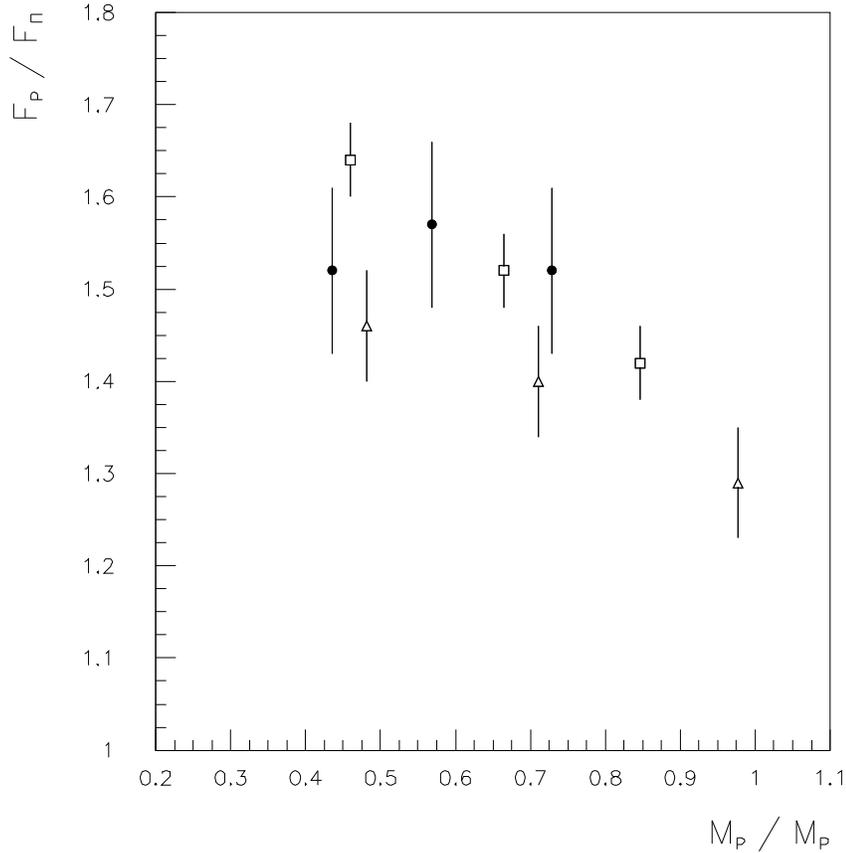}}

\vskip -0.5 true cm

\caption[1]{\footnotesize
The ratio $F_P(g^2_0,K_l,K_h)/F_P(g^2_0,K_u,K_u)$
as a function of $M_V(g^2_0,K_u,K_u)/M_P(g^2_0,K_l,K_h)$.
Open symbols are for $n_f=2$ \cite{GC} with $K_l$ somewhat below $K_s$ (squares)
and $K_l$ somewhat above $K_s$ (triangles). Filled circles are for $n_f=0$
and $K_l = K_s$ \cite{fb4}.
                                                       }\label{f_fdyn}
}
%%\endinsert
\end{figure}

We want to go somewhat beyond the interpretation of the
data presented in ref.~\cite{GC}.
For the hadron mass spectrum, dynamical fermion simulations have yielded
results that are completely compatible with the quenched spectrum
{\it at the same value
of the lattice spacing}. It is in fact quite plausible that one should compare
at a fixed value of the cutoff, since if one has a small effect of quark
loops on a certain observable it is natural that also the lattice artifacts
do not feel much of the presence of dynamical fermions.
 
In order to test to what extent
this statement is correct for HL decay constants,
we compare the numbers given in the tables in ref.~\cite{GC} with the quenched
data of ref.~\cite{fb4} at $\sqrt{\Sigma} \sim 0.38$.
The systematic uncertainty on $Z_A$ is eliminated by looking at the ratio
of HL--decay constants to the pion decay constant\footnote{
We used the normalization eq.~(\ref{za}). Inserting eq.~(\ref{zaKron}) would
result overall in quite different numbers, but the agreement of the
quenched QCD with the full QCD results would be just as good.
}.
 A further uncertainty
originates from the masses of the quarks.  In fig.~\ref{f_fdyn},
we show the quenched result
for the strange quark mass and the results with $n_f=2$ for the two sea quark
masses that straddle that mass.
The heavy quark mass varies along the $x$--axis where the mass of the meson
is given in units of $m_{\rho}$.
 
Remembering that the quenched mesons have a light--quark mass in between the
two (independent) simulations with dynamical fermions,
there is complete agreement between  the two sets of data.
We may take this as positive evidence - albeit with a precision of maybe
10\% - that there is no effect of dynamical fermions on the decay constants.
Of course, this is a check at {\it one} relatively large value of the lattice spacing
and one cannot exclude a somewhat different result at a smaller value of $a$.
We should remember further that the light quarks have masses around
the strange quark mass in this test.
 
Further tests with smaller $a$ and smaller quark masses will become
possible  soon.
\newpage
 
\section{~The Leptonic Decay Constants $f_B$, $f_{B_s}$ \label{s_fb}}
 
Computations of properties of  B--mesons are considerably more difficult than
the ones discussed in the previous section because of the large mass of the
b--quark (cf. section \ref{s_bL}). At present, appropriate values of the
lattice spacing, i.e. such that $m_b a << 1$,  have not been reached.
 
Therefore, the estimation of $f_B$  starts from the limiting value obtained
in the approximation of neglecting terms of order $O(m_b^{-1})$. Next, one
investigates the mass dependence of HL decay constants for varying masses
of the heavy quark, where those masses are around the mass of the
charm quark. Finally, one matches the two results through some interpolation
and arrives at an estimate of  $f_B$. We describe these steps
in the following three sections.
 
\subsection{Static Approximation \label{s_SA}}
Using a static propagator eq.(\ref{statprop}) for a B--meson,
the (bare) axial vector
correlation function does not depend on the mass of the heavy
quark. So, the combination $F_P\sqrt{M_P}$ does not depend on the mass
of the quark in the bare regularized theory.
Since we are working with an effective action for the heavy quark, we have to
restore the relation to QCD by matching the effective theory to QCD
\cite{Zstat1,Zstat2}. It is most transparent to require directly that
the
correlation function eq.~(\ref{spec}) is the same in the effective theory and
in QCD up to terms of order $O(m_h^{-1})$ \cite{Zstat1}. To 1--loop order
perturbation theory this
is achieved after accounting for a (linearly divergent) subtraction of the
binding energy and a renormalization of the axial vector current by a
factor\cite{Zstat1,Zstat2,Zstat3}
\bes
Z_{\mbox{stat}}(a~m_h) &=& 1 + \twiggle{\alpha} ~
[\frac{1}{\pi} \log (a~~m_h) - 2.372]~,
              {\mbox{for}}~c_{SW}=0~, \label{zstat} \\
Z_{\mbox{stat}}(a~m_h) &=& 1 + \twiggle{\alpha} ~
[\frac{1}{\pi} \log (a~~m_h) - 1.808]~,
              {\mbox{for}}~c_{SW}=1~.
\ees
In the static approximation one thus computes
\be
\fhat^{\mbox{stat}} = Z_{\mbox{stat}}(a~m_b)
<0|\bar{q}_h\gamma_0\gamma_5q_l(0)|P>|_{\mbox{stat}}~,
\ee
where, because of the physical interest, we have renormalized at $m_b$.
$\fhat^{\mbox{stat}}$ contains the full information
on the decay constant that can be obtained
in the static approximation in the form\cite{SVPW}
\bes
\fhat^{\mbox{stat}} &=& \lim_{ M_P \rightarrow \infty} \fhat(M_P)
                             \label{fhatstat}\\
\fhat(M_P) &=& F(M_P) \sqrt{M_P}
 \left (\frac{\alpha_s(M_P)}{\alpha_s(M_B)} \right)^{2/\beta_0}~;
~~\beta_0=11-\frac{2}{3}n_f~,
\label{fhat}
\ees
where  $n_f$ is the number of dynamical
quark flavors (Here  the leading logarithms have been summed using the
renormalization group equation and the difference between the mass of the
meson and the mass of the quark has been neglected).
 
As a first overview of the effort that went into the computation of
$\fhat^{\mbox{stat}}$, we list in
table~\ref{t_fstat} the different published computations of
$f^{\mbox{stat}} = a^{-3/2} \fhat^{\mbox{stat}} / \sqrt{m_B}$. Here,
one has to note, that part of the scatter in the table originates from
different ways of setting the scale and other systematic errors, which we
attempt to analyze  below.
 
% ----------------------------------------------------------------------
\begin{table}[htb]
 
\centering
\begin{tabular} { c c c c c }
Ref.   & $f^{\mbox{stat}}_{B_d}$[MeV]  & $f^{\mbox{stat}}_{B_s} /
f^{\mbox{stat}}_{B_d} $
 & $a^{-1}_{\sigma}$[GeV]  & $c_{SW}$\\
\hline
\cite{BPHSM} & $<\sim 260$    &  & 1.9 & 0 \\
\cite{Allt1}& $310\pm 25\pm 50$          &                         & 1.9 & 0 \\
\cite{fb1}  & $366\pm 22 \pm55$          & 1.10 & 1.9 & 0 \\
\cite{fb3}  & $230\pm 22 \pm26$ & 1.16 & 1.4 - 2.7 & 0\\
\cite{APEfstat} & $350\pm 40\pm30$          &     $1.14\pm 0.04$   & 1.9 & 0 \\
\cite{Eicham} & $319\pm  \times\frac{Z}{0.79}
     \times(\frac{a^{-1}}{1.75GeV})^{3/2}$                &        & 1.6 & 0 \\
\cite{BLS}  & $235\pm20 \pm21 $ & $1.11\pm 0.04$              & 2.9 & 0 \\
\cite{APEfstat} & $370\pm 40$          &                 & 1.9 & 1 \\
\cite{UKQCDfb} & $253^{+16~+105}_{-15~~-14}$ & $1.14^{+0.04}_{-0.03}$
         & 2.6 & 1 \\
\hline
 
\end{tabular}
\caption[t_fstat]{\footnotesize
A compilation of published values of the decay constants in the B--system
as they were obtained in the {\it static approximation}.
The fourth column gives the value of the cutoff
estimated from the string tension as
$a^{-1}_{\sigma}$[GeV] = 0.420~/$\sqrt{\Sigma}$ .
         }\label{t_fstat}
\end{table}
% ----------------------------------------------------------------------
 
{\bf Ground state domination \\}
The question of how to extract the ground state meson decay constant
has been discussed quite extensively over the last years. The problem is
described in section~\ref{s_RA}. Here, we summarize what is known in
particular for mesons in the static approximation.
 
Following the original suggestion of Eichten \cite{Eich} to compute
$\fhat^{\mbox{stat}}$, it was noted by Boucaud et al. \cite{BPHSM}, that
when one employs the static approximation,
the statistical fluctuations increase rapidly with the euclidean time
separation $x_0$.  Therefore, with local operators one never observes a
significant plateau in the effective mass.
For that reason, they could only give an upper bound for the decay constant.

Subsequently, nontrivial wave functions were used\cite{Eich90,Allt1,fb1} and
$\fhat^{\mbox{stat}}$
could first be estimated in \cite{Allt1,fb1}. Quite some  effort  went
into the construction of good wave functions for static--light mesons.
Nevertheless,
in all computations that have been done so far, one sees only
relatively short plateaus in the
effective masses. Ref.~\cite{Jap} stimulated the discussion on this  point.
At  present the best solution to the problem is given in
ref.~\cite{Eicham}. There,  the
smearing wave function is determined through the diagonalization
of a matrix correlation
function. One thus obtains an optimal wave function out of a space
of linear combinations of a number of basis functions.
As a result, out of all computations, the plateaus in the effective
masses are longest and most convincing in ref.~\cite{Eicham}. Unfortunately,
only a result at one value of $a$ is presently published with that method.
 
A point of discussion is also how to fit the two correlation functions
$C^{I,I}$ and $C^{I,loc}$ in order to extract the matrix element of the local
axial vector current.
The analysis of refs.~\cite{Eicham,BLS} is based on a combined fit
to $C^{I,I}$ and $C^{I,loc}$. The appropriate range of the fits is determined
from their $\chi^2$. Since a large correlation matrix has to be
determined
from a quite limited statistical ensemble, this criterion may be
misleading\cite{fb3,covar}.

The other groups determine the (linearly divergent) binding
energy $\tilde{M}_P$ 
from the correlators $C^{I,I}$, where $I$ labels the smearing wave
function. These correlation
functions have a positive spectral representation and  the corresponding
effective mass is easy to interpret. Also,  because it is the correlation
of two smeared fields, the unwanted excited states are more
strongly suppressed than in $C^{I,loc}$. The analysis then proceeds
in the following way:
either one uses the binding energy
as extracted from $C^{I,I}$, performs a fit to
$C^{I,loc}$ constraint by this energy  and then obtains the decay constant
from the amplitudes in the two fits\cite{fb1} or one determines the
height of the plateau of the ratio  $C^{I,loc}/C^{I,I}$
and combines that with the amplitude of the fit to $C^{I,I}$. For further
variations on the theme, see ref.~\cite{fb3,fb4}.
Concerning this analysis method, one may argue\cite{BLS} 
that the determination of the
binding energy from
$C^{I,I}$ is somewhat uncertain because there are larger statistical errors
in $C^{I,I}$ than in $C^{I,loc}$.
 
Altogether, some amount of ambiguity remains and one may suspect that
some of the published results have an underlying systematic error which
might be even as large as the statistical error.

We note, however, that there are nontrivial checks on the results.
In particular, at $\beta=6.0, c_{SW}=0$, there are five computations
\cite{Allt1,fb1,APEfstat,fb3,BLS}. Ref.~\cite{APEfstat} finds that the fits
in \cite{Allt1} started somewhat early, introducing a small systematic error.
Ref.~\cite{fb3} extends the calculation of ref.~\cite{fb1}.
Therefore, we compare \cite{APEfstat,fb3,BLS}. The interesting point is
 that three
different types of
wave functions were used. Refs.~\cite{APEfstat,BLS} used a wave function
which is a three--dimensional cube for the light quark, implemented in
Coulomb gauge. The sizes of the cubes were
5--7 lattice spacings\cite{APEfstat} and 9 lattice
spacings\cite{BLS}. In obvious notation we label them by $C5 - C9$ .
Ref.~\cite{fb3} used two different, gauge covariant, wave
functions. One has an approximately gaussian shape (`$G$') and the other one an
approximately exponential shape (`$E$'). We can compare the results from the
three
different wave functions directly at one value of the light quark hopping
parameter $K=0.1540$.
The results quoted are
$<0|\bar{q}_h\gamma_0\gamma_5q_l(0)|P>_G=0.36(3)$,
$<0|\bar{q}_h\gamma_0\gamma_5q_l(0)|P>_E=0.37(2)$ \cite{fb3} and
$<0|\bar{q}_h\gamma_0\gamma_5q_l(0)|P>_{C5}=0.41(5)$,
$<0|\bar{q}_h\gamma_0\gamma_5q_l(0)|P>_{C7}=0.38(3)$ \cite{APEfstat}.
So the decay
constants are in good agreement for these four  different wave functions.
Ref.~\cite{BLS} obtains a somewhat low value
$<0|\bar{q}_h\gamma_0\gamma_5q_l(0)|P>_{C9} = 0.31(4)$ due to the
analysis method discussed above.
Moreover, it is  interesting to note that the systematic effect that
arises from
a fit to $C^{G,loc}(x_0)$ at too small a value of $x_0$ is to decrease the
value of the decay constant, while for $C^{E,loc}(x_0)$ this results in an
overestimate of the decay constant. So the agreement that is found at large
values of $x_0$ is nontrivial and suggests that $x_0$ is large enough in the
analysis of ref.~\cite{fb3}. Also in the computations with the action of
Sheikholeslami and Wohlert\cite{UKQCDfb,APEfstat},
two different wave functions are seen to give consistent results at
$\beta=6.0$.
These consistency checks indicate that the problem of contaminations from
excited states is not too severe.

\begin{figure}[htb]
%%\topinsert
\vbox{
\vskip 0 true cm

\def\fpsangle{0}
\fpshskip=0.8 true cm

\centerline{
\fpsxsize=12.5 true cm
\fpsbox[30 30 275 285]{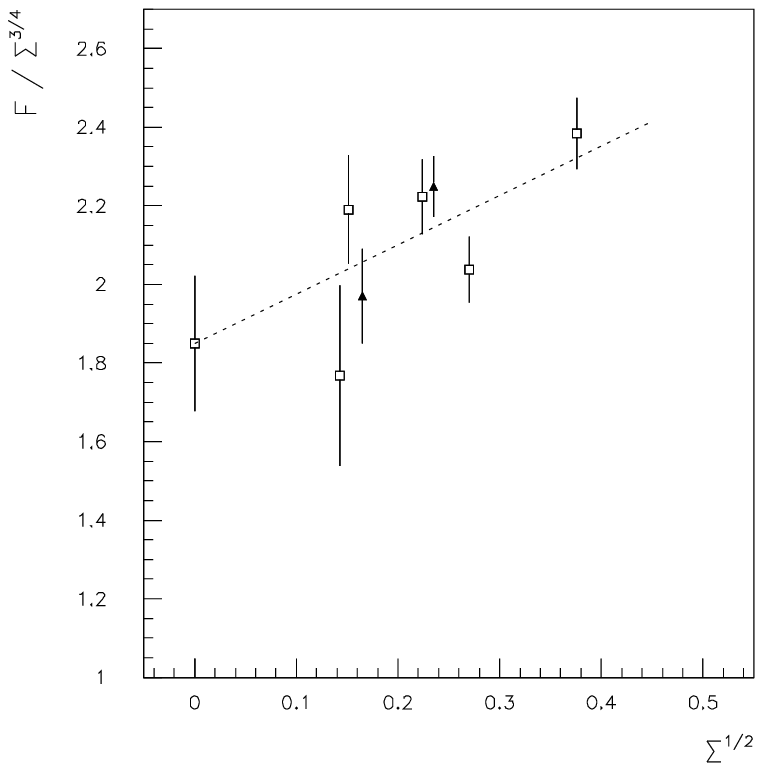}}

\vskip -0.5 true cm

\caption[1]{\footnotesize
$\fhat^{\mbox{stat}}$ in units of the string tension as a function of the 
lattice spacing.  Data of refs.~\cite{fb3,Eicham,APEfstat,BLS,UKQCDfb} are shown
and have been combined, when they were available at overlapping values of the
lattice spacing. Open boxes correspond to the Wilson action and filled triangles
to the improved action.                     }\label{f_fstat}
}
%%\endinsert
\end{figure}

{\bf Finite size effects} of $\fhat^{\mbox{stat}}$ were investigated in
refs.~\cite{fb3,Eicham}. We can already infer from fig.~\ref{f_f_fs}, that
they should be negligible once $L\sqrt{\Sigma}=3$ (the figure shows that
the finite size effects decrease as the heavy quark mass increases). This is
in agreement with the study of refs.~\cite{fb3,Eicham}.

{\bf Lattice spacing effects} are a significant remaining source of
a systematic error.
In order to investigate it, we consider  $\fhat^{\mbox{stat}}$ in units of the
string tension.
The data of refs.~\cite{fb3,Eicham,APEfstat,BLS,UKQCDfb} is shown
in fig.~\ref{f_fstat}. Here, the renormalization constant was obtained by
using the perturbative expansion in terms of $\twiggle{\alpha}$.
Overall, the lattice spacing dependence appears not to be very strong.
This statement refers precisely to the ratio
$\fhat^{\mbox{stat}}/\Sigma^{3/2}$:
In most of the early computations, the mass of the $\rho$--meson was used
to set the scale.
Due to the strong dependence of $M_{\rho} / \sqrt{\Sigma}$ (with $c_{SW}=0$)
on the lattice spacing, there is a strong dependence on $a$ when the scale is
set through  $m_{\rho}$. This explains why the earlier computations,
working at $\sqrt{\Sigma} > 0.2$ and setting the scale through
$m_{\rho}$, obtained quite
large values of $\fhat^{\mbox{stat}}$ (see table~\ref{t_fstat}).
 
The data with $c_{SW}=0$ is extrapolated to zero lattice spacing as shown by
the dashed line: $\hat{f}^{\mbox{stat}} / \sigma^{3/4} =1.85(17) $.
One can see at a glance that the fit does not
have a good $\chi^2$. However, it is obvious from the figure that
this is not due to deviations from the assumed linear dependence on the
lattice spacing. Rather, it is due to the scatter of the data itself.
The most likely explanation for the scatter is incomplete ground state
domination in some of the data points. A systematic error of the order of half
of the statistical errors could account for the scatter in fig.~\ref{f_fstat}.
 
Since a better procedure is lacking at the moment, we account for this
scatter by taking in addition to the
statistical error of the extrapolated value a systematic error of the same
size. In the future, this systematic error should be removed by a further
improvement of the trial wave functions.
 
We further estimate the error of the missing higher order terms in the
perturbative expansion of $Z_{stat}$ in the following way. We allow for a
2--loop term $5~\twiggle\alpha$ in $Z_{stat}$ (a coefficient as large as five
is certainly a conservative estimate). With this term the whole extrapolation
is repeated. The change in the extrapolated value gives the systematic error
$\pm 0.08$.
 
The data with the
improved action ($c_{SW}=1$, triangles) are fully consistent with the
$c_{SW}=0$ points. We must, however, remember that in the comparison between
the two sets
of data an $O(\twiggle{\alpha}^2)$--uncertainty exists and the $a$--dependence
could be quite different. We should only compare the extrapolated values.
With the two points at hand, such an extrapolation is not possible\footnote{
Clearly, there is no evidence that the $a$--dependence in {\it this}
observable is reduced compared to $c_{SW}=0$.
}.
 
Altogether we obtain
\be
 \hat{f}^{\mbox{stat}} / \sigma^{3/4} =1.85(17)(17)(8)~.
\ee
    As in the case of $f_D$, it is better to change from a prediction in units of
the string tension to one in units of $f_{\pi}$, since a phenomenological
value for the string tension is based on assumptions. Furthermore, one may hope
that effects of dynamical fermions are reduced when one normalizes to a decay
constant. So we combine
$\hat{f}^{\mbox{stat}} / \sigma^{3/4}$ with $f_{\pi} / \sqrt{\sigma} =
0.270(12)$ to
\be
\frac{\hat{f}^{\mbox{stat}}}{  f_{\pi}^{3/2}} \times(0.132~\hbox{GeV})^{3/2} =
0.63(6)(6)(3) \hbox{GeV}^{3/2}~. \label{fstat_fpi}
\ee
This corresponds to $f_B= 276(26)(24)(12)~\hbox{MeV}  + O(m_N/m_B)$.
The errors quoted are (in order of appearance) the statistical errors including
the extrapolation, the systematic error due to possible contaminations by
excited states, and a $5\twiggle{\alpha}^2$ term in the renormalization.
(For the latter, it is not justified to assume a cancellation between the higher
orders in $Z_{stat}$ and $Z_A$ because these renormalizations are of quite
different origin.)
 
We may also change to physical units through the mass of the $\rho$--meson.
Here, the extrapolation of fig.~\ref{f_mrho_sig} resulted in
$m_{\rho}/\sqrt{\sigma}=1.81(4)$. Therefore,
\be
\frac{\hat{f}^{\mbox{stat}}}{  m_{\rho}^{3/2}} \times(0.77~\hbox{GeV})^{3/2} =
0.51(5)(5)(2)\hbox{GeV}^{3/2} \label{fstat_rho}
\ee
is a further estimate corresponding to
$f_B= 224(22)(21)(10)~\hbox{MeV}  + O(m_N/m_B)$.
 
The difference between these two estimates is most likely an effect of the
quenched approximation\cite{GF11_f}.
Consequently,
the difference between eq.~(\ref{fstat_fpi}) and eq.~(\ref{fstat_rho})
should once again be regarded as a systematic error.

\subsection{Mass Dependence\label{s_MD}}
In order to get an estimate of the terms of order $O(m_N/m_B)$, one may
examine the mass dependence of the decay constant in the mass region around
the mass of the D--meson\cite{fb3,euro,BLS,UKQCDfb,fb4,APEdyn}.
We  do not consider the results of \cite{fb3,euro} further at this point,
because they
were (partly) influenced by data with contaminations from excited states.
In addition, the subsequent simulations are more precise.
 
Ref.~\cite{BLS} uses the normalization eqs.~(\ref{zaKron},\ref{ZApertKron})
for the axial vector current.
At a fixed value of the lattice spacing, they observe that $\fhat(M_P)$ for
$m_P=M_P/a=$ 1GeV -- 3GeV joins smoothly with  $\fhat^{\mbox{stat}}$ obtained
at that value of the cutoff. This is 
interpreted as showing that this normalization is
the correct one in the sense of having small $a$--effects. This
behavior at a fixed value of $a$ is, however, automatic and
is not necessarily connected to a reduction of $a$--effects for
intermediate values of the mass. Rather, with this normalization,
the expression for $\fhat(M_P)$
 becomes identical to $\fhat^{\mbox{stat}}$ if
one takes the mass of the quark in lattice units to infinity (apart from the
renormalization constants that differ slightly). Therefore, $\fhat(M_P)$
automatically joins smoothly with the static point.
Fig.~\ref{f_fd_sig} {\it shows} that strong $a$--effects are present
when this
normalization is used. These effects are also visible in fig.~5.18 of
ref.~\cite{BLS}. We conclude that the belief that
an application of eqs.~(\ref{zaKron},\ref{ZApertKron}) essentially eliminates
the dependence on $a$ in the difficult region where  $m_h$ becomes close to
$a^{-1}$ is not justified.
 
Instead, one needs to extrapolate\cite{fb4} to $a=0$ at a fixed value of $m_P$
as it is shown in fig.~\ref{f_fd_sig} for the case of the D--meson.
This both requires systematic calculations at different values of the bare
coupling and increases the statistical error of the final result. However,
the latter is
necessary, since it is only in this way that the errors
due to the discretization are covered.
Comparing different normalization prescriptions
or results from different actions is indicative at most. It can not replace
the systematic study of the $a$--dependence with one action and one
normalization condition.
 
We have supplemented the extrapolations of \cite{fb4} with the raw data of
\cite{BLS,APEdyn}. The extrapolations are illustrated in fig.~\ref{f_fp_sig}
for the largest and the smallest value of the mass of the heavy quark that was
used. All that was said in section~\ref{s_LSDR} remains true for these other
values of the mass. We only include the points with
$\sqrt{\Sigma}\sim0.4$ at the smallest values of the mass.
This corresponds to a cut on the mass of the heavy
quark in lattice units of about $m_h a < 3/4$.

\begin{figure}[htb]
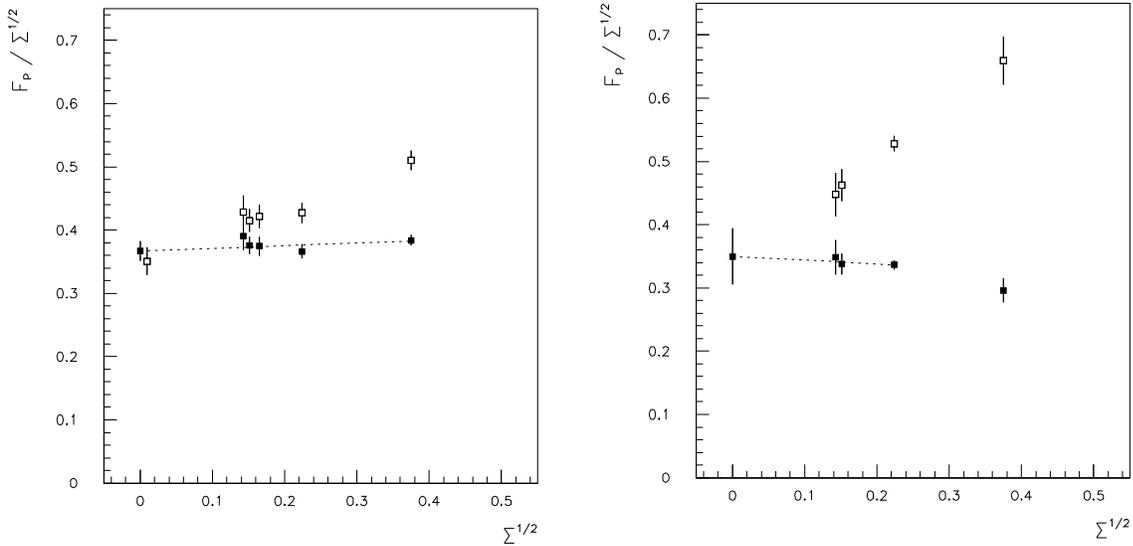

%%\topinsert
\vbox{
\vskip 0 true cm

\def\fpsangle{0}
\fpshskip=0.8 true cm

\centerline{
\fpsxsize=10.5 true cm
\fpsbox[120 0 440 300]{fp_sig.ps1}
}
\vskip -9.95 true cm
\centerline{
\fpsxsize=10.5 true cm
\fpsbox[-120 0 200 300]{fp_sig.ps2}
}

\vskip -0.5 true cm

\caption[1]{\footnotesize
The lattice spacing dependence of HL--decay
constants. Data points as in fig.~\ref{f_fd_sig}.
              }\label{f_fp_sig}
}
%%\endinsert
\end{figure}

Changing again to a prediction in terms of $f_{\pi}$,
we finally arrive at the mass--dependence of $\hat{f}(m_P)$ depicted in
fig.~\ref{f_fhat}.
 
\subsection{Interpolation\label{s_I}}

\begin{figure}[htb]
%%\topinsert
\vbox{
\vskip 0 true cm

\def\fpsangle{0}
\fpshskip=0.8 true cm

\centerline{
\fpsxsize=12.5 true cm
\fpsbox[30 30 275 285]{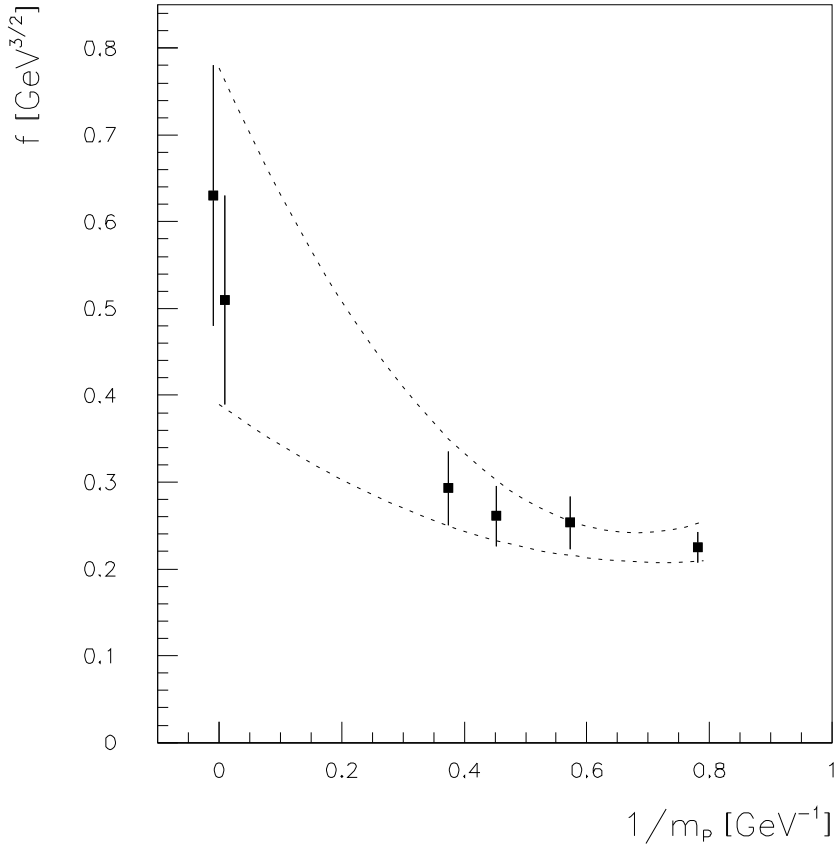}}

\vskip -0.5 true cm

\caption[1]{\footnotesize
$\hat{f}$ as a function of $1/m_P$. The two points at $1/m_P=0$ represent
eq.~(\ref{fstat_fpi}) and eq.~(\ref{fstat_rho}). All data has been extrapolated 
to the continuum allowing for a linear dependence on the lattice
spacing. 
              }\label{f_fhat}
}
%%\endinsert
\end{figure}

In principle, we would now like to fit the results at finite values of
$m_P$ together with the result in the static approximation to a power series
in $1/m_P$. Unfortunately, we have significant  systematic errors
(included in fig.~\ref{f_fhat}), in particular for the
result in the static approximation.
Thus, it is not appropriate to perform such a fit. We just connect
the upper and the lower ends of the error bars with functions
$\hat{f}(m_P) = c_0~ + c_1 ~ m_P^{-1} + c_2 ~ m_P^{-2}$ instead.
 
At $m_P=m_B$ we read off from the error band
\be
 f_B \equiv f_{B_{d}}= 180(46) ~\hbox{MeV}~. \label{fb}
\ee

One may compare this result to\cite{UKQCDfb}
$
f_{B}/f_{\pi} \times 132 \hbox{MeV} = 186^{+35}_{-21}\hbox{MeV}
$,
a value obtained with the improved action at one value of the lattice spacing
$a^{-1}_{\sigma}=2.7 \hbox{GeV}$.
 
The value quoted here for $f_B$ is obtained in the quenched approximation.
The discussion in section~\ref{s_FQCD} suggests that a similar value would
result in full QCD.
 
Finally, the columns 3 of table~\ref{t_fd} and table~\ref{t_fstat} may be
summarized to
\be
f_{B_{s}} / f_{B_{d}} = 1.10-1.18~.
\ee
 
\subsection{Comparison to QCD Sum Rule Estimates\label{s_QSR}}
 
There are a number of QCD sum rule (SR) computations of HL decay constants.
These results are summarized in table~\ref{t_QSR} and
 are compared to the lattice gauge theory results in the quenched
approximation.
 
% ----------------------------------------------------------------------
\begin{table}[htb]
 
\centering
\begin{tabular} { l l l l l }
Ref. & Method & $f_D/f_{\pi}$  & $f_B/f_{\pi}$  & $f_B^{stat}/f_{\pi}$ \\
\hline
\cite{AlEl,Nari,Rein} &  Laplace SR & 1.1 -- 1.5 & 0.9 -- 1.5 & \\
\cite{DoPa,Nari} & Hilbert Moments SR & 1.5 -- 1.9 & 1.1 -- 1.6 & \\
\cite{Ne92,BBBD} & Laplace SR in HQET &         & 1.1 -- 1.9 & 1.5 -- 2.3\\
 eqs. (\ref{fD},\ref{fstat_fpi},\ref{fb}) & LGT quenched approximation &
 1.1 --1.4  & 1.0 -- 1.7 & 1.7 -- 2.5\\
\hline
\end{tabular}
\caption[t_splitt]{\footnotesize
Estimates for HL decay constants.
         }\label{t_QSR}
\end{table}
% ----------------------------------------------------------------------

We do not want to discuss details of
the sum rule approach here. The interested
reader may, for instance,  consult the recent review ref.~\cite{CNP}.
We point out, however, that the error estimates in
table~\ref{t_QSR} are obtained
from a variation of the input parameters. They do not include uncertainties
due to the truncation of the operator product expansion or the question of
the applicability of perturbation theory in the regime where it is used.
The latter question  is a  particularly relevant one, since in the limit
of a large heavy quark mass, the only relevant dynamical
scale is the r.m.s.--radius of the light quark wave function.  At such
 a low energy scale, it is very questionable
whether perturbation theory can be applied.
 
An essential difference between the QCDSR estimates and the lattice gauge
theory approach is that the latter can be systematically improved. We
elaborate further on this point in the following.
 
\subsection{Potential Improvements\label{s_PI}}
The present best estimates for $f_{B}$ still contain a significant uncertainty.
As we have explained in the introduction, the value of  $f_{B}$ is very
important in the  phenomenology of the CKM--matrix. A higher precision is
clearly desirable.
 
To this end, it is  important to note  that the overall
computational effort that went into eq.~(\ref{fb}) is quite small, when one
takes
a few months on todays massively parallel computers as a scale.
Nevertheless, the investigations that have been done,
have  lead to eq.~(\ref{fb})
and, even more important, a semiquantitative understanding of the different
sources of systematic errors (it goes without saying that the mean values have
changed due to this). We therefore have a reasonable understanding of how to
reduce the errors further.
 
In order to estimate what can be done realistically, let us assume that a
parallel computer performing at 20~Gflops is
accessible to a group of physicists for a good
fraction of the time.
 
On such a computer with the existing algorithms, the computation of all
correlation functions on one configuration of $32^4 \times 96$ points
at $\sqrt{\Sigma}=0.1$ takes roughly 6 hours.
Therefore, one can obtain data points for figs.~\ref{f_fd_sig},~\ref{f_fp_sig}
which have the average precision seen in the plots and
which are at a 30-40\% smaller value of the lattice spacing.
This enables one to add points at masses as high as about 4~GeV in
fig.~\ref{f_fstat}
with the present precision. At the same time, one can reduce the errors of
the points in the range up to 4~GeV because
a new point, closer to the continuum, is added in each extrapolation\footnote{
It would also be useful to add points at $\sqrt{\Sigma}\sim 0.3$ to study the
$a$--dependence further.\\
Note that we assume that the computations are performed at $\sqrt{\Sigma}~L
\sim 3$. With increased precision, the size of finite size effects must be
checked, of course. This does not pose a problem since it can be done at a
relatively large value of the lattice spacing.
}.
In addition, the values in the static approximation can be significantly
improved by reducing the uncertainties due to contributions by excited states.
The latter point is already being addressed at Fermilab\cite{EichDa}.
 
Combining this information, an overall reduction of the error of $f_B$
(within the quenched approximation) by a
factor of 2 seems plausible within a period of a year or two.
\newpage
 
\section{~B--Parameter\label{s_Bp}}
 
In the Standard Model, the mixing of $B_0$ and $\bar B_0$ is mediated
by box diagrams with two $W$--boson exchanges. This corresponds to
a short distance operator $O^{\Delta b =2}$, which in the continuum
 and at distance scales that are large compared to
the interaction range of the weak interaction has the form
\be
O^{\Delta b =2}(x) = (\bar{d}(x)\gamma_{\mu}(1-\gamma_5) b(x)) 
       (\bar{d}(x)\gamma_{\mu}(1-\gamma_5) b(x))~. \label{Odeltab}
\ee 
Its matrix element is conventionally denoted by
\be
B_B^{RGI} = \alpha_s(\mu)^{-6/33} B_B(\mu)
\ee
with
\be
\frac{8}{3} B_B(\mu) f_B^2 M_B^2 = < B_0| O^{\Delta b =2}(0) | \bar B_0>~,
          \label{Bpar}
\ee
and $\mu$ denoting the renormalization scale.
In the lattice regularization (Wilson action), the operator 
$O^{\Delta b=2}_{lat}(0)$ is given to lowest order in $\alpha$ by 
the naive operator
eq.~(\ref{Odeltab}). To first order in $\alpha$, it receives admixtures
from operators of the 
same flavor but different chiral structure\cite{Bpert}. This mixing is possible
because of the breaking of chiral symmetry in the Wilson formulation and
has been taken into account to first order in $\alpha$ so far.

The B--parameter eq.~(\ref{Bpar}) at scale $\mu = a^{-1}$ can be computed
directly from the  ratio of three--point and two--point correlation
functions
\be
\frac{ \sum_{\vec x}\sum_{\vec y} < {\cal M}(x) O^{\Delta b =2}_{lat}(0)
                                     {\cal M}^{\dagger}(y)>
}
     { 8 \sum_{\vec x} <{\cal M}(x) A^{b,d}_0>
         \sum_{\vec y} <{\cal M}(y) A^{b,d}_0>
}
 \longrightarrow B_B(\mu = a^{-1})~, \label{Bratio}
\ee
where ${\cal M}$ denotes any interpolating field for a $b \bar d$ meson.
In eq.~(\ref{Bratio}) the time separations $-x_0$ and $y_0$ are to be
taken large enough such that only the lowest state contributes, i.e. 
such that one obtains the on--shell matrix element. 
Analogously to the case of two--point functions, this condition is checked 
by looking for a joint plateau in $x_0$ and $y_0$ over some range of these 
variables. Such a plateau
appears to be reached already at moderate
$-x_0, y_0$ \cite{BDHS,RajBpar,euro}. Presumably, the reason is  that the
contributions of excited states to eq.(\ref{Bratio}) are similar to the ground
state contribution. 
 
Abada et al.\cite{euro} have computed the B--parameter for meson masses
between 1.5~GeV and 3.5~GeV and with a cutoff $a^{-1}\sim3.7$~GeV.
Since the mass dependence is very weak, an extrapolation to $m_B$
seemed justified. Also the B--parameter for the $D$ meson was quoted:
\bes
 B^{RGI}_{D} &=& 1.05(8) \\
 B^{RGI}_{B} &=& 1.16(7).
\ees
The B--parameters turned out to depend only
weakly on the mass of the light quark
in the meson. For the ratio of the B--parameter of the strange
meson to that of the  nonstrange meson ref.~\cite{euro} obtained (independently
of the mass of the heavy quark)
\be
 B_{P_s} / B_{P_d} = 1.02(2)~.
\ee
We point out that the above quoted errors do not contain the
$O(\alpha^2)$ uncertainty in the renormalization (which includes mixing)
of the operator.
An estimate of the $O(a)$ terms in this quantity has not been given either.
As discussed in the previous section, the latter uncertainty needs to be
studied through systematic computations at different values of the lattice
spacing. Before such a study is completed, the numbers quoted in this section
have to be considered as rough estimates.
\newpage
 
\section{~The Size of $1/m_h$ Corrections to the Heavy Quark Limit
\label{s_corr}}
 
An important nonperturbative question in heavy quark physics is to determine
for which values of the  heavy quark mass the predictions that one obtains
in the heavy quark limit, i.e. for $m_h \rightarrow \infty$,
 become accurate. Of course, this is not a question with
a precise answer: if we have a quantity like $\hat{f}$
 with an expansion in the inverse
heavy quark mass $m_h^{-1}$
\be
\hat{f}(m_h^{-1}) = \hat{f}^{\rm stat}(1 + \hat{f}_1 m_h^{-1} + ...) ~~,
\ee
the mass scale $\hat{f}_1$, that characterizes the size of
the corrections is nonuniversal;  it depends on the quantity considered.
Nevertheless, it is of interest for applications of the HQET 
to see, whether -- in examples -- scales
like $\hat{f}_1$ are of the order of the size of the QCD scale
$\Lambda_{QCD}$ as it is frequently assumed.
 
Fig.~\ref{f_fhat} gives a rough estimate  $\hat{f}_1 \sim 1$GeV.
Although a number of this order has been quoted in almost all recent
publications \cite{euro,fb3,BLS,UKQCDfb},  it is evident from our discussion
in section~\ref{s_fb} that $\hat{f}_1$ has very large uncertainties. The main
uncertainty is due
to the present error  of the static value.
 
Hence, it is very useful to consider a quantity for which the heavy quark
limit is known. Such a quantity was first discussed in refs.~\cite{Tsuk,euro}.
Besides the pseudoscalar decay constant, one also considers  the vector meson
decay constant $f_V$, which is conventionally defined as
\be
<0|V_{\mu}(0)|V> = \epsilon_{\mu} F_V^{-1} M_V^{3/2} / \sqrt{2}~.
\ee
Here, $V_{\mu}$ denotes the (renormalized) heavy light vector current,
$|V>$ is a momentum zero vector meson state with polarization vector
$\epsilon_{\mu}$ and $M_V$ is the mass of the vector meson.
Due to the spin symmetry, the combination $U(M)=F_V F_P / M$,
with $M=(M_P+3M_V)/4$, becomes one
in the heavy quark limit\cite{SVPW}:
\be
U(M)\equiv F_V F_P / M =
( 1 + \frac{2}{3\pi}\alpha(M) + ...)(1+U_1/M  + ...)~.
\ee
Since $U(M)$ is the ratio of two very similar quantities,
the  finite lattice spacing effects in $U(M)$
may be reduced compared to the ones in $\hat{f}$ and also excited
state contributions are expected to be less of a problem.
Using masses around the mass of the $D$--meson,
Baxter et al.\cite{UKQCDfb} determined
$U(M)$ with the improved action. Interpreting their results in terms
of $U_1$, one has $U_1 \sim 0.5$~GeV in  agreement with
the rough results\cite{Tsuk,euro} that were obtained with the
Wilson action.
 
So scales of about 0.5~GeV to 1~GeV are found for the coefficients
of the $1/m_h$ terms in the two known examples.
This suggests that the heavy quark limit
only gives  the correct picture for charm quarks on a qualitative level,
while it appears to be a quite reasonable approximation for $b$--physics.
In the important application of $b \rightarrow c$ transitions,
corrections of order $O(1/m_c)$ may be of practical relevance.
Although these conclusions are obtained entirely in
the quenched approximation, it is to be expected that this {\it qualitative}
result  also holds true for full QCD.
 
Since the $1/m_c$--corrections are nonuniversal,
the simple examples considered above give us only a rough idea
about the quality of the approximations
for different processes. In the end,
the relevant quantities have to be determined directly from QCD.

\section{~The $B-\bar{B}$ Threshold, String Breaking and Hybrid
         Mesons \label{s_T}}
Following ref.~\cite{fb3}, we now review, how to obtain information about
the breaking of the QCD--string from the binding energy $\twiggle{M}_p$
 of a static B--meson
computed in the quenched approximation.
Moreover, it is  of interest to compare
 the $B-\bar{B}$ threshold  with the lowest
level of a hybrid $b-\bar{b}$ meson\cite{hybrid,MichAa}
in order to estimate whether such $b-\bar{b}$ mesons are broad
or narrow resonances.
This is done at the
end of the section.

In full QCD simulations,  the breaking of the
QCD string, i.e. the flattening of the heavy quark potential at large distances,
has been sought after for some time.
No effect was found in the most serious effort~\cite{MTC}. In the following,
we   estimate
the distance $r_b$, where the full QCD potential
flattens off using only quantities calculated in the quenched
approximation~\cite{fb3}.
 
Consider a large Wilson loop $W({\vec r},x_0)$,  with~~$ x_0>>r$, in full QCD.
It has a representation in terms of the eigenvalues of the QCD-hamiltonian
(or transfer matrix):
\be
W({\vec r},x_0) = \sum_{n \geq 0}
 |c^W_n({\vec r})|^2
             \exp( - V_n({\vec r}) x_0) \quad . \label{Wloop}
\ee
Here, $\exp( - a V_n({\vec r}))$ are the eigenvalues of the transfer matrix in the
corresponding charged sector of the Hilbert space:
the states in this sector transform according
to the 3-representation at  position $\vec{0}$ and according
to the $\bar{3}$-representation at  position ${\vec r}$ under gauge
transformations.
The same states contribute in the spectral
 decomposition of the correlation function of static--light meson fields
${\cal M}^{J}({\vec x},x_0)$
\bes
H({\vec r},x_0) &=& <  {\cal M}^{J}({\vec 0},x_0)~
[ {\cal M}^{J}({\vec r},x_0)]^\dagger
{\cal M}^{J}({\vec r},0)~
[ {\cal M}^{J}({\vec 0},0)]^\dagger    >  \nonumber \\
&=&
\sum_{n \geq 0}
 |c^H_n({\vec r})|^2
             \exp( - V_n({\vec r}) x_0 ) \quad .    \label{H}
\ees
The ground state potential $V({\vec r}) \equiv V_0({\vec r})$ can therefore
be called a static quark
potential or a static meson potential. Physically, the
first interpretation is sensible at relatively short distances, whereas
the second one is the appropriate language for large distances (compared
to the confinement scale). This can even be put into a quantitative
relation: we expect $ |c^H_0({\vec r})|^2 << |c^W_0({\vec r})|^2$ at
 small distances and $ |c^H_0({\vec r})|^2 >> |c^W_0({\vec r})|^2$ at large
distances.
 
Furthermore, at large distances, the potential will approach
a constant up to nonleading terms (Yukawa-type interactions), because the
correlation function factorizes:
\be
\log[H({\vec r},x_0)] = 2 \log[ C^{J,J}(x_0) ]
+ O( \exp[ - m_{\pi}r] x_0/r )\quad .    \label{factor}
\ee
Simulation results of full QCD~\cite{MTC}
 indicate that the QCD-potential is
approximated rather well by the quenched potential out to relatively
large distances,
up to about $r \sim 0.7$$fm$.
At very large distances, on the other hand, we expect
that $H({\vec r},t)$
is represented with some accuracy
by the quenched approximation. We must
switch from one correlation function to the other when using the
quenched approximation, since we have to put the breaking of the
string in by hand. Obviously, the quenched approximation does not
give us much information  about the intermediate regime.
 
The (in $r=|\vec r|$) asymptotic
behavior of the correlation function $H({\vec r},t)$
 is given by the
mass in the static approximation.
So one defines\cite{fb3} the string breaking distance  $r_b$ by
\be
V(r_b) = 2 \tilde{M}_{P},
\label{Rb}
\ee
with $V(r)$ being the quenched potential including the self-energy term that
cancels in eq.(\ref{Rb}).
$r_b$ defined in this way, gives an upper bound
to the distance where the potential starts deviating significantly from
the form of eq.(\ref{pot}).

\begin{figure}[htb]
%%\topinsert
\vbox{
\vskip 0 true cm

\def\fpsangle{0}
\fpshskip=0.8 true cm

\centerline{
\fpsxsize=12.5 true cm
\fpsbox[30 30 275 285]{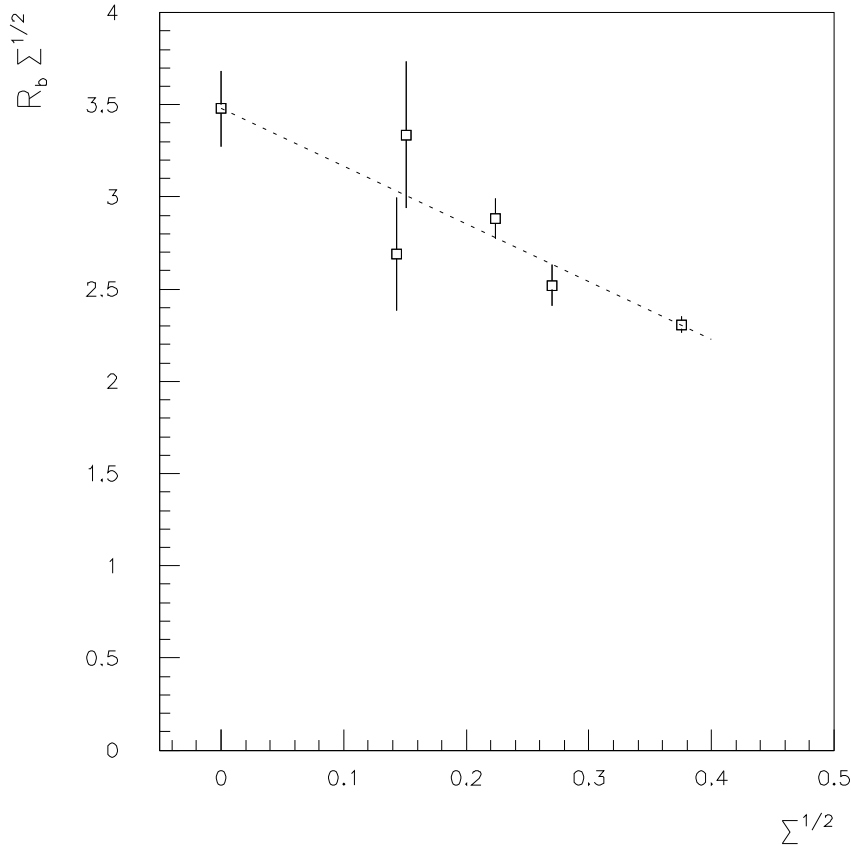}}

\vskip -0.5 true cm

\caption[1]{\footnotesize
Continuum extrapolation of $r_b$ for light--quark mass zero.
Eq.~(\ref{rb}) is used with data for $\tilde{M}_{P}$
 from refs.~\cite{fb3,Eicham,BLS}.
              }\label{f_rb_sig}
}
%%\endinsert
\end{figure}

From the explicit parametrization of the potential eq.(\ref{pot})
one can calculate $r_b$ in lattice units:
\be
R_b=(\tilde{M}_P-\frac{1}{2} V_0)/\Sigma +
\sqrt{[(\tilde{M}_P-\frac{1}{2} V_0)/\Sigma]^2+
\frac{\pi}{12\Sigma} } ~~. \label{rb}
\ee
Extending the numerical evaluation of ref.~\cite{fb3},
we have included the data of refs.~\cite{Eicham,BLS} and extrapolated
$r_b \sqrt{\Sigma}$ linearly to the continuum (see fig.~\ref{f_rb_sig}).
The resulting value in the continuum limit is (for vanishing light--quark mass)
\be
r_b \sqrt{\sigma} = 3.5(2)(2)~,
\ee
where the first error is statistical and the second is an estimate of
the uncertainty due to the extrapolation form (In this case, the
$a$--dependence is rather strong). With $\sqrt{\sigma}\sim 420$~MeV,
this corresponds to $r_b=1.7(2)$~fm.
 
Such distances are difficult to reach in a calculation of the potential
including dynamical fermions.
It should be noted, however, that as a result of the
investigation of \cite{fb3},
 the screening of the potential is expected
to appear at distances that are only weakly dependant on the
dynamical quark mass. The mechanism can hence  be studied
with relatively large (dynamical) quark masses.
 
In the above presentation, we have translated the information that is
present in the threshold energy $2\tilde{M}_P$ into the string breaking
distance $r_b$, where the divergent self--energy
of the static quark is cancelled. In a certain approximation, to be discussed
below, the
difference between the
$B-\bar{B}$ threshold and the mass of a hybrid $b-\bar{b}$ meson\cite{MichAa}
is also free of this divergence. As the position of the $b-\bar{b}$
energy level relative to the threshold determines the width of the  $b-\bar{b}$
resonance, this quantity is an interesting  observable.

A hybrid $b-\bar{b}$ meson is a $b-\bar{b}$ bound state, with quantum
numbers that do not occur in a naive potential model. Such ``exotic''
bound states
are an interesting possibility due to QCD. Describing the b--quarks through
a nonrelativistic potential model, exotic states are possible through
nontrivial angular momentum configurations of the gluon field which generates
the potential. Such excited potentials  have been computed from
lattice gauge theory\cite{hybrid}. The nonrelativistic Schr\"odinger equation
with those potentials can be solved and thus one gets an estimate for the
energy levels of  hybrid $b-\bar{b}$ mesons.  As the (excited)
potential between
static quarks is used,
the energy levels contain (twice) the self energy of a static quark.
The difference of an energy level to $2\tilde{M}_P$ is free of this divergent
contribution and represents a well defined
 estimate of whether the state is above
or below the open $b$ threshold. Of course, the estimate is obtained within
the quenched and the potential model approximations.
 
Using the result of ref.~\cite{hybrid} one estimates for the lowest exotic level in
the $b-\bar{b}$ system
\be
[E_{1^{-+}} - V(r_0)] / \sqrt{\sigma} \sim 2.0 ~~
\ee
and for the threshold
\be
[E_{B\bar{B}} - V(r_0)] / \sqrt{\sigma}  =  2 \tilde{M}_P - V(r_0) / \sqrt{\sigma}
= 2.5(4)~~,
\ee
where we have arbitrarily normalized to $V(r_0)$.
So, the exotic state appears to be $\sim 200$~MeV {\it below} the threshold.
Adding the information that the wave function at the origin is very
small due to the flat potential\cite{hybrid},
the state is expected to have a very small width.
Note, however, that even 
within the model used here, the uncertainty is so large
that a coincidence of the threshold and the hybrid level is just at the
edge of the error bar.
 
Previously, the position of the
$B \bar{B}$ threshold relative to the mass of the
hybrid meson has been estimated in the following way\cite{hybrid}: 
The difference of
$E_{1^{-+}}$ and the mass of the lowest  $b-\bar{b}$ bound state
$m_{\Upsilon}$ was estimated with the above model. Then the experimental
value for $2m_B - m_{\Upsilon}$  gives the position of the threshold relative
to $E_{1^{-+}}$. In this way one arrives at the conclusion that the $1^{-+}$
state is $\sim 250$~MeV {\it above} the threshold.
 
Remaining consistently in the quenched approximation, the opposite seems to
emerge, giving new hope for the existence of a narrow, interesting state.
Further progress about this problem can most likely be made within the
framework of non-relativistic QCD.
\newpage
 
\section{~The Beauty Spectrum \label{s_BS}}
 
The spectrum of bound states of a b--quark with light quarks
represents one of the possible predictions of QCD.
It has been explored to
some  extent  using the static approximation. Estimates of the corrections of
order $O(m_N/m_B)$ do not  exist and are in fact not of primary interest at this
point, since the computations within the static approximation need considerable
improvement.  For any state,  the self energy of the static quark must be cancelled
to obtain a finite result. This is done by considering energy differences to
the mass of the B--meson .
 
Due to the spin symmetry in the static approximation\cite{IsWi},
there is a degeneracy of
vector-- with pseudoscalar mesons and
axialvector-- with scalar mesons.  We denote the remaining splitting
between scalar and pseudoscalar by $\Delta_S$.  Of further interest
is the splitting between the $\Lambda_b$ and the $B$, denoted by
$\Delta_{\Lambda}$.  Also the first radially excited state
with splitting $\Delta_{2s}$ of the $B$
has been investigated\cite{Eicham}  and an estimate of the difference
between the $B_s$ and the $B \equiv B_d$ was given\cite{APEfstat}.
We now comment on some technical points of these computations.

The computation of the mass splittings between states with different
quantum numbers  is straight forward in principle.
i) One chooses (in addition to  eq.~(\ref{meson}) ) an interpolating field
with the
desired quantum numbers.  ii) One searches for the exponential
decay of the correlation functions and computes the mass splitting
as the difference, or one takes  the ratio of the two correlation
functions directly and searches for a plateau in the effective mass
of the ratio.
 
The scalar meson is a p-wave in  the non-relativistic
quark model. It is therefore natural to insert a p-wave smearing function
into eq.~(\ref{smear}) in order to obtain 
an interpolating field for a scalar meson. 
For gauge covariant wave functions, this can be
achieved by applying a covariant derivative to the symmetric wave function.
This approach was tried in ref.~\cite{fb3}, but the correlation function was
found to be very noisy. The results quoted in ref.~\cite{fb3} were obtained
by replacing  $\gamma_0 \gamma_5 \rightarrow 1$ in eq.~(\ref{meson})  and
using a spatially symmetric wave function, that means the
lower components of the light--quark field are used to obtain the
parity change relative to the pseudoscalar. Ref.~\cite{Eicham} does obtain
results with a p-wave smearing function in Coulomb gauge. Presumably, the
reason is that in a fixed gauge, the detailed form of the
smearing function can be chosen;  ref.~\cite{Eicham}  uses the wave function of
a semi--relativistic potential model for that purpose.
 
The interpolating field for the $\Lambda_b$ has been
chosen as\cite{Bochmass,fb3}
\be
{\cal B}^{I,J}_\alpha (x) = \sum_{a,b,c,\beta,\gamma}
\epsilon_{abc} ~ (h^I (x))^a_\alpha ~
(u^J (x))^b_\beta ~ (C\gamma_5)_{\beta \gamma} ~
(d^J (x))^c_\gamma \quad .
\ee
Here a,b,c ($\alpha,\beta,\gamma$)
denote color (Dirac) indices.
The field ${\cal B}^{I,loc}_\alpha (x)$ is to be interpreted as
an extreme di--quark trial wave function for the baryon:
the two light--quark fields
are taken at the same point.
Otherwise, ${\cal B}^{I,J}_\alpha (x)$
amounts to a more general wave function ansatz.
In ref.~\cite{Bochmass},  the di--quark option was chosen,
while ref.~\cite{fb3} explored both possibilities.
An interesting feature of the correlation functions of
${\cal B}^{I,J}_\alpha (x)$ is that the  di--quark interpolating field
${\cal B}^{I,loc}_\alpha (x)$ is much less effective in exciting
a $\Lambda_b$ state than the field ${\cal B}^{I,I}_\alpha (x)$,
where all quarks have an independent spatial wave function\cite{fb3}.
% ----------------------------------------------------------------------
\begin{table}[htb]
 
\centering
\begin{tabular} { c c c c c c}
Ref.   & $\Delta_{\Lambda}$[GeV]  & $\Delta_S$[GeV]  & $\Delta_{2s}$[GeV]  &
$M_{B_s} -  M_{B_d}$[MeV] & $a^{-1}_{\sigma}$[GeV] \\
\hline
\cite{Bochmass} & $0.72^{+0.16 + 0}_{-0.16 -0.13}$  &  &  &
                         $71^{+13 + 0}_{-13 -16}$ & 1.9  \\
\cite{fb3}      & $\sim 0.6$ & $\sim 0.35$ &  & & 1.3 - 2.7 \\
\cite{Eicham} &  & $\sim 0.4$ & $\sim 0.4$ & & 1.6  \\
\cite{APEfstat} &  &  & & 70 - 140 & 1.9  \\
\hline
\end{tabular}
\caption[t_splitt]{\footnotesize
A compilation of mass splittings
as they were obtained in the {\it static approximation}.
The last column gives the value of the cutoff.
         }\label{t_splitt}
\end{table}
% ----------------------------------------------------------------------
 
Determining the energy of a (radially) excited state is a more difficult task.
For that purpose, a matrix correlation function needs to be considered.
From such a matrix correlation function, a quantity which is
analogous to the local mass (cf. sect.~\ref{s_RA}) can be constructed
and it was shown that this quantity converges exponentially (in $x_0$) to
the mass splitting\cite{LuWo}. Thus the situation is -- at least in principle --
the same as  for the splitting  between two states with different
quantum numbers.
 
The computation of the 2S - 1S splitting in ref.~\cite{Eicham} has
not exactly been done  in this way.
Rather,  starting from a $2\times2$ matrix correlation function (where the two
smearing functions are again obtained from the
semi--relativistic potential model and should lead to relatively good
interpolating fields for the lowest two levels), an optimal
wave function was found for the ground state first.
Then the correlation function
was projected onto the part orthogonal to that {\it approximate} ground state.
The excited state mass was determined from that projected correlation function.
Since there always remains a small contribution from the exact ground state
in the projected correlation function, one will at very large $x_0$ determine
again the mass of the ground state.
Although this is in principle a problem of the computation ref.~\cite{Eicham},
we do not think that this effect
is numerically important at the moment. Furthermore, it can easily be
corrected in the future.

The main feature of all the computations\cite{Bochmass,fb3,Eicham}
of the mass splittings is that
the plateaus in the effective masses are less convincing than for the
pseudoscalar state. So the results are less precise
and it has not been possible to study
systematic errors due to $a$--effects\footnote{ 
Finite volume effects have 
been studied within a model  for $\Delta_S$ and $\Delta_{2s}$ \cite{Eicham}.}.
Therefore,  these estimates are qualitative
at the moment.  Where this was done in the literature,
we will quote numbers with error bars below, but we stress
that  they do not include a realistic estimate of systematic errors.
The results
of the various investigations are listed in table~\ref{t_splitt}.

In addition to the observables that we considered so far, 
Bochicchio et al.\cite{Bochmass}
estimated the
vector -- pseudoscalar splitting at order $1/m_h$. The computation is done
starting from the static approximation and including the $\vec{\sigma}\vec B$
term of eq.~(\ref{NRQCD}) perturbatively. 
The result at $a^{-1}_{\sigma} = 1.9$~GeV  is
$m^2_V - m^2_P = (0.19^{+0.04 + 0}_{-0.04 -0.07})\hbox{GeV}^2$, which is
 significantly below the experimental
$m^2_{B^*} - m^2_B \sim m^2_{D^*} - m^2_D \sim 0.55$~GeV$^2$. Lacking
estimates of the systematic errors, it is premature
to speculate from where this difference originates. It poses an
interesting problem for future investigations.
\newpage
 
\section{~Further Lattice Investigations}
There are a number of interesting investigations of other physical
observables in the context of HL hadrons. Let us mention just the ones
which we believe have the potential to contribute to
the analysis of experimental data as mentioned in the introduction.
 
Semileptonic decay form factors for $D$--decays have been studied
\cite{SLD,euroSL} and
exploratory studies of $B$--decays have been
performed\cite{euroSL}.
Computations of the Isgur--Wise--function are attempted\cite{IW} and
the nonperturbative amplitude for the decay $B \rightarrow K^* \gamma$
has recently been estimated\cite{BKG}.
 
We  mention these investigations for completeness.
Compared to the leptonic decay constants there is scarcely a study of the
systematic uncertainties available at the moment. In particular, the size
of finite
lattice spacing effects are not known at present, but we would like to
point out that the feasibility of such computations is clearly demonstrated.
Thus,
 there is considerable potential in these computations.
An initial  understanding of the systematic
uncertainties of these quantities should develop
within the next year or two.
After that stage, these computations might represent predictions of a
well--tested model, quenched QCD. As such, they might already
be a valuable contribution to the analysis of experimental data.
 
The necessary computations in full QCD will be
facilitated by what can be learned from the quenched approximation.
Nevertheless, it is  not possible to predict exactly when first principle
calculations of these hadronic matrix elements, i.e. reliable
calculations in full lattice QCD, can be done.
\newpage
 
\section{~Summary}
We have given an overview  of the present status of
QCD lattice simulations that involve the $b$ quark.
The  discussion was centered around the most easy quantity, the
vacuum-to-one-meson matrix element. That matrix element gives us
the leptonic decay constant and in the case of the $B$--meson it is
the primary unknown in the process of extracting the CKM angles from
experimental data.
 
Unfortunately, the $B$--meson is particularly difficult to
treat in a lattice simulation. This  is simply due to the large gap in the
relevant scales in the problem.
These scales are i) the confinement scale which determines the physical
size of a $B$--meson and thus is the scale that is relevant for
finite size effects and ii) the mass
of the $b$--quark itself. The latter needs to
 be small compared to the cutoff such that the quark can propagate without
large distortions due to the finite cutoff.
These two scales cannot be accommodated on today's lattices.
 
A way out is to consider the observables of interest in two unphysical
regimes, obtaining the physical observables through a matching of the two.
The first regime is the static limit $m_b \rightarrow \infty$.
In this limit, the scale  $m_b$ itself becomes irrelevant and results can
-- in principle -- easily be obtained. The second regime is $m_h \sim m_c$,
where one is at the edge of the possibility of treating
the heavy quark correctly while still keeping finite size effects small.
We have emphasized that it is very important to systematically
perform simulations with different values of the cutoff $a^{-1}$
and fixed
value of $m_h$ in this regime and extrapolate the results to the continuum
limit $a \rightarrow 0$. The uncertainties due to the
finite cutoff can only be taken into account  in this way.
The necessity to extrapolate $a \rightarrow 0$ introduces quite large
errors, even in the quenched approximation.
Nevertheless, it is  straightforward
to reduce these uncertainties using the knowledge
which has been acquired over the last years.
 
The precise studies are still restricted to the
quenched approximation. It is quite interesting, however,
that a systematic comparison of full QCD results with
the quenched approximation in the range $m_h \sim m_c$ shows no effect
of  sea quarks with a mass of about $m_s$.
 
Other quantities like the B-parameter and mass splittings have not been
studied very systematically yet.
Particularly concerning  the B-parameter, this deficiency needs to be
filled.
 
In addition, we  explained how one can get
information  on the breaking of the QCD--string from simulations
in the quenched  approximation by comparing the open $B$ threshold
in the static approximation with the static potential.
\newpage

\appendix
\section{~%Appendix A:
Renormalization of Vector Currents}
For HL--mesons  the axial vector current and the vector
current have  been normalized  in different ways\cite{Lepa,Kron,LM},
as discussed for the axial vector current
 in sect.~\ref{s_R}. In the case of
 the local vector current, the relativistic
normalization is
\be
V_\mu^{f,f'}(x) = Z_V(g^2_0,K_f,K_{f'}) ~ \bar{q}_f(x) \gamma_{\mu}
 q_{f'}(x)~ \label{zv}
\ee
with the one--loop expression\cite{KaSm,Za,LM}
\be
Z_V(g^2_0,K_f,K_{f'}) = 1 - 0.17408 ~\tilde{g}^2 + O(\tilde{g}^4) +
O(am_f) + O(am_{f'}) \label{ZVpert}~.
\ee
The non--relativistic normalization of Kronfeld, Lepage and Mackenzie
reads
\be
V_\mu^{f,f'}(x) = \tilde{Z}_V(g^2_0,K_f,K_{f'})~ \sqrt{{1\over 2K_f}-3\bar{u}}~
            \sqrt{{1\over 2K_{f'}}-3\bar{u}}~~\bar{q}_f(x) \gamma_{\mu}
            q_{f'}(x)~, \label{zvKron}
\ee
with
\be
\tilde{Z}_V(g^2_0,K_f,K_{f'}) = 1 - 0.0656~ \tilde{g}^2 + O(\tilde{g}^4) +
O(am_f) + O(am_{f'}) \label{ZVpertKron}~.
\ee
It is claimed in the literature  that in the region where $a m_f$ is of order
one, the normalization eqs.~(\ref{zvKron},\ref{ZVpertKron}) (and
eqs.~(\ref{zaKron},\ref{ZApertKron})) represent the ``correct'' normalization
of the
currents. Here, ``correct''  has to be interpreted as being the
normalization with small lattice artifacts.
 
The partially conserved vector current
\be
\hat{V}_\mu^{f,f'}(x) =
 \frac{1}{2} [
    \bar{q_f}(x)  (\gamma_{\mu}-1)U_{\mu}(x)
     q_f(x+a\hat{\mu})
  +  \bar{q}_f(x+a\hat{\mu}) (1+\gamma_{\mu})
     U^{\dagger}_{\mu}(x)q_f (x)]  \label{zvcons}
\ee
has the same normalization in the two approaches, however\cite{GBD}.
 
We consider\cite{euroSL,GBD} the ratio
\be
R_V(K_f,K_{f'}) = \frac{ \sum_{\vec x}< \bar{q}_{f'}(x) \gamma_{k}  q_{f}(x)
                          ~~ \hat{V}_k^{f,f'}(0) > }
                       { \sum_{\vec x}< \bar{q}_{f'}(x) \gamma_{k}  q_{f}(x)
                          ~~ \bar{q}_f(0) \gamma_{k}  q_{f'}(0)> }~.
\ee
Away from the short distance lattice artifacts this gives one definition
of the renormalization constant of the local current.
In the relativistic normalization one obtains for this ratio
\be
R_V(K_f,K_{f'}) = Z_V(g^2_0,K_f,K_{f'}) =
1 - 0.17408 \tilde{g}^2 + O(\tilde{g}^4) +
O(am_f) + O(am_{f'})~, \label{zvpt}
\ee
while the non--relativistic normalization predicts a strong
mass dependence
\be
R_V(K_f,K_{f'}) = (1 - 0.0656 \tilde{g}^2)~
\sqrt{{1\over 2K_f}-3\bar{u}}~
            \sqrt{{1\over 2K_{f'}}-3\bar{u}}
               + O(\tilde{g}^4) +
O(am_f) + O(am_{f'}) \label{zvKMpt}
\ee
In fig.~\ref{f_rv}, the Monte Carlo data for $R_V$ is compared to the above two
expressions. It is seen that at small $m=1/(2K_f)+1/(2K_{f'}) - 1/K_{c}$
the predictions are essentially the same. At growing quark masses,
the prediction of the non--relativistic normalization
deviates much further from the Monte Carlo results
than the one from the relativistic normalization. In fact, the latter
describes the mass--dependence very well. One also sees that the difference
between the 1--loop expression for $R_V$ and the nonperturbative results
decreases significantly as the lattice spacing is reduced (from the right
figure to the left, the lattice spacing changes by a factor $\sim 0.6$).

\begin{figure}[htb]
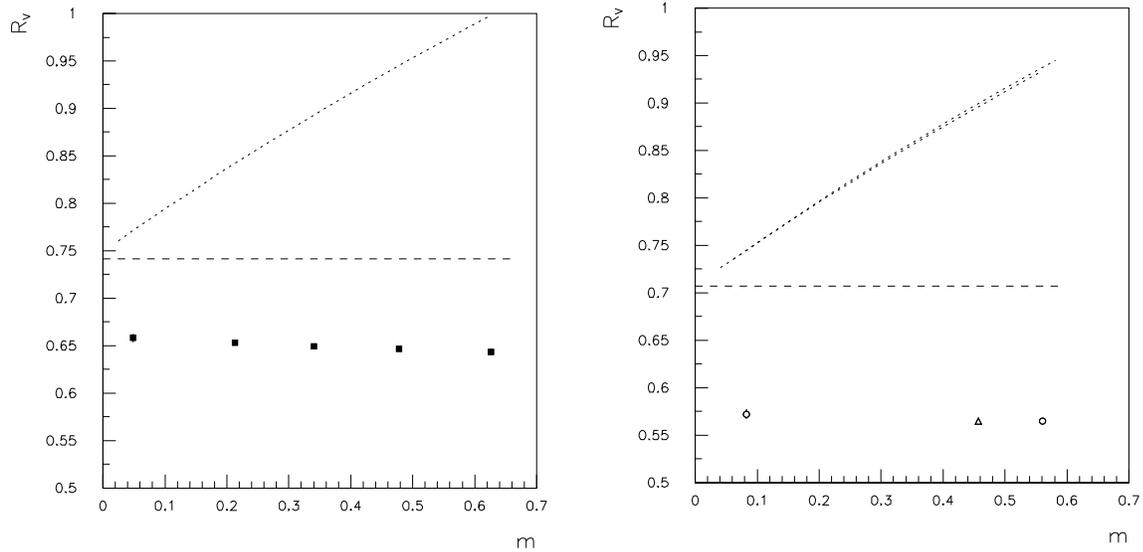

%%\topinsert
\vbox{
\vskip 0 true cm

\def\fpsangle{0}
\fpshskip=0.8 true cm

\centerline{
\fpsxsize=10.5 true cm
\fpsbox[120 0 440 300]{rv.ps1}
}
\vskip -9.95 true cm
\centerline{
\fpsxsize=10.5 true cm
\fpsbox[-120 0 200 300]{rv.ps2}
}

\vskip -0.5 true cm

\caption[1]{\footnotesize
$R_V$ compared to eq.~(\ref{zvpt}) (dashed line) and eq.~(\ref{zvKMpt})
(dotted line). The left hand figure is for $\beta=6.4$ \cite{euroSL}
and the right hand figure for
$\beta=6.0$ \cite{GBD}. }
              \label{f_rv} 
}
%%\endinsert
\end{figure}

Finally, let us mention  that for the improved action, one obtains a
much better agreement of $R_V$ with 1--loop perturbation theory.
This and the fact that other nonperturbative definitions of $Z_V$ give
significantly different results\cite{Heat,imprres}, 
suggests that the difference
between the dashed curve and the data in  fig.~\ref{f_rv} is dominantly
an $O(a)$ lattice artifact rather than an effect of truncating perturbation
theory at that order. Indeed, the numbers are in agreement with
an almost linear correction in $a$, but does, of course, not rule out
more complicated corrections.
 
This example shows that the nonrelativistic normalization, in general,
does not  reduce
lattice artifacts that are due to large quark masses $m_f a \sim 1$.
Rather, in the example, the lattice artifacts are
increased compared to the relativistic normalization.
We mainly conclude from this example that
lattice artifacts have to be removed by systematically performing
the limit $a \rightarrow 0$.
 
\newpage
 
{\bf Acknowledgements}
 
I would like to thank C. Alexandrou, S. G\"usken, F. Jegerlehner and
K. Schilling  for an enjoyable and
productive collaboration during which 
I learned about most of the topics covered in this review.
 
I have profited from discussions on these topics with many colleagues.
I would like to mention A. Ali, E. Eichten, J. Labrenz, M. L\"uscher,
G. Martinelli,  O. P\`ene,  C. Sachrajda and S. Sint.\\
I acknowledge furthermore the hospitality of the INT Summer Institute 
``Phenomenology and Lattice QCD'', Seattle, Waschington, USA, where this work
was started. 
 
Moreover, my thanks go to C. Allton, Y. Iwasaki, A. Ukawa and  D. Weingarten
for sending me their data through electronic mail.
 
Above all I thank my wife, Dorothy, for her great patience and for proofreading
this manuscript.


\newpage
\addcontentsline{toc}{section}{References}

\begin{thebibliography}{999}


\bibitem{CKM}{
             N. Cabibbo, Phys. Lett. {\bf 10} (1963) 513; \\
             M. Kobayashi and
             K. Maskawa, Prog. Theor. Phys. {\bf 49} (1973) 652.}

\bibitem{LeRo}{H. Leutwyler and M. Roos, Z. Phys. C25 (1984) 91.}

\bibitem{PDG}{
             Particle Data Group, J. J. Hern\'{a}ndez et al., Phys.
             Lett. B239 (1990) 1.}
\bibitem{Wolf}{L. Wolfenstein, Phys. Rev. Lett. 51
               (1983) 1945.}

\bibitem{formf}{M. Bauer, B. Stech and M. Wirbel, Z. Phys. C29 (1985) 637;
             C34 (1987) 103;\\
             N. Isgur, D. Scora, B. Grinstein and M. B. Wise, Phys. Rev. D39
             (1989) 799;\\
             N. Isgur and D. Scora, Phys. Rev. D40 (1989) 1491.}

\bibitem{IsWi}{N. Isgur and M.B. Wise, Phys. Lett. B232, (1989) 113;\\
               Phys. Lett. B237 (1990) 527.}

\bibitem{Mann}{M. Neubert and V. Rieckert, Nucl. Phys. 
               B382 (1992) 97; M. Neubert, Phys. Lett. 264B (1991) 455,
               Phys. Rev. D46 (1992) 2212; for a summary see e.g.
               T. Mannel, in \cite{bfac}.}

\bibitem{bfac}{ECFA Workshop on a European B--Meson Factory, B-Physics
               Working Group Report, DESY preprint, DESY 93-151.}

\bibitem{Luke}{M.E. Luke, Phys. Lett. 252B (1990) 447}

\bibitem{Argus_vcb}{H. Albrecht et. al (The Argus Collaboration),
               preprint DESY 92-146, October 1992.}

\bibitem{AlLo}{A. Ali and D. London, preprint DESY 93-022, February 1993.}


\bibitem{SLD}{
             M.Crisafulli et al., Phys. Lett. 223B (1989) 90;\\
             V.Lubicz, G.Martinelli and C.T.Sachrajda,
                Nucl. Phys. B356 (1991) 310;\\
             V.Lubicz, G.Martinelli, M.McCarthy and
                C.T.Sachrajda, Phys.Lett. 274B (1992) 415;\\
            C.Bernard, A.El-Khadra and A.Soni, Phys. Rev. D43 (1992) 2140;\\
            C.Bernard, A.El-Khadra and A.Soni, Phys. Rev. D45 (1992) 869.} 
\bibitem{euroSL}{A. Abada et al., preprint LPTENS 93/14.}

\bibitem{FaNe}{A.F. Falk and M. Neubert, Phys. Rev. D47(1993)2965.}

\bibitem{CP}{P. Franzini, Phys. Rep. C173 (1989) 1;\\
             E. A. Paschos and U. T\"urke, Phys. Rep. C178 (1989) 145.}


\bibitem{Roos}{
              J. Maalampi and M. Roos, Particle World 1 (1990) 148.}
  
\bibitem{Lusi}{
              M. Lusignoli, L. Maiani, G. Martinelli and L. Reina, 
             Univ. di Roma,
                Preprint n.792 (1991).}


\bibitem{Allt1}{C.R. Allton et al., Nucl. Phys. B349(1991)598.}

\bibitem{fb1}{
              C. Alexandrou, S. G\"usken, F. Jegerlehner, K. Schilling
              and R. Sommer, Phys. Lett. B256 (1991) 60.}

\bibitem{Ne92}{M. Neubert, Phys. Rev. D45 (1992) 2451.}

\bibitem{BBBD}{E. Bagan, P. Ball, V.M. Braun and H.G. Dosch,
               Phys. Lett. B278 (1992) 457.}

\bibitem{Wils}{K. G. Wilson, in {\it New Phenomena in Subnuclear
 Physics}, Erice 1975, Plenum, New York (1977).}

\bibitem{LQCD}{For an introduction see for example\\
          M. Creutz, Quarks, gluons and lattices
              (Cambridge 1983);\\
              P. Hasenfratz, {\it Lattice Quantum Chromodynamics},
       in ``Schladming 1983, Proceedings, Recent Developments In High Energy
        Physics'', 283;\\
        I. Montvay and G. M\"unster, Quantum Fields on a Lattice,
               Cambridge University Press (1993) (to appear).}

\bibitem{SW}{B. Sheikholeslami and R. Wohlert,
             Nucl. Phys. B 259 (1985) 572.}

\bibitem{LuTM}{M. L\"uscher, Comm. Math. Phys. 54 (1977) 283.}

\bibitem{Symm}{K. Symmanzik, Cutoff Dependence in Lattice $\Phi_4^4$ Theory, 
               Lecture given at Carg\`ese (1979), {\it in}
               Recent Developments in Gauge Theories, ed. G't Hooft et al. 
               (Plenum, New York, 1980).}

\bibitem{Reis}{T. Reisz, Nucl. Phys. B318 (1989) 417.}

\bibitem{LWimpr}{M. L\"uscher and P. Weisz, Comm. Math. Phys. 97 (1985) 59.}

\bibitem{imprres}{G. Martinelli, C.T. Sachrajda and A. Vladikas,
                   Nucl. Phys. B 358 (1991) 212.}

\bibitem{su2}{R. Sommer, preprint
             DESY 93-062, Nucl. Phys. B (in press).}
\bibitem{KaSm}{L. H. Karsten and J. Smit, Nucl.Phys. B183 (1981) 103.}

\bibitem{Boch}{M. Bochicchio, L. Maiani, G. Martinelli, G. C. Rossi and 
               M. Testa,   Nucl. Phys. B262 (1985) 331.}
\bibitem{Lepa}{ G.P. Lepage in {\it ``Lattice 91"}, Nucl. Phys. B
              (Proc. Suppl.) 26 (1992) 45.}
\bibitem{Kron}{A. Kronfeld,  Nucl. Phys. B (Proc. Suppl.)
           30 (1993) 445.}
\bibitem{LM}{G.P. Lepage and P. Mackenzie, Phys. Rev. D48(1993)2250.}
\bibitem{Za}{R. Groot, J. Hoek and and J. Smit, Nucl. Phys. B237(1984)111.}

%%\bibitem{RoTe}{G.C. Rossi and M. Testa; for the original work see \cite{Boch}.}

 
\bibitem{LMold}{G.P. Lepage and P.B. Mackenzie, Nucl. Phys. B(Proc. Suppl.) 20
            (1991) 173. }
\bibitem{grsu3}{M. L\"uscher, R. Sommer, P. Weisz and U. Wolff, A Precise 
               Determination of the Running Coupling
               in the SU(3) Yang-Mills Theory, DESY 93-114 (1993).}
\bibitem{Paris}{G. Parisi,
               {\it in}\/: High-Energy Physics --- 1980,
                XX. Int. Conf., Madison (1980), ed. L. Durand and L. G. Pondrom
               (American Institute of Physics, New York, 1981.)
\bibitem{Heat}{
      G. Heatlie,  G. Martinelli, C. Pittori, G.C. Rossi and C.T. Sachrajda, 
       Nucl. Phys. B352 (1991) 266.}

\bibitem{Step}{S. G\"usken in proceedings of the 1989 International
 Symposium {\it ``Lattice '89"}, Nucl. Phys. B (Proc. Suppl.)
           17 (1990) 361}

\bibitem{ga}{S. G\"usken, U. L\"ow, R. Sommer, K. Schilling, K.-H. M\"utter
 and A. Patel, Phys. Lett. B227 (1989) 266.}

\bibitem{fb3}{C. Alexandrou, S. G\"usken, F. Jegerlehner, K. Schilling
              and R. Sommer, preprint PSI-PR-92-27; Nucl. Phys. B (in press).}
\bibitem{fb4}{C. Alexandrou, S. G\"usken, F. Jegerlehner, K. Schilling
             and R. Sommer, preprint DESY 93-179.}
\bibitem{Eich90}{ E. Eichten, G. Hockney, and H. B. Thacker,                   
                  Nucl. Phys. B (Proc. Suppl.) 20 (1991) 500.}

\bibitem{UKQCDfb}{R.M. Baxter et al. Edinburgh preprint 93/526.}

\bibitem{ape1}{P. Bacilieri et al., Nucl. Phys. B317 (1989) 509.}

\bibitem{guido}{New, unpublished results from the APE--collaboration, do
               indeed show plateaus at large values of $x_0$.
               I thank Guido Martinelli for communicating these findings
               prior to publication.}


\bibitem{HMC}{S. Duane, A.D. Kennedy, B.J.
              Pendleton and  D. Roweth, Phys. Lett. B195 (1987) 216.}
\bibitem{GIKP}{S. Gupta, A. Irb\"ack, F. Karsch and B. Petersen,
               Phys. Lett. B242 (1990) 437.}

\bibitem{To91}{D. Toissant, Nucl. Phys. B (Proc. Suppl.)
           26 (1992) 3.}
\bibitem{Latt_mass}{A. Ukawa, Nucl. Phys. B (Proc. Suppl.)
           30 (1993) 1.}
\bibitem{APE}{The APE group, Phys. Lett. B 214,  Phys. Lett. B258 (1991)                 
                 195, Nucl. Phys. B (Proc. Suppl.) 26 (1992) 399,
              Nucl. Phys. B378 (1992) 616.}

\bibitem{GF11_mass}{F. Butler, H. Chen, J. Sexton , A. Vaccarino and
                   D. Weingarten, Phys. Rev. Lett. 70(1993) 2849.}

\bibitem{qu_chir}{C.W. Bernard, M.F.L. Golterman, Phys. Rev. D46(1992)853;\\
                 S.R. Sharpe, Phys.Rev.D46(1992)3146.}

\bibitem{Lues_string}{M. L\"uscher, Nucl. Phys. B180 (1981) 317.}

\bibitem{BS}{G.S.~Bali and K. Schilling, 
               Phys. Rev. D46 (1992) 2636; the values quoted here
               have been reanalysed by G. Bali using eq.(\ref{pot}). }

\bibitem{MTC}{K.D. Born {\it et al.},
              Nucl. Phys. B (Proc. Suppl.) 20 (1991) 394,
              Nucl. Phys. B (Proc. Suppl.) 26 (1992) 268.}

\bibitem{UKQCD}{C.R. Allton et al. Edinburgh preprint 92/507.}

\bibitem{UkFS}{S. Aoki, M. Fukugita, N. Ishizuka, Y. Kuramashi, H. Mino, 
               M. Okawa, A. Ukawa and T. Umemura, unpublished study of
               finite size effects in hadron masses; talk by A. Ukawa
               at Schloss Ringberg 1992.
               }
\bibitem{ape_ks}{S. Cabasino et al., Phys. Lett. B258 (1991) 202;\\
                P. Bacilieri et al., Nucl. Phys. B343 (1990) 228.}

\bibitem{stag_ks}{R. Gupta et al., Phys. Rev. D43 (1991) 2003.}

\bibitem{hemcgc_ks}{K.M. Bitar et al., Nucl. Phys. B (Proc. Suppl.) 20 
                 (1991) 362.}
\bibitem{fmiou_ks}{N. Ishizuka et al., Nucl. Phys. B (Proc. Suppl.) 26 
                 (1991) 284 }

\bibitem{Shar}{S. Sharpe, Nucl. Phys. B (Proc. Suppl.) 26 (1992) 197.}

\bibitem{APEfstat}{C.R. Allton et al., preprint LPTENS 93/12.}

\bibitem{APEdyn}{C.R. Allton et al. (the APE group), private communication
                 by C.R. Allton.}

\bibitem{NRQCD}{ G.P. Lepage and  B.A. Thacker in {\it ``Field Theory on
the Lattice"}, Nucl. Phys. B (Proc.
 Suppl.) 4 (1988) 199.}

\bibitem{PT_NRQCD}{ C. Davies and B. Thacker in{\it ``Lattice 91"}, 
                Nucl. Phys. B (Proc. Suppl.) 26 (1992) 375 and 378.}

\bibitem{MaMaSa}{G. Martinelli, L. Maiani and C. Sachrajda, Nucl. Phys.
                 B368 (1992) 281.  }
\bibitem{Mack}{P. Mackenzie,  Nucl. Phys. B (Proc. Suppl.)
           30 (1993) 35.}

\bibitem{Latt93}{Proceedings of ``Lattice 93'', to appear in Nucl. Phys. B
                 (Proc. Suppl.).}

\bibitem{Zstat1}{Ph. Boucaud, C. L. Lin, and O. Pene, Phys. Rev. D40 (1989) 1529
             + erratum;\\
                 Ph. Boucaud, J. P. Leroy, J. Micheli, O. Pene, and
                 G. C. Rossi, CERN preprint CERN-TH-6599-92.}


\bibitem{Zstat2} {E.Eichten and   B. Hill, Phys.Lett.B240(1990)193.}

\bibitem{Eich}{ E. Eichten, in {\em Field Theory on the Lattice},
                   Nucl. Phys. B (Proc. Suppl.) 4 (1988) 147.}

\bibitem{EF}{E. Eichten and F. Feinberg, Phys. Rev. D 23 (1981) 2724.}

\bibitem{GC}{K. M. Bitar et al., 
             preprint FSU-SCRI-93-110.}

%%\bibitem{WDLW}{R.M. Woloshyn, T. Draper, K.F. Liu and W. Wilcox,
%%            Phys. Rev. D39(1989)978.}

\bibitem{BDHS}{C. Bernard, T. Draper, G. Hockney and
               A. Soni, Phys. Rev.D38(1988)3540.}

\bibitem{GMPMP}{M.B. Gavela et al., Nucl. Phys. B306 (1988) 677;\\
               M.B. Gavela et al., Phys. Lett. 206B (1988) 113.}


\bibitem{GL}{T.A. De Grand and D.R. Loft, Phys. Rev. D38 (1988) 954.}

\bibitem{euro}{A. Abada et al, Nucl. Phys. B376 (1992) 172.}

\bibitem{BLS}{C. Bernard, J. Labrenz, and A. Soni, University of 
              Washington preprint UW/PT-93-06.}
 
\bibitem{HH}{H. Hamber, Phys. Rev. D39 (1989) 896.}

\bibitem{fb2}{ C. Alexandrou, S. G\"usken, F. Jegerlehner, K. Schilling,
               and R. Sommer, Nucl. Phys. B374 (1992) 263.}

\bibitem{GaLe}{see e.g. J. Gasser and H. Leutwyler, Phys. Rep. C87 (1982) 77.}

\bibitem{MiTe} {C. Michael and M. Teper, Nucl. Phys. B314(1989)347.}

\bibitem{UKQCDmg} {G.S. Bali, et al. (UKQCD--coll.), Phys. Lett. B309
                   (1993) 378.}

\bibitem{BLS_wf}{C. Bernard, J. Labrenz and A. Soni,
                 Nucl. Phys. B (Proc. Suppl.) 20 (1991) 488.}

\bibitem{GF11_f}{F. Butler, H. Chen, J. Sexton A. Vaccarino and
                   D. Weingarten,  
               preprint IBM/HET 93-3. We would like to thank D. Weingarten
                 for communicating preliminary numbers to us prior to 
                 publication.}

\bibitem{KS}{The euclidean staggered fermion action is discussed  e.g. in
             H.S. Sharatchandra, H.J. Thun and P. Weisz, Nucl. Phys. B192
             (1981); it is based on the hamiltonian formulation given in
              L. Susskind, Phys. Rev. D16 (1977) 3031.}

\bibitem{Greg}{G. Kilcup, private communication.}

\bibitem{Zstat3} {A. Borelli, R. Frezotti, E. Gabrielli and
                 C. Pittori, CERN preprint TH-6587 (1992).}

\bibitem{SVPW}{E. V. Shuriak, Nucl. Phys. B198 (1983) 83;\\
        M. A. Shifman and M. B. Voloshin, Sov. J. Nucl. Phys. 45 (1988) 292;\\
        H. D. Politzer and M. B. Wise, Phys. Lett. B206 (1988) 681;
             Phys. Lett. B208 (1988) 504;\\
        X. Ji and M.J. Musolf, Phys. Lett. B257 (1991) 409;\\
        D.J. Broadhurst and A.G. Grozin, Phys. Lett. B274 (1992) 421;\\
        M. Neubert, Phys. Rev. D46 (1992) 1076.}

\bibitem{BPHSM}{Ph. Boucaud,  O. Pene, V.J. Hill, C.T. Sachrajda and G.
                Martinelli, Phys. Lett. 220B (1989) 219.}

\bibitem{Eicham}{A. Duncan et al., Nucl Phys. B(Proc. Suppl.)30 (1993) 433;\\
                 A. Duncan, E. Eichten and H. Thacker, Phys. Lett.
                 B303 (1993) 109.}  
                 
\bibitem{Jap}{S. Hashimoto and Y. Saeki, Hiroshima University
         preprint HUPD-9120 and Nucl. Phys. B (Proc. Suppl.) 26 (1992) 381.}

\bibitem{covar}{C. Michael,  Liverpool Preprint LTH
                  321.}

\bibitem{AlEl}{ T. M. Aliev and
        V. L. Eletsky, Sov. J. Nucl. Phys. 38 (1983) 936.}

\bibitem{Nari}{ S. Narison, Phys. Lett. B198 (1987) 104.}

\bibitem{Rein}{ L. J. Reinders, H. Rubinstein and
       S. Yazaki, Phys. Lett. B104 (1981) 305; \\
      Phys. Rep. C127 (1985) 1;\\
      L. J. Reinders, Phys. Rev. D38 (1988) 947.}

\bibitem{DoPa}{ C. A. Dominguez and
           N. Paver, Phys. Lett. B197 (1987).} 

\bibitem{CNP}{P. Colangelo, G. Nardulli and N. Paver in \cite{bfac}.}

\bibitem{EichDa}{E. Eichten in the Proceedings of ``Lattice 93'', 
                  to appear in Nucl. Phys. B
                 (Proc. Suppl.).}

\bibitem{Bpert}{G.Martinelli, Phys. Lett. 141B (1984) 395;\\
                C. Bernard et al., Phys. Rev. D36 (1987) 3224.}

\bibitem{Tsuk}{R. Sommer, C. Alexandrou, S. G\"usken, F. Jegerlehner
         and  K. Schilling, Nucl. Phys. B (Proc. Suppl.) 26 (1992) 387.}

\bibitem{RajBpar}{R. Gupta, D. Daniel, G. Kilcup, A. Patel and S. Sharpe,
                Phys. Rev. D47 (1993) 5113.}

\bibitem{hybrid}{S.J. Perantonis and C. Michael, Nucl. Phys. B 347 (1990)854.}

\bibitem{MichAa}{For a nice review see: 
                 C. Michael, in {\it ``QCD 20 Years Later''}, eds. P.M. Zerwas
                 and H.A. Kastrup, World Scientific, Aaachen 1993.}
  
\bibitem{Bochmass}{M. Bochicchio, G. Martinelli, C. R. Allton, C. T. Sachrajda,
                      and D. B. Carpenter, Nucl. Phys. B 372 (1992)403.}


\bibitem{LuWo}{M. L\"uscher and U. Wolff, Nucl. Phys.~B339 (1990) 222.}

\bibitem{IW}{S.P. Booth et al., Edinburgh preprint 93/525;\\
             C. Bernard, Y. Shen and A. Soni,    preprint BUHEP/93--13.}
\bibitem{BKG}{C. Bernard, P. Hsieh and A. Soni, Washington University
             preprint Wash. U. HEP/93--35
             K.C. Bowler et al., Edinburgh preprint 93/528.}
\bibitem{GBD}{R. Gupta, T. Bhattacharya and D. Daniel,
              Los Alamaos preprint LA UR--93--3580. } 

} 
\end{thebibliography}
\end{document}